\documentclass[vecphys]{svmult} 
 
\usepackage{amssymb}   
\usepackage{makeidx}         
\usepackage{graphicx}        
\usepackage{multicol}        
\usepackage[bottom]{footmisc}
 
\begin{document} 
\newcommand{\beq}{\begin{equation}} 
\newcommand{\eeq}{\end{equation}} 
\newcommand{\beqa}{\begin{eqnarray}} 
\newcommand{\eeqa}{\end{eqnarray}} 
\newcommand{\ds}{{\sf DarkSUSY}} 
   \def\esim{\mathrel{\rlap{\raise2pt\hbox{$\sim$}} 
    \lower1pt\hbox{$-$}}}         
\def\lsim{\mathrel{\rlap{\lower4pt\hbox{\hskip1pt$\sim$}} 
    \raise1pt\hbox{$<$}}}         
\def\gsim{\mathrel{\rlap{\lower4pt\hbox{\hskip1pt$\sim$}} 
    \raise1pt\hbox{$>$}}}         
\newcommand{\bsg}{${\rm b}\to {\rm s}\gamma$} 
\title*{Saas-Fee Lecture Notes: Multi-messenger Astronomy and Dark Matter}
\author{Lars Bergstr\"om}  
\institute{The Oskar Klein Centre\\ 
Department of Physics\\AlbaNova, Stockholm University\\ 
SE-106 91  Stockholm, Sweden\\ 
\texttt{lbe@fysik.su.se}}


\maketitle 

 \setcounter{tocdepth}{5}
 \tableofcontents

\mainmatter

\section{Preamble}\label{sec:preamble} 
Astrophysics, and more specifically astroparticle physics, has been going  
through tremendous progress during the last two decades. Still, one of  
the main problems, that of the nature of the dark matter, remains unsolved.  
With the help of  accelerator experiments (at CERN:s LHC in particular, which started operation in 2010 and which is currently gathering an impressive integrated  
luminosity)  
we could soon hope to get a first indication of the mass scale for the  
new physics that is associated with dark matter. However, to actually  
prove that a particle discovered at accelerators has the right properties  
to constitute the astrophysical dark matter, complementary methods are needed.  
The fact that a candidate for dark matter is electrically neutral (as  
not to emit nor absorb light - that is what we mean with the term ``dark'')   
can plausibly be determined at accelerators. However, the coupling of the dark matter particles  
to other matter needs to be weak, and the lifetime of the  
dark matter particle needs to be at least of the order of the age of the  
universe. This cannot  be tested at accelerators - the dark matter particles would leave the detector in some 100  
nanoseconds. There could be very useful information still gathered at the LHC, as possibly decays of more massive states  
in the ``dark sector''  would be observable, and the missing energy could be estimated. 
 
Fortunately,  through  
observations of various types of messengers - radio waves, microwaves,  
IR, optical and UV radiation, X-rays, $\gamma$-rays  
and  neutrinos, there is great hope that we could get an 
independent indication of the mass scale of dark matter. This variety  
of possible methods of  indirect detection  
methods is a part of  multimessenger astronomy, and it is the  
second way by which we approach the dark matter problem. In particular, for models where the dark matter particles are involved in breaking the electroweak symmetry of the Standard Model, so-called WIMP models (for weakly interacting massive particles), prospects of detection in the near future look promising. We will  
look in some detail on the properties of  WIMP candidates, where that fact that they are massive means that they move non-relativistically in galactic halos, and form so-called  cold dark matter (CDM). One thought earlier that neutrinos could be the dark matter, but they would constitute  hot dark matter (HDM), which is not favoured 
by observations. Due to free-streaming motion, they would only form very large structures first, which then fragment into smaller scales, like galaxies. This scenario does not agree with observations, as it gives too little power on small scales. Of course, one may also consider an inbetween scenario, warm dark matter, usually consisting of having a sterile neutrino (i.e., with no direct Standard Model couplings) in the keV mass region. These may perhaps have some virtue of explaining  possible anomalies in dark matter distribution on the very smallest scales, but reliable methods are so far lacking to probe the dark matter distribution, and its couplings to baryons, on these scales. 
 
As a third approach, ingenious experiments for direct detection  
employing solid state devices, liquid noble gases etc, can be used to  
tell us about other important properties of dark matter, like the  
spin-dependent or spin-independent cross section of dark matter particle  
scattering on nucleons. Once signals start to be found (and there are  
some, however not undisputed ones, already), an exciting puzzle will present itself,  
putting all these pieces of information together. For indirect detection, astrophysical backgrounds that could mask or mimic dark matter signatures will often be a great challenge to overcome. It should therefore be useful to the reader to 
study also the accompanying articles by Felix Aharonian and Chuck Dermer in this volume - not the least to understand the very interesting aspects of those processes in their own right. 
 
In this set of lectures, I will treat all of the dark matter-related aspects in some  
detail, and also cover some other current problems of astroparticle  
physics and cosmology. The  sections in these lectures correspond roughly  
to  the lectures at the Saas-Fee School in Les Diablerets  
in March, 2010, i.e., 
\begin{itemize} 
\item  The particle universe: introduction, cosmological parameters.  
\item  Basic cross sections for neutrinos and $\gamma$-rays; IceCube.  
\item  Density of relic particles from the early universe. 
\item  Dark matter: Direct and indirect detection methods; the galactic  
centre \& other promising DM sources.  
\item  Neutrinos and antimatter from dark matter, Sommerfeld enhancement.  
\item  Supersymmetric dark matter, DarkSUSY. 
\item  Particular dark matter candidates (WIMPS, Kaluza-Klein particles,  
sterile neutrinos\ldots).  
\item  Diffuse extragalactic $\gamma$-rays, Primordial black holes, Hawking  
radiation.  
\item Gravitational waves.  
\end{itemize} 
The order has been slightly changed (cf.~the Table of Contents), and in many cases I have updated the material since the time of the lectures,  
referring to   
important developments (actually, quite a number of them) that have appeared after time of the School. This is of course mandatory  
in a field that evolves so rapidly. For the more basic parts 
of this review, I have relied heavily on the Springer/PRAXIS textbook by Ariel Goobar and myself 
\cite{BGbook}. Also material from various reviews I have written over the last few years \cite{ds,lb_bertone_book,grenoble,cta} has come to use, but also a lot of new material. With these lecture notes, I hope to convey at least some of the excitement I feel for this topic, which relates to some of the outstanding questions still with us in particle physics and cosmology.

\section{The Particle Universe: Introduction}\label{ch:1}

\subsection{Introduction}\label{sec:intro0} 
One of the most impressive achievements of science is the development of 
a quite detailed  
understanding of the physical properties of the universe,  
even at its earliest stages. Thanks to  the fruitful interplay between  
theoretical analysis, 
astronomical observations and laboratory experiments we have today  
very successful `Standard Models' of both particle physics and  
cosmology. The Standard Model of particle physics involves matter  
particles: quarks which always form bound states such as neutrons and 
protons, and leptons like the electron which is charged and  
therefore can make up neutral matter when bound to nuclei  
formed by neutrons and protons. There are 
also neutral leptons, neutrinos, which do not form bound states but 
which play a very important role in cosmology and particle  
astrophysics as we will see throughout these lecture notes. The other important 
ingredients in the Standard Model of particle physics are the  
particles which mediate the fundamental forces: the photon, the  
gluons and the W and Z bosons. 
 
The Standard Model of cosmology is the hot big bang model, 
which states that the universe is not infinitely old but rather came 
into existence some 13.7 billion years ago. There may have been a short  
period with  extremely rapid expansion, inflation, which  
diluted all matter, radiation and other structures  
(like magnetic monopoles) that might  
have existed before inflation. When inflation ended, there was a rapid heating 
(or, thus, rather re-heating) which meant a re-start of expansion,  
now governed by the relativistic degrees of freedom of our universe, i.e.,  
radiation.  The expansion started out in a  
state which after this small fraction of a second was enormously compressed  
and very hot (the relation between the density and the temperature can be  
determined by near-equilibrium thermodynamics in this epoch, when the expansion was ``slow'' and adiabatic). No bound states could exist because of the 
intense heat which caused immediate dissociation even of protons 
and neutrons into quarks if they  were formed in the  quark-gluon 
plasma. Subsequently, the universe expanded and  
cooled, making possible the formation of a sequence of ever more  
complex objects: protons and neutrons, nuclei, atoms, molecules,  
clouds, stars, planets,\ldots. As we will see, the observational  
support for the big bang model is overwhelming, but it contains new elements,
of dark matter and dark energy, that were not entirely expected. The key  
observations are: 
\begin{list}{$\bullet$}{\rightmargin 1cm} 
 \item The present expansion of the universe. 
 \item The existence of the cosmic microwave background radiation,  
 CMBR,  
i.e. the relic radiation from the hot stage of the early universe, and measurements of the temperature variations therein.
\item The presence of related structure in the late-time distribution of galaxies, so-called ``baryon acoustic oscillations" (BAO). 
\item Supernova cosmology that measures the expansion history, with the surprising result that the cosmic expansion is accelerating (Nobel Prize to S. Perlmutter, B. Schmidt and A. Riess, 2011). 
\item The successful calculations of the relative abundance of light 
elements in the universe, which accurately agrees with what would be synthesized in  
an initially hot, expanding universe.
\item The concept of cosmological inflation, which successfully predicted the geometric flatness of the universe, (thus that the average density is near the critical  
density, i.e., $\Omega_{tot}=1$ to an excellent approximation)  and gave an explanation of the form of the nearly scale invariant, gaussian  temperature
fluctuations.
\item The discovery of dark matter, pioneered by Zwicky in the 1930's, has stood the test of time and is now an established piece of the cosmological standard model. This is what the main part of these lecture notes will be about. Dark energy, in its simplest form just a constant vacuum energy, is the other part which explains $\Omega_{tot}=1$  and the accelerated expansion of the universe.
\end{list} 
 
Several of these observations have been awarded the Nobel Prize, the latest thus being the prize for the  
discovery of the accelerated expansion of the universe through supernova observations.

As another piece of evidence in favour of the big bang scenario, can be taken the fact that the oldest  
objects found in the universe -- globular clusters of stars and some  
radioactive isotopes -- do not seem to exceed an age around 13 billion 
years. This gives strong evidence for a universe with a finite age, such as  
the big bang model predicts. 

In some areas, there are new pieces of information to await. For instance, one of the main objectives of the  
Planck satellite, which will present cosmological data in early 2013, is to search for 
non-gaussian features, which could tell us more about the mechanism of inflation.
 
Although there are still many puzzles and interesting details  
to fill in, both in  
the  Standard Model of particle physics and in the big bang model,  
they do remarkably well in describing a majority of all phenomena 
we can observe in nature. Combined, they allow us to follow the history of our  
universe back to only about 10$^{-10}$ seconds after the big bang  
using established physical laws that have been checked in the laboratory.  
Extending the models, there are inflationary  
scenarios that describe the evolution back to  10$^{-43}$ seconds after the  
big bang! 
 
Behind this remarkable success are the theories of General  
Relativity and Quantum Field Theory, which we use in these  
lecture notes.  
However, many fundamental aspects of the laws of  
nature remain uncertain and are the subject of present-day research.  
The key problem is, as it has been for many decades,  
to find a valid description of quantized gravity, 
something which is needed to push our limit of knowledge even closer  
to (and maybe eventually explaining?) the big bang itself. 
 
In this section we will review some of the most striking observational  
facts about our universe. 
\subsection{Basic Assumptions} 
A basic concept in modern cosmology is that of the ``Copernican principle", i. e. the supposition that the universe 
on the average is homogeneous and isotropic. Although this is definitely not true on galactic scales and smaller, the distribution 
of matter seems to become more and more smooth on large scales, and on the largest scales we can observe, probed by the CMBR, 
isotropy and homogeneity seems to be fulfilled. The inhomogeneities seem to be $10^{-5}$ or smaller, apart from a dipole component in the CMB radiation, which however has a natural  
interpretation in terms of motion of our galaxy towards other massive galaxies. Given isotropy and homogeneity, the most general line element is the one found  
by Friedmann, Lema\^\i tre , Robertson and Walker (FLRW), 
$$ds^2=dt^2-a^2(t)\left({dr^2\over  
 1-kr^2}+r^2d\theta^2+r^2\sin^2\theta  
 d\phi^2\right).$$ 
Measurements on the cosmic microwave background gives (and inflationary theories predicted) $k = 0$, i.e., a geometrically flat universe on large scales, to good accuracy. (There have been suggestions that some of the features of the homogeneous and isotropic model can be alternatively explained if we live in an inhomogeneous universe with large ``bubbles'' of atypical density. Although a logical possibility, combined constraints from galaxy surveys, supernova data, and the CMBR mean that we would have to live at a fine-tuned location near the centre of such a bubble \cite{edvard}. We will thus not consider these scenarios.) 
 
The scale factor $a(t)$ follows equations first derived by Friedmann  from Einsteins equations in general relativity: 
$$   H(t)^2\equiv   \left({\dot a\over a}\right)^2={8\pi G_N\over 3}\rho_{tot}.\label{eq:friedmann}$$ 
Here $G_N$ is Newton's gravitational constant, and $\rho_{tot}$ is the total average energy density of the universe. The time-dependent 
Hubble parameter $H(t)$, has a value today which is known as the Hubble constant, 
$$   H(t_{0})\equiv H_0=h\cdot100\ {\rm kms}^{-1}{\rm Mpc}^{-1}.$$ 
This defines the dimensionless quantity $h\sim 0.7$, which has to be given by measurement. 
  
The equation which determines the acceleration of the scale factor is also derived from Einstein's equations: 
$$     {2\ddot a\over a}+\left({\dot a\over a}\right)^2=-8\pi  
G_Np,$$ 
with $p$ being the total pressure. 
\subsection{Energy and  Pressure}  
In general, there are several components contributing to the energy density,
at least matter, radiation and dark energy, where the simplest possibility is a constant vacuum energy - the modern version of Einstein's cosmological constant: 
$$    \rho_{tot}=\rho_{m}+\rho_{rad}+\rho_\Lambda.$$ 
For an isotropic and homogeneous model, the non-zero elements of the energy-momentum tensor are 
\begin{eqnarray*} 
T^{ij}&=p\delta_{ij}\\ 
T^{i0}&=0 \\ 
T^{00}&=\rho_{tot} 
\end{eqnarray*} 
and there is for each component contributing to $p$ and $\rho_{tot}$ a relation 
$$p_i=w_i\cdot\rho_i$$ 
called the equation of state, which enables one to make predictions for the time evolution of the expansion of the universe and for the relative weights of the different energy components. For non-relativistic matter, the pressure is proportional $(v/c)^2$, and therefore negligible, $p=0$, i.e. $w_M=0$. For radiation on the other hand, $p=\rho/3$, so $w_R=1/3$. What is the equation of state for vacuum energy? This is easy to motivate from symmetry reasons (as was done already by Lema\^\i tre  in the 1930’s). The energy momentum tensor has to be proportional to the only available rank-2 tensor in empty space-time, namely the Minkowski metric tensor in the cosmic rest frame: 
$$ T^{\mu\nu}_\Lambda= \rho_\Lambda\left( 
\begin{array}{cccc} 
        1 & 0 & 0 & 0  \\ 
        0 & -1 & 0 & 0  \\ 
        0 & 0 & -1 & 0  \\ 
        0 & 0 & 0 & -1 
\end{array}\right)=\left( 
\begin{array}{cccc} 
        \rho_\Lambda & 0 & 0 & 0  \\ 
        0 & -\rho_\Lambda & 0 & 0  \\ 
        0 & 0 & -\rho_\Lambda & 0  \\ 
        0 & 0 & 0 & -\rho_\Lambda 
\end{array}\right).  $$ 
This form is thus dictated by the requirement of Lorentz invariance. Comparing with the general form of the energy-momentum tensor which has $\rho$ and $þ$ in the diagonal, 
we thus see that the equation of state is  $p=-\rho$, i.e., $w_\Lambda=-1$. 
The vacuum energy thus acts as a fluid with negative pressure.

\subsection{Contributions to Vacuum Energy}  
How do we describe the contents of the universe, including vacuum energy? Based on its success in particle physics, we try to do it by using quantum field theory, with its particles and fields.
A field is a dynamical quantity which is defined in all points of space and at all times.  
Particles are the lowest excitations of the fields. A particle is characterized by the mass $m$, spin $s$, charge $Q$, and maybe other internal quantum numbers.

The lowest excitations of the field, carrying energy $E$ and three-momentum $p$ can be quantized as harmonic oscillators fulfilling, in the cosmic rest frame (the reference frame where the CMBR looks maximally isotropic), the mass shell condition 
$$p_\mu p^\mu = m^2,$$ where the four momentum $$p^\mu =(E,{\mathbf p})$$ and $$p_\mu =(E,{\mathbf -p}).$$ For each possible momentum mode, there will, as for the simple quantum mechanical harmonic oscillator, be a zero-point energy 
$$ E_i=\omega(p_i)\left( n+\frac{1}{2}\right)_{n=0}= 
\sqrt{p_i^2+m^2}\left( n+\frac{1}{2}\right)_{n=0}=\frac{1}{2}\sqrt{p_i^2+m^2}.$$ 
 
However, for a given field, these have to be summed for all modes, meaning that there will be a huge zero-energy density 
$$ \rho_\Lambda= \frac{1}{2}\frac{1}{(2\pi)^3}\int d^3 p\sqrt{p^2+m^2}.$$

The highly divergent integral has to be cut-off at some large energy scale, and the first guess is the Planck mass, thus 
$$ \rho_\Lambda= \frac{1}{2}\frac{1}{(2\pi)^3}\int^{m_{Pl}} d^3 p\sqrt{p^2+m^2}\sim m_{Pl}^4.$$ 
Unfortunately, this is orders of magnitude too large, and is the same disastrous result one would get by using simple dimensional analysis. Namely, 
what is the natural scale of $\rho_\Lambda$? We see here that it is governed by the cut-off mass scale when new physics appears, and dimensional analysis gives that in units where $c=1$ so that length is proportional to an inverse mass, energy per unit volume becomes $[\rho_\Lambda] = [M^4]$. The only mass scale in gravity is $m_{Pl}$, thus  
$$\rho^{th}_\Lambda\sim m_{Pl}^4.$$ 
Unlike other guesses in physics based on dimensional analysis, this is a terrible prediction. The present-day vacuum energy density of the universe is given by measurements of supernovae and the CMBR and is  (using $k=0$) 
$$\rho^{obs}_\Lambda\sim 10^{-122}m_{Pl}^4 << m_{Pl}^4\sim \rho_\Lambda^{th}.$$ 
 
To go back to our field theory result, the zero-point energy is really a consequence of the quantum mechanical commutator between the field and its canonical momentum. However, for  fermions, anticommutators are used, meaning the sign of the vacuum energy is changed. So, introducing the fermion number $F = 1$ for fermions, $F = 0$ for bosons, one gets 
 
$$ \rho_\Lambda = \sum (-1)^F\frac{1}{2}\frac{1}{(2\pi)^3}\int^{m_{Pl}} d^3 p\sqrt{p^2+m^2}.$$ 
 
Remarkably, if there are as many fermionic degrees of freedom as bosonic, and they pairwise have the same mass, the vacuum energy would vanish. Examples of theories having this property are supersymmetric theories, with unbroken supersymmetry. 
However, since we do not see 0.511 MeV scalar electrons (selectrons), supersymmetry has to be broken. Therefore large effects of the zero-point energy remain, and $\rho_\Lambda\sim m_{SUSY}^4$ with $m_{SUSY}$ (1000 GeV, say) the scale of SUSY breaking. Better, but still enormously much ”too high”. 
 
In summary, we have encountered one of the most severe problems of cosmology and particle astrophysics: Why is the cosmological constant so small, but still not zero? (By the way, nobody found a good reason that it should be exactly zero, anyway\ldots) Supersymmetry alleviates the problem somewhat, but since supersymmetry is broken there remains large mass terms still giving a value some 50-60 orders of magnitude different from the observed value. 
 
In cosmology the cosmological constant has a dramatic effect. Since it is related to the energy density of the vacuum, and the vacuum is growing in size due to the expansion, it will eventually dominate completely.  
Matter is on the other hand more and more diluted and becomes less and less important, and radiation is also diluted plus red-shifted:  
$$\rho_m\sim(1+z)^3,\ \ \ \rho_r\sim (1+z)^4,\ \ \ \rho_\Lambda\sim (1+z)^0.$$ 
We see that in the early universe (large redshifts), vacuum energy was irrelevant. Today matter and vacuum energy are almost equal (why now?). In the future, the expansion rate will grow exponentially, as we will see in the section on inflation, Sec.~\ref{sec:inflation}.   
 
To explain the smallness of $\Lambda$ some people resort to (almost) non-scientific reasoning: the anthropic principle, or the landscape of string theory vacua. There the argument goes roughly like this: There exist an amazingly large number of different vacua, i.e., ground states, of the theory, and maybe all of these are realized somewhere in nature. But of course, those with very large values of $\Lambda$ would accelerate so quickly that structure would not form in the universe and therefore no life could appear. But since we exist, we have to do so in one of the very few universes where life did evolve. Of course, this sounds more like postdicting the properties of our universe rather than predicting them, which perhaps just shows the desperation in front of the problem of the size of the cosmological constant. 
 
Let us have another look at Planck-mass phenomena. Consider the scattering of a photon on an electron, Compton scattering (we will treat this in detail later, see Fig.~\ref{fig:thomson}). 
The relation between the incident and outgoing wavelength as a function of scattering angle is given by 
$$\lambda^\prime - \lambda= \frac{2\pi\hbar}{m_ec}\left(1-\cos\theta\right)=\frac{2\pi}{m_e}\left(1-\cos\theta\right)\equiv \lambda_c\left(1-\cos\theta\right).$$ 
 
Here $\lambda_c$ is called the Compton wavelength of the particle 
(the electron in this case). We will see in section~\ref{ch:8} the expression for the Schwarzschild radius (the radius which marks the limit of where light can leave the black hole) 
 
$$r_s=\frac{2G_NM}{c^2}=2G_NM$$ 
(we use here and onwards units such that $c=\hbar=1$). Thus, the Compton radius decreases with mass, but the Schwarzschild radius increases with mass. When are the two equal,  
i.e., how big must the mass be for the Compton radius to be smaller than the Schwarzschild radius? This is when quantum gravity should be important.  
(All details of particle properties are smeared out by quantum fluctuations on the order of the Compton wavelength or less, so for $\lambda_c > r_s$ the black hole properties should be unnoticeable.) We see 
$$\frac{\lambda_c}{r_S}=\frac{\pi}{G_NM^2}\sim \frac{m_{Pl}^2}{M^2}.$$ 
Thus, when the mass of an elementary particle is larger than the Planck mass, its Compton radius is smaller than its Schwarzschild radius, which implies that we need quantum gravity! None exists yet, but perhaps string theory is the best bet for such a fundamental theory at the Planck scale? 
For an electron, $\lambda_c/r_S\sim 10^{45}$, so quantum gravity effects are completely negligible at the particle level. The same is true for all other Standard Model particles. 
 
\subsection{Summary of Observations}  
To end this section where the main theoretical lines for describing the universe have been laid out, we summarize what we know about the cosmological parameters of the universe from the impressive recent measurements. 
Analyses combining high-redshift supernova luminosity distances, 
microwave background fluctuations (from the satellite WMAP) and baryon acoustic oscillations (BAO) in the galaxy distribution give tight constraints \cite{wmap10} on the present mass density of matter in the universe. This is usually expressed in the ratio   
 
$$\Omega_M=\rho_M/\rho_{\rm c},$$  
 
normalized to the critical density,  
 
$$\rho_{\rm c}=3H_0^2/(8\pi G_N)=h^2\times 1.9\cdot 10^{-29} \ {\rm 
g\, cm}^{-3}.$$ 
 
The value obtained  for the 7-year WMAP data\cite{wmap10} for cold dark matter for the (unknown) particle $X$ making up the dark matter is $\Omega_Xh^2 = 0.113\pm 0.004$, which is around 5 times higher  than the value obtained for baryons, $\Omega_Bh^2 = 0.0226\pm 0.0005$.  Here $h=0.704\pm 0.014$ is the derived \cite{wmap10} present value of the Hubble constant in units of 
$100$~km~s$^{-1}$~Mpc$^{-1}$.  
In addition, the WMAP data is consistent within a percent with a flat universe ($\Omega_{\mathrm tot}=1$)and a value for the dark energy component, e.g. the cosmological constant $\Lambda$, of $\Omega_\Lambda = 0.73\pm 0.02$.    
 
One of the main problems for cosmology and particle physics is to explain the measured density of dark matter, and to give candidates for the identity of the dark matter particles. The fact that dark matter is definitely needed on the largest scales (probed by WMAP), on galaxy cluster scales (as pointed out by Zwicky already in 1933 \cite{zwicky}, and verified by gravitational lensing and the temperature distribution of X-ray emitting gas) all the way down to the smallest dwarf galaxies, means that solutions based on changing the laws of gravity seem less natural. In particular, the direct empirical proof of the existence of dark matter given by the "Bullet Cluster" \cite{Clowe:2006eq} is very difficult to circumvent, as the X-ray signal from the baryonic matter and the gravitational lensing signal from dark matter are clearly separated. 
 
Although the existence of a non-zero cosmological constant (or some similar form of dark energy)  in the present-day universe came as a big surprise to most cosmologists and particle physicists, the most successful models of evolution in the universe do make use of a similar effect in models of inflation, as we will see in the next section.

\section{Relic Density of Particles}\label{ch:3} 
There are several important examples of freeze-out in the early universe, for 
instance at the synthesis of light elements one second to a few 
minutes after the big bang, and the microwave photons from the  
``surface of last  
scattering'' several hundred thousand years later. 
Before we calculate freeze-out,  
it is convenient to introduce a formalism which 
considers freeze-out in general: that is, what happens when a particle 
species goes out of equilibrium. A rigorous treatment has to be based 
on the Boltzmann transport equation in an expanding background, but here 
we give a simplified treatment (see, for example \cite{BGbook}
for a more complete discussion). 
 
There are several different contributions to $\Omega={\rho\over \rho_{\rm c}}$, like  
radiation $\Omega_{R}$, matter $\Omega_{M}$ and vacuum energy  
$\Omega_{\Lambda}$.  
 
The equations of motion for the matter in the universe are given by 
the vanishing of the covariant divergence of the energy-momentum tensor 

\beq  
T^{\alpha\beta}_{\ \ ;\beta}= 0 
\eeq 
This gives, for the FLRW metric,  
\beq\label{eq:ddt} 
{d\over dt}(\rho a^3) = -p{d\over dt}a^3 
\eeq 
which shows that the change of energy in a comoving volume element is 
equal to minus the pressure times the change in volume.  
This can be rewritten as  
\beq \label{eq:a3} 
a^3{dp\over dt}={d\over dt}[a^3(\rho + p)] 
\eeq 
which  can be interpreted as a conservation law for the  
entropy in a volume $a^3(T)$.
For radiation, where $p = \rho/3$, (\ref{eq:ddt}) gives  
$\rho\sim a^{-4}$.  
Note that all particles fulfilling $mc^2\ll k_BT$ have the  
equation of state of radiation. 
 
The Friedmann equation is 
\beq\label{Fried}
H^2(t) = {8\pi G_N\rho\over 3} 
\eeq 
where as a good approximation only the relativistic species contribute 
appreciably to $\rho$.  
Note that the Hubble parameter $H(t)$ has units of 1/(time).  
This means in our units that it has  
dimensions  
of mass.  
The age of the universe at a given time $t$ is simply of the order of $H^{-1}(t)$, at least when the  
scale factor increases as a power of $t$. 
 
We now treat schematically the thermodynamics of the expanding universe.  
We assume, which is true if reactions between different species of particles are rapid enough, that we can use the thermodynamical  
quantities, 
temperature $T$, pressure $p$, entropy density $s$, and other 
quantities, at each time $t$ to 
describe the state of the universe. The  constituents have 
number density $n$ and typical relative velocities $v$, and  
scattering or annihilations cross-section $\sigma$, 
 meaning that the interaction rate per particle 
 $\Gamma$ is given by $$\Gamma = n\sigma v.$$  
The  
condition that the interactions maintain equilibrium is that the interaction 
rate is larger than the expansion rate of the universe: 
\beq 
\Gamma \gg H 
\eeq 
Typically, the number density of particles decreases faster with  
temperature 
and therefore with time than the Hubble parameter does. This means that  
at  
certain 
epochs some of the particle species will leave thermodynamic equilibrium. Their number 
density will be ``frozen'' at some particular value which then only 
changes through the general dilution due to the expansion. This ``freeze-out'' of particles is an important mechanism which explains the particle content of the universe we observe today. 
 
\newcommand{\pp}{{\bf p}} 
 
Using relativistic statistical mechanics in the cosmic rest frame, 
the distribution 
function $f_i(\pp)$ for particle species of type $i$ is 
\beq\label{MB} 
f_i(\pp)={1\over e^{{(E_i-\mu_i)\over T}}\pm 1}\label{eq:distfunc} 
\eeq 
with $E_i = \sqrt{\pp^2+m_i^2}$  the energy, $\mu_i$ is the chemical  
potential 
and $T$  the temperature (we put $k_B =1$). The minus sign  
is 
for particles that obey Bose-Einstein statistics (bosons) and the plus sign 
is for particles obeying the exclusion principle and therefore Fermi-Dirac 
statistics (fermions). To a good approximation 
 the chemical potentials can be neglected in 
the very early universe.

We denote by $g_i$  
the number of internal degrees of freedom of 
 particle $i$.  
  The photon has two polarization states and therefore 
$g_\gamma = 2$. The neutrinos only have one polarization state, giving 
$g_\nu =1$, electrons and muons have $g_{e,\mu}=2$ (and the same  
numbers for the antiparticles). 
 
With these definitions, the number density for species $i$ is 
\beq\label{eq:equil} 
n_i = {g_i\over (2\pi)^3}\int f_i(\pp)d^3p,\label{eq:numbdens} 
\eeq 
and its energy density is 
\beq\label{eq:rhoi} 
\rho_i={g_i\over (2\pi)^3}\int E_i(\pp)f_i(\pp)d^3p. 
\eeq 
The expression for the   pressure 
 is  
\beq\label{eq:pp} 
p_i={g_i\over (2\pi)^3}\int {|\pp|^2\over 3E_i(\pp)}f_{i}(\pp)d^3p. 
\eeq 
 
In the nonrelativistic limit $T/m \ll 1$ we can solve the integrals 
analytically, and the result both for Fermi-Dirac and Bose-Einstein  
particles is 
\beq 
n_{NR}=g_i\biggl({mT\over 2\pi}\biggr)^{{3\over 2}}e^{-{m\over  
T}},\label{eq:nnr} 
\eeq 
\beq 
\rho_{NR}=m\cdot n_{NR}, 
\eeq 
and 
\beq 
p_{NR}=T\cdot n_{NR} \ll \rho_{NR} 
\eeq 
For nonrelativistic matter, $\langle E\rangle = m + 3T/2$.  
 
In the ultrarelativistic approximation, $T/m \gg 1$, the integrals 
can also be performed with the results 
\beq\label{BE} 
\rho_{R}={g_i\over 6\pi^2}\int_0^\infty {E^3dE\over e^{E\over T}\pm1}= 
\cases{{\pi^2\over 30}g_iT^4,&Bose-Einstein\cr 
{7\over 8}\biggl({\pi^2\over 30}g_iT^4\biggr),&Fermi-Dirac,} 
\eeq 
\beq 
n_{R}=\cases{{\zeta(3)\over \pi^2}g_iT^3,&Bose-Einstein\cr 
{3\over 4}\biggl({\zeta(3)\over \pi^2}g_iT^3\biggr),&Fermi-Dirac,}\label{eq:eq} 
\eeq 
with $\zeta(x)$ is the Riemann zeta function, $\zeta(3) = 1.20206...$ 
The average energy  
$\rho/n$ for a relativistic particle is  
\beq 
\langle E\rangle_{BE} \sim 2.7T\label{eq:avgbe} 
\eeq 
and 
\beq 
\langle E\rangle_{FD} \sim 3.15T\label{eq:avgfd} 
\eeq 
For photons, with the mass $m_{\gamma}=0$, and $g_\gamma=2$, 
the expression for $\rho_\gamma(T) \sim T^4$ is the famous 
 Stefan Boltzmann 
law for electromagnetic black-body radiation. 
 
The total 
contribution to the energy and number density of all kinds of particles in 
the early universe is to a good approximation (since 
the energy and number density of a nonrelativistic 
species is exponentially suppressed),  
 
\beq\label{eq:rhor} 
\rho_{R}(T) = {\pi^2\over 30}g_{\rm eff}(T)T^4 
\eeq 
\beq\label{eq:pr} 
p_{R}(T) = {1\over 3}\rho_{R}(T)={\pi^2\over 90}g_{\rm eff}(T)T^4 
\eeq 
where  $g_{\rm eff}(T)$ counts the  
total number of internal degrees of freedom  
(such as spin, colour, etc.) of  
the 
particles                                                                               whose mass 
fulfill $m\ll T$, and which are in thermodynamic equilibrium with the ``primordial cosmic soup'' of particles in the early universe. The expression for $g_{\rm eff}(T)$ has the factor $7/8$ for  
fermions. 
 
As an example, we calculate $g_{\rm eff}(T)$ for a temperature of, say, 
1 TeV 
when all the particles of the Standard Model were relativistic and 
in thermal equilibrium. The total number of internal degrees of freedom of  
the fermions is 90 and for the gauge and Higgs bosons 
28, so the total expression for $g_{\rm eff}$ is 
\beq 
g_{\rm eff}(T\sim1\,{\rm TeV})=28+{7\over 8}\cdot 90= 106.75 
\eeq

If we insert the expression for the energy density into the Friedmann equation (\ref{Fried}) we  
get for the radiation-dominated epoch in the early universe 
\beq 
H^2={8\pi G\over 3}\rho_{R}={8\pi G_N\over 3}{\pi^2\over 30}g_{\rm eff}T^4= 
2.76{g_{\rm eff}T^4\over m_{Pl}^2} 
\eeq 
or 
\beq\label{H} 
H=1.66\sqrt{g_{\rm eff}}{T^2\over m_{Pl}}\label{eq:exprateth} 
\eeq 
This is a very important formula governing the physics of the  
early universe. 
 
For radiation domination, it can be shown that 
\beq 
a(t)\sim \sqrt{t} 
\eeq 
deriving from the equation of state $p=\rho/3$. For matter domination, that is, for  
$p\sim 0$, one has 
\beq 
a(t)\sim t^{2\over 3}. 
\eeq 
 
So for radiation domination,  
\beq 
H={\dot a\over a} = {1\over 2t} 
\eeq 
and the  
time temperature relation becomes 
\beq\label{eq:timevstemp} 
t=0.30{m_{Pl}\over \sqrt{g_{\rm eff}}T^2}\sim \left({\ {\rm 1\ MeV}\over T} 
\right)^2\  
{\rm  
sec} 
\eeq 
 
We now have to determine which particles are in thermal equilibrium at 
a given temperature, so that we can calculate $g_{\rm eff}(T)$. The entropy 
$S(V,T)$ is introduced through 
\beq\label{eq:ds} 
dS(V,T)={1\over T}\left[d(\rho(T)V)+p(T)dV\right] 
\eeq 
this gives (see \cite{BGbook}) 
\beq\label{eq:ss} 
S(V,T)={V\over T}(\rho(T)+p(T)) 
\eeq 
and from the conservation of the energy-momentum tensor follows
\beq\label{eq:momcons} 
{d\over dt}\biggl({a^3\over T}[\rho(T)+p(T)]\biggr) =0. 
\eeq 
Identifying the volume $V$ with $a^3(t)$ and comparing with (\ref{eq:ss})  
we find the law of conservation of entropy in the  
volume 
$a^3(t)$. Sometimes it is more useful to work with the entropy density 
$s(T)$ rather than the total entropy $S(V,T)$ within the volume $V$. The  
definition is thus: 
\beq\label{eq:smalls} 
s(T)\equiv {S(V,T)\over V}={\rho(T)+p(T)\over T} 
\eeq 
 
In the early universe, both the energy density and the pressure  
were dominated by relativistic particles with the equation of state 
$p=\rho/3$. Using (\ref{eq:smalls}) and the relativistic expressions for the energy density and the pressure, gives 
density 
$s$ 
\beq 
s={2\pi^2\over 45}g^s_{\rm eff}T^3 
\eeq 
where $g^s_{\rm eff}$ is defined in a similar way to $g_{\rm eff}$. 
 
Since $s$ and $n_\gamma$ both  vary as $T^3$ there is a simple  
relationship between them. With 
\beq 
n_\gamma = {2\zeta(3)\over \pi^2}T^3 
\eeq 
we find
\beq 
s={\pi^4\over 45\zeta(3)}g^s_{\rm eff}n_\gamma\sim 1.8g^s_{\rm eff}n_\gamma 
\eeq 
 
Following \cite{BGbook} we now consider a case of great interest for the dark matter problem. 
Suppose that there exists some kind of unknown particle $\chi$, with 
antiparticle $\bar\chi$, that can annihilate each other and be pair created 
through processes $\chi +\bar\chi\leftrightarrow X+\bar X$, where $X$ stands 
for any type of particle to which the $\chi$s can  
annihilate. The supersymmetric neutralino is a charge-less, Majorana particle and is its  
own antiparticle (just as the photon is its own antiparticle). The  
formalism is very similar in this case. In particular, a neutralino  
can annihilate with another neutralino giving other,  
non-supersymmetric particles in the final state. We further 
assume (which is usually an excellent approximation in the early universe) 
that the $X$ particles have zero chemical potential and that they 
are kept in thermal equilibrium with the photons and the other light particles 
when the temperature was much larger than the rest mass of $\chi$ (the $X$ particles can be quarks, leptons etc.) 
 
How will the number density $n_\chi$ evolve with time (and therefore with 
temperature)? It is clear that in exact thermal equilibrium the number of 
$\chi$ particles in a comoving volume $N_\chi=a^3n_\chi$  
will be given by the equilibrium value $n_\chi^{EQ}(T)$ (see  
(\ref{eq:eq})).  
(In exact thermal equilibrium the rate for the process 
$\chi +\bar\chi\leftrightarrow X+\bar X$ is the same in both directions.) 
If at a given temperature $T$ the  number density $n_\chi(T)$ is larger than the equilibrium 
density the reaction will go faster to the right. Thus, the $\chi$ 
particles will annihilate faster than they are created. The depletion rate 
of $\chi$ should be proportional to $\sigma_{\chi\bar\chi\to X\bar X}|{\bf v}| 
n_\chi^2$ (quadratic in the density, since it should be proportional to 
the product of $n_\chi$ and $n_{\bar\chi}$, and these according to our assumptions are equal).  
However, $\chi$ particles are also created by the inverse process, with  
a 
rate proportional to $(n_{\chi}^{EQ})^2$. We have 
thus heuristically derived the basic equation that governs the number density
of species $\chi$, also as it starts to depart from  
equilibrium: 
\beq 
{dn_\chi\over dt}+3Hn_\chi=-\langle\sigma_{\chi\bar\chi\to X\bar X}|{\bf v}| 
\rangle [n_\chi^2-(n_\chi^{EQ})^2]. 
\eeq 
The left-hand side derives from ${1\over a^3}{d\over dt}[n_\chi a^3]$, 
and the term  
proportional to $3H$ expresses the dilution 
that automatically comes from the Hubble expansion. 
The quantity $\langle\sigma_{\chi\bar\chi\to X\bar X}|{\bf v}| 
\rangle$ stands for the thermally averaged cross section times velocity. 
This averaging is necessary, since 
the annihilating particles have random thermal velocities and directions. 
Summing over all possible annihilation channels gives 
\beq 
{dn_\chi\over dt}+3Hn_\chi=-\langle\sigma_{A}|{\bf v}| 
\rangle [n_\chi^2-(n_\chi^{EQ})^2], 
\eeq 
where $\sigma_A$ is the total annihilation cross section. 
 
Using the time-temperature relation  equation\,~(\ref{eq:timevstemp})  
(for radiation dominance) 
\beq 
t=0.30{m_{Pl}\over T^2\sqrt{g_{\rm eff}}}, 
\eeq 
this can be converted to an equation for how $n_\chi$ evolves with decreasing temperature. Introducing the dimensionless variable $x=m_\chi/T$, and 
normalizing for convenience $n_\chi$ to the entropy density, 
\beq 
Y_\chi={n_\chi\over s},
\eeq 
we find after some intermediate steps (that you may want to reproduce yourself)
  
\beq\label{eq:dm} 
{dY\over dx}=-{m_\chi m_{Pl}c_{\rm eff}\over x^2}\sqrt{{\pi\over 45}} 
\langle\sigma_A|{\bf v}|\rangle (Y_\chi^2-(Y_\chi^{EQ})^2) 
\eeq 
where 
\beq 
c_{\rm eff}={g^s_{\rm eff}\over \sqrt{g_{\rm eff}}}. 
\eeq 
After rearranging we find, 
\beq\label{eq:Y} 
{x\over Y_\chi^{EQ}}{dY\over dx}=-{\Gamma_A\over H}\biggl[\biggl( 
{Y_\chi\over Y_\chi^{EQ}}\biggr)^2-1\biggr], 
\eeq 
where $\Gamma_A =n_\chi^{EQ}\langle\sigma_A|{\bf v}|\rangle$. 
This equation can be solved numerically with the boundary condition that 
for small $x$, $Y_\chi\sim Y_\chi^{EQ}$. This is because at high temperatures, much larger than $m_\chi$, the 
$\chi$ particles were in thermal equilibrium with the other particles. 
We see from (\ref{eq:Y}) that the evolution conveniently is governed by the 
factor $\Gamma_A/H$, the interaction rate divided by the Hubble expansion rate. 
 
The solutions to these equations have to be obtained numerically in the  
general case to find the temperature $T_f$ and therefore the value of  
$x_{f}$ of freeze-out and the 
asymptotic value $Y_\chi (\infty)$ of the relic abundance of the 
species $\chi$. There are, however, some simple limiting cases.  
If the species $\chi$ is relativistic at freeze-out, then 
$Y_\chi^{EQ}$ is not changing with time during the period of freeze-out, and 
the resulting $Y_\chi (\infty)$ is just the equilibrium value at freeze-out, 
\beq 
Y_\chi (\infty)=Y_\chi^{EQ}(x_f)={45\zeta(3)\over 2\pi^4}  
{g_{eff}\over g^s_{\rm eff}(x_f)} 
\eeq 
where $g_{eff}=g$ for bosons and $3g/4$ for fermions. 
A particle that was relativistic at freeze-out is called a hot relic, or hot dark matter. 
A typical example is the neutrino. The present mass density of a hot relic 
with mass $m$ is  
\beq 
\Omega_\chi h^2=7.8\cdot 10^{-2}{g_{eff}\over g^s_{\rm eff}(x_f)} 
\biggl({m_\chi\over 1\ {\rm eV}}\biggr) 
\eeq 
Note that today the motion of a particle with 
mass greater than the small number $T_0$ = 2.73 K = $2.4\cdot 10^{-4}$ eV is 
of course non-relativistic and therefore the contribution to the energy 
density is dominated by its rest mass energy. 
A Standard Model neutrino has $g_{eff} = 2\cdot 3/4 = 1.5$ and decoupled at a  
few MeV  
 when $g^s_{\rm eff}=g_{\rm eff}=10.75$.  We find 
\beq 
\Omega_{\nu\bar\nu}h^2 = \frac{\sum_i m_{\nu_i}}{\left(93\ {\rm eV}\right)}.
\eeq 
As we will see, present estimates of the neutrinos masses, based on the observation of neutrino oscillations, give a sum much less than 1 eV, which means 
that neutrinos are not the main form of dark matter. On the other hand, we are now rather certain that they do contribute a small fraction of nonbaryonic dark mtter. In a sense a (small) part of the dark matter problem is solved!

\subsection{The WIMP Miracle}
This analysis has been valid for hot relics, or hot dark matter.  
For cold relics 
(particles that were non-relativistic at freeze-out) the equation 
(\ref{eq:Y}) has to be found numerically. Then one finds that for massive particles in the mass range between, say, 10 GeV and a few TeV, $x_f \sim 1/20$, and moreover to a good approximation the relic density only depends on the cross section times velocity,
\beq
\Omega_X h^2\simeq 0.11\times\frac{2.8\cdot 10^{-26}\ {\rm cm}^3{\rm s}^{-1}}{\langle\sigma_A|{\mathbf v}|\rangle}.\label{eq:wimp}
\eeq
 Another striking result is that, if one gives typical gauge gauge couplings to the particle $X$, and a mass of typical weak interaction magnitude (100 -- 300 GeV, say), then $\langle \sigma_Av\rangle$ is such that the resulting $\Omega_X h^2\sim 0.11$. This is the essence of what is sometimes called the ``WIMP miracle''.
 
As can be understood, the value of $x_f$ when $Y_\chi$ leaves the equilibrium 
curve is lower for a smaller cross section $\sigma_A$. This is because of the fact that in thermal equilibrium, massive particles will eventually be exponentially suppressed. That is, more weakly 
interacting particles decouple earlier, and since the equilibrium curve for 
a nonrelativistic species drops  fast with increasing $x$,  
more weakly coupled particles will have a higher relic abundance. 
 
Going through the numerical analysis one finds that a hypothetical neutrino 
with mass $m_\nu\sim 3$ GeV would also  have about the right mass to close the  
universe. On the other hand, the range between 90 eV and 3 GeV is  
cosmologically disallowed for a stable neutrino (it would overclose the universe). 
There are strong arguments from large-scale structure formation that favour 
cold relics over hot relics, so such a neutrino would in principle be a good 
dark matter candidate. Data from the LEP accelerator at {\sc CERN} did, 
however, exclude any ordinary neutrino with a mass in the GeV range.  

\subsection{Coannihilations}
There are instances when the simple treatment discussed here has to be 
improved. One example is for instance the possibility that entropy may be generated by other particles than those of the Standard Model, before, at, or after decoupling. Another example, which for instance appears in some supersymmetric models, is that of coannihilations. This was first discussed in \cite{coannih},
here we follow the more detailed treatment in \cite{coann}.

We will here outline the procedure developed in \cite{coann,ds} which is 
used in \ds\ \cite{joakim_ds}. For more details, see \cite{coann,ds}. \ds\ is a free {\sc Fortran} package that can be used to compute a variety of dark matter related
quantities, such the relic density and the scattering and annihilation rates to many different channels. It was developed for computations in the Minimal Supersymmetric extension to the Standard Model (MSSM), but it is modular and can be adapted to most WIMP models.

We consider annihilation of $N$  particles with mass 
$m_{i}$ and internal degrees of freedom $g_{i}$.  For convenience, we may order them such that 
$m_{1} \leq m_{2} \leq \cdots \leq m_{N-1} \leq m_{N}$. For the 
lightest particle (which is the dark matter candidate, if a symmetry is guaranteeing the stability, like what is called $R$-parity for supersymmetry, see 
section \ref{ch:susy}), we use both the notation $m_1$ and $m_\chi$. 
 
All heavier particles will 
eventually decay to the lightest, stable, and therefore we add the number densities up, 
$$ n= \sum_{i=1}^N n_{i}.$$ 
The scattering rate of particles on  
particles in the 
thermal background ``soup'' is generally much faster 
than the annihilation rate, 
since the background particle densities of Standard Model particles, $n_{SM}$ is much larger than each of 
the  particle densities in the dark sector $n_i$. The important SM particles
are, as we have seen, those that are relativistic and cold dark matter 
particles (WIMPs) are nonrelativistic, and thus suppressed by 
the Boltzmann factor.
Thus, the $n_i$ distributions remain in 
thermal equilibrium during their (``chemical'') freeze-out. 

We then get
\begin{equation} \label{eq:Boltzmann2} 
   \frac{dn}{dt} = 
   -3Hn - \langle \sigma_{\rm{eff}} v \rangle 
   \left( n^2 - n_{\rm{eq}}^2 \right) 
\end{equation} 
where 
\begin{equation} \label{eq:sigmaveffdef} 
   \langle \sigma_{\rm{eff}} v \rangle = \sum_{ij} \langle 
   \sigma_{ij}v_{ij} \rangle \frac{n_{i}^{\rm{eq}}}{n^{\rm{eq}}} 
   \frac{n_{j}^{\rm{eq}}}{n^{\rm{eq}}}. 
\end{equation} 
with 
\begin{equation} 
   v_{ij} = \frac{\sqrt{(p_{i} \cdot p_{j})^2-m_{i}^2 m_{j}^2}}{E_{i} E_{j}}. 
\end{equation} 
 
Using the Maxwell-Boltzmann approximation for the velocity 
distributions one can derive the following expression for the 
thermally averaged annihilation cross section \cite{coann} 
\begin{equation} \label{eq:sigmavefffin2} 
   \langle \sigma_{\rm{eff}}v \rangle = \frac{\int_0^\infty 
   dp_{\rm{eff}} p_{\rm{eff}}^2 W_{\rm{eff}} K_1 \left( 
   \frac{\sqrt{s}}{T} \right) } { m_1^4 T \left[ \sum_i \frac{g_i}{g_1} 
   \frac{m_i^2}{m_1^2} K_2 \left(\frac{m_i}{T}\right) \right]^2}\,. 
\end{equation} where $K_1$ ($K_2$) is the modified Bessel function of the second kind of order 1 (2), $T$ is the temperature, $s$ is the usual Mandelstam variable and 
\begin{eqnarray} \label{eq:weff} 
   W_{\rm{eff}} & = & \sum_{ij}\frac{p_{ij}}{p_{\rm{eff}}} 
   \frac{g_ig_j}{g_1^2} W_{ij} \nonumber \\ 
   & = & 
   \sum_{ij} \sqrt{\frac{[s-(m_{i}-m_{j})^2][s-(m_{i}+m_{j})^2]} 
   {s(s-4m_1^2)}} \frac{g_ig_j}{g_1^2} W_{ij}. 
\end{eqnarray} 
Here, 
\begin{equation} 
    p_{ij} = 
   \frac{\left[s-(m_i+m_j)^2\right]^{1/2} 
   \left[s-(m_i-m_j)^2\right]^{1/2}}{2\sqrt{s}}, 
\end{equation} 
and the invariant annihilation rate is (see \cite{coann}) 
\begin{equation} \label{eq:Wijcross} 
   W_{ij} = 4 p_{ij} \sqrt{s} \sigma_{ij} = 4 \sigma_{ij} \sqrt{(p_i 
\cdot p_j)^2 - m_i^2 m_j^2} = 4 E_{i} E_{j} \sigma_{ij} v_{ij} 
\end{equation} 
and, finally, the effective momentum  
\begin{equation} \label{eq:peff} 
    p_{\rm{eff}} = p_{11} = \frac{1}{2} 
   \sqrt{s-4m_{1}^2}. 
\end{equation} 
Since $W_{ij}(s) = 0 $ for $s \le (m_i+m_j)^2$, the terms in 
(\ref{eq:weff})  are non-negative. 
For a two-body final state, $W_{ij}$ is given by the expression 
\begin{equation} \label{eq:Wij2body} 
   W^{\rm{2-body}}_{ij} = \frac{|\bf k|}{16\pi^2 g_i g_j S_f \sqrt{s}} 
   \sum_{\rm{internal~d.o.f.}} \int \left| {\cal M} \right|^2 
   d\Omega , 
\end{equation} 
that after some manipulations leads to (\ref{eq:dm}).
Here ${\bf k}$ is the final center-of-mass momentum, $S_f$ is a symmetry 
factor equal to 2 for identical final particles.

So, 
what could the dark matter be? It turns out that in particle physics, there 
are hypothetical particles, like supersymmetric partners of ordinary 
particles, that have the right interaction  
strength and mass range to be 
promising dark matter candidates. In particular, the neutralino 
 has all the properties of a good dark matter candidate. 
Since it is electrically neutral it does not emit or  
absorb radiation which makes it `dark' (invisible matter  
is thus a better term than dark matter).  The couplings of neutralinos are  
generally of   
weak interaction strength, but the large number of possible annihilation  
channels, which depends on the unknown supersymmetry breaking  
parameters, makes an exact prediction of mass and relic abundance  
uncertain. Scans of parameter space show, however, that a neutralino  
in the mass range between 30 GeV and a few TeV 
 could give a relic density close to  
the critical density. We will later in these notes have much more to 
say about this.
 
\subsection{Inflation}\label{sec:inflation} 
 
An important ingredient in today's cosmology is, as mentioned, the concept  
of inflation, which was introduced by Alan Guth in the early 1980's \cite{guth} and later improved by Albrecht and Steinhardt \cite{albrecht}, and Linde \cite{linde} (see also Sato \cite{sato}).
This is nowadays a vast field of research, and we will only go through
some very basic concepts and some of the first ideas. For a thorough treatment, see the book by Mukhanov \cite{mukhanov_book}.
  
Einstein's equations including a cosmological constant read 
\beq 
R_{\mu\nu}-{1\over 2}g_{\mu\nu}{\cal R}=8\pi GT_{\mu\nu} 
+\Lambda g_{\mu\nu}.
\eeq 
Here we have put the $\Lambda$ term on the right hand side, which shows that a cosmological term acts as a stress-energy tensor,  albeit
with the unusual equation of state $p_{vac}=-\rho_{vac}$.  
(We have already used that one may trivially include 
vacuum energy in the term proportional to $G$, with  
$\rho_{\Lambda}= \Lambda/(8\pi G)$.)  The equation of state implies that the entropy  
density according to (\ref{eq:smalls}) is $s \sim \rho+p = 0$. This means that, when vacuum energy dominates, the entropy vanishes. 
This can be understood from statistical  
mechanics. Entropy is related to the total number of degrees of  
freedom, and the vacuum (at least if it is unique) is just one state, that is 
 only one degree of freedom. Of course, the entropy that was in a  
 patch before inflation will still be there after inflation -- but it  
 will be diluted by an exponentially large factor due to the expansion. 
 
In the situation when the 
constant  
vacuum energy dominates the expansion, the Friedmann equation (\ref{eq:friedmann}) becomes 
very simple: 
 
\beq 
H^2=\biggl({\dot a\over a}\biggr)^2={\Lambda\over 3} 
\eeq 
or 
\beq 
H={\dot a\over a}=\sqrt{{\Lambda\over 3}}={\rm const} 
\eeq 
with the elementary (de Sitter) solution 
\beq 
a\sim e^{Ht} .
\eeq 
In inflation, the expansion 
rate is constant, which causes  an exponential growth of the scale factor.

In many models of inflation, the phase transition involving a scalar field, 
the inflaton field, took place at temperatures around  
the hypothetical  Grand Unification scale $T_{GUT}\sim 10^{15}$ GeV, at 
the corresponding Hubble time  
$H^{-1}\sim 10^{-34}$ sec.  If the universe stayed in the 
inflationary state for a short time, say $10^{-32}$ sec, 
many e-folds of inflation 
took place.  When inflation stopped, 
the huge vacuum energy of the inflaton field went into creation of ordinary particles, 
and a reheating of the universe took place.  
The reheating temperature is of the order of  
the temperature of the phase transition,
$T_{RH}\sim 10^{15}$ GeV if the inflaton 
is strongly enough coupled to ordinary matter, as it is in many successful models 
of inflation. 
 
Let us see what happened to a small region with radius of 
for example $10^{-23}$ cm before  
inflation. The entropy within that volume was only around $10^{14}$, 
but after inflation the volume of the region has increased by a factor given by 
the cube of the scale factor, 
$(e^{100})^3=10^{130}$. Thus, after the entropy generated  
by reheating, 
the total entropy within the inflated region had grown to around $10^{144}$. 
Entropy was generated because the equation of state changed from 
$p=-\rho$ to $p=\rho/3$, meaning that the entropy density 
$s\sim p+\rho$ increased dramatically. 
 
This huge entropy increase solves many problems of cosmology. 
The ``horizon problem'' -- i.e., how come that regions of the universe that are too far from each other to be causally connected today, still have exactly the same microwave background temperature -- is solved since our whole observable universe  
arose from a very small thermalized  volume  before inflation, and the smooth region after inflation had sufficient entropy to encompass 
our observable universe. 
 
During inflation the energy density, and the negative pressure, of the universe were constant, whereas 
the scale factor $a$ increased exponentially. This means  
that the total $\Omega$ after inflation was exponentially close to unity. (Like a balloon which would inflate to be as big as the Earth would locally look very flat.) Thus,  
the present value should also be equal to unity with an accuracy of many decimal places, perhaps the most important successful prediction of inflation. 

Even if $\Omega=1$ is predicted, there is nothing  
 telling us the subdivision of $\Omega$ into  
contributions from radiation, matter and vacuum energy. As we have  
noted, however, the `natural'  contribution of $\Omega_{\Lambda}$ is 
either extremely small or extremely large. Only during very brief  
epochs can $\Omega_{\Lambda}$ be of similar magnitude as the matter 
contribution $\Omega_{M}$. This is actually a still unsolved problem, why is it that the energy density in matter $\rho_M$ is about the same as  $\rho_\Lambda$ today? 
 
The period of inflation and reheating is strongly non-adiabatic, since 
there was an enormous generation of entropy at reheating. After the end 
of inflation, the universe  `restarted' in an adiabatic phase with the 
standard conservation of $aT$, and it is because the universe automatically  
restarts from very special 
initial conditions given by inflation that the horizon and flatness 
problems are solved. 

It is instructive to see how inflation can be produced in field theory. 
A Lagrangian density of the  
form 
\beq 
{\cal L}={1\over 2}\partial^\mu\phi\partial_\mu\phi -  
V(\phi)\label{eq:nuttx} 
\eeq 
can be shown to give a contribution to the energy-momentum tensor $T^{\mu\nu}$ of 
the form 
\beq\label{eq:tmunuscalar2} 
T^{\mu\nu}=\partial^\mu\phi\partial^\nu\phi - {\cal  
L}g^{\mu\nu}.\label{eq:nutty} 
\eeq 
 
For a homogeneous state, the spatial gradient terms vanish, meaning that 
$T^{\mu\nu}$ becomes that of the perfect fluid type. If one would keep  
the gradient terms, one sees that they are divided by $a(t)^2$, which  
means that after a short period of inflation they are exponentially  
suppressed. The resulting $\rho$ and $p$ are  
\beq 
\rho={\dot\phi^2\over 2}+V(\phi) 
\label{eq:scalerho} 
\eeq 
and 
\beq 
p={\dot\phi^2\over 2}-V(\phi), 
\label{eq:scalep} 
\eeq 
and we see that the equation of state $\rho=-p$ will be fulfilled if we can neglect the kinetic term  $\sim \dot\phi^2$ (this is called ``slow-roll'' inflation). 
 
The exact equations of motion of $\phi$ can be derived from the condition of  
vanishing covariant divergence of the energy-momentum tensor,  
$T^{\mu\nu}_{\,\,\,\,;\nu}=0$, which gives  
\beq 
\ddot\phi+3H\dot\phi+V'(\phi)=0 \label{eq:eqmotion} 
\eeq 
This is similar to the equation of motion of a ball in a potential  
well with friction $\sim 3H\dot\phi$, and can be solved  
by elementary methods. We assume that at very high temperatures,  
$\phi=0$ gives the locations of the minimum of the potential. Temperature dependent 
terms in the effective potential then generate another  minimum  for  
$\phi=\phi_{vac}\ne 0,$ 
an example of what is called spontaneous symmetry breakdown. To produce a long enough period of inflation  
and a rapid reheating after inflation, the potential $V(\phi)$ has as mentioned to  
be of the ``slow-roll'' type, with the field spending a long time on the nearly flat, horizontal  part of the potential. 
In the beginning, on the almost horizontal slow `roll' towards a deep  
potential well, $\ddot\phi$ can be neglected, and the slow-roll equation of  
motion 
\beq 
3H\dot\phi+V'(\phi)=0,\label{eq:slowroll4} 
\eeq  
together with the Friedmann equation 
\beq 
H^2={8\pi G_N\over 3}\left[{1\over 2}\dot\phi^2+V(\phi)\right],\label{eq:fried_scal} 
\eeq 
which during slow roll, when $\dot\phi^2$ is small,  can be approximated by 
\beq 
H^2={8\pi G_N\over 3} V(\phi).
\eeq 
One can from this get an expression for the number $N_{\phi}$ of $e$-folds of the scale factor, 
\beq 
N_{\phi}\equiv\log\left({a_{2}\over a_{1}}\right)=\int Hdt\sim\int_{\phi_{1}}^{\phi_{2}}{V(\phi)\over V'(\phi)}d\phi.\label{eq:nphi} 
\eeq 
Thus, for a large growth of the scale factor, $V(\phi)$ has to be very 
flat ($V'(\phi)\sim 0$). This may be unnatural except perhaps in some supersymmetric theories  
where `flat directions' can occur because of the pattern of  
supersymmetry breaking. In a situation of such a slow roll of the inflaton 
field, the exact form of the potential does not matter so much, and the relevant 
physics can be expressed in terms of the so-called slow-roll parameters 
\beq 
\varepsilon=-{\dot H\over H^2}=4\pi G{\dot\phi^2\over H^2}={1\over 16\pi G}\left({V'\over V}\right)^2\label{eq:slowroll1} 
\eeq  
\beq 
\eta={1\over 8\pi G}\left({V''\over V}\right)={V''\over 3H^2}\label{eq:slowroll2} 
\eeq 
where the second equation in (\ref{eq:slowroll1}) comes from taking the 
derivative of (\ref{eq:fried_scal}) and inserting into (\ref{eq:eqmotion}). 
The variable $\varepsilon$ is a measure of the change of the Hubble expansion  
during inflation; for inflation to happen at all, $\varepsilon < 1$ is 
needed.

In the picture of the rolling ball, reheating corresponds to  
oscillations in the potential well. Thus, for enough entropy to be  
generated the well has to be rather steep. The problem of constructing a  
suitable potential is to simultaneously have it flat near $\phi=0$ and 
steep near $\phi=\phi_{min}$. 
 
A way to avoid a phase transition, and in fact the simplest model of inflation 
is the chaotic inflation model of Linde \cite{linde_chaotic}.  
It relies on the fact 
that the key ingredient for inflation to occur is that the field is rolling 
slowly, so that the energy density is nearly constant during a 
sufficient number of e-foldings of the scale factor. Since the rolling is 
damped by the presence of the term proportional to $H$ in  
(\ref{eq:eqmotion}), and $H$ according to the Friedmann equation is 
given by the height of the potential (if kinetic terms can be neglected), 
inflation will be possible for any positive, power-law potential  
$V\left(\phi\right)$, for example 
the simplest  $V\left(\phi\right)={1\over 2} m^2\phi^2$, as long as the field values start 
out large. As Linde has argued, this may not be unreasonable since these 
initial values may be given stochastically (``chaotically'')  
at the Planck epoch,  
and those regions 
where the field values are large start to inflate rapidly dominating the  
volume of the universe.  There are also constructions relying on the existence of more than one scalar field, keeping the same general features but with changes in the details.
 
Since the value of the total energy density $\Omega=1$ is found   
observationally in current measurements of 
the CMBR anisotropy which yield $\Omega = 1.003 \pm 0.010$, the most natural explanation seems to 
be that the universe has indeed gone through a period of inflation. The nearly
gaussian and adiabatic temperature fluctuations measured by COBE and WMAP are a strong indication  of the correctness of the basic quantum fluctuation mechanism active during inflation \cite{mukhanov}.
 
An important test of this mechanism may be produced by the upcoming  
measurements from the Planck satellite of the even more detailed pattern of temperature 
fluctuations in the cosmic microwave background radiation.  
Inflation predicts a nearly  
but not perfect scale-invariant spectrum of fluctuations (which is when the index of scalar fluctuations 
$n_s=1$), and present measurements from WMAP 
give $n_s\sim 0.96$, in excellent agreement. Inflation could also have generated gravitational (tensor) waves during the phase transitions which would give a particular pattern (``B-modes'') in the microwave sky. However, the amplitude of tensor to scalar fluctuations depends rather strongly on the model. It will be interesting to see whether the Planck satellite, when cosmological data are released in early 2013, will detect such a B-mode pattern. 

There are constructions showing, after some initial difficulties, how inflation can also be embedded into supergravity \cite{kt}, or string theory \cite{kklt}.

\section{Basic Cross Sections for Neutrinos and $\gamma$-rays}\label{ch:2} 
Among the various messengers from the Galaxy and structures further away, neutrinos and $\gamma$-rays have the advantage that they follow straight lines  
(or to be more exact, geodesics; the deviations from straight lines can in almost all cases be safely neglected - exceptions are given for particles 
originating or travelling very near black holes). This is in contrast to all other cosmic rays, electrons, protons, nuclei, and antimatter (positrons, antiprotons and some antinuclei 
like antideuterons). Neutrons would in principle also travel rectilinearly apart from effects of their magnetic moment. However, their finite lifetime (of the order of 10 minutes in the rest frame) means that for energies less than a few TeV which is the energy range we will generally be concerned with, they cannot travel over astrophysical distances. They $\beta$-decay to a proton, and electron and an (anti-)neutrino 
 
Although neutrinos and $\gamma$-rays (high-energy photons) are both encompassed in the Standard Model of particle physics and therefore in principle should interact with similar strengths 
given by gauge couplings, this is in practice not so. The reason is the difference that the photon is a massless, spin-1 gauge particle, i.e., a mediator of a force (the electromagnetic force, i.e., it couples to electrons and protons, and all other particles with electric charge) while the neutrino is a spin-$1/2$ matter particle which in turn interacts through a weak forces mediated by the heavy $W$ and $Z$ bosons. The large, important difference of   
masses between weak bosons and the photon is due to the hitherto experimentally  unverified, but hopefully soon to be verified mechanism,  the Higgs mechanism. This breaks the gauge group of the Standard Model, leaving only the electromagnetic $U(1)_{em}$ unbroken and therefore the photon massless. It means that for energies up to 1 TeV or so, neutrinos have very small cross section, which however rises with energy, until the centre-of-mass energy is of the order of the $W$ and $Z$ masses, at which point neutrinos start to react roughly as strongly as photons.   
Let us now discuss in some more detail how some simple particle cross sections are computed. 
\subsection{Estimates of  Cross Sections}\label{subs:estimates} 
 
The calculation of collision and annihilation cross sections, and decay 
rates of particles, is an important task in particle physics. Here 
we will present only a brief outline of how this is done, and focus on 
`quick-and-dirty' estimates which may be very useful in cosmology and  
astrophysics. For the local microphysics in the FLRW model, only 
three interactions  -- electromagnetic, weak and strong  -- between  
particles need to  
be considered. The gravitational force is completely negligible  
between individual elementary particles -- for instance, the gravitational force  
between the proton and the electron in a hydrogen atom is around 
$10^{40}$ times weaker than the electromagnetic force. However,  
gravity, due to its coherence over long range, still needs to 
be taken into account through its influence on  
the metric. This means that the dilution of number densities due  
to the time dependence of the scale factor $a(t)$ has to be taken into  
account. In the next section we will see how this is done.  
 
Let us begin with the interaction strengths. The strength of the electromagnetic  
interaction  is governed by  the electromagnetic coupling  
constant $g_{em}$, which is simply the electric charge. As usual, we take the  
proton charge $e$ as the basic unit and can thus write 
\beq 
g_{em}=Qe 
\eeq 
where $Q$ is the charge of the particle in units of the proton charge 
(for a $u$-quark, for example, $Q_{u}=+2/3$). In our system of units, 
\beq 
{e^2\over 4\pi}\equiv\alpha_{em} 
\eeq 
where $\alpha_{em}$ is the so-called fine structure constant which has  
the value of around $1/137$ at low energies.\footnote{This coupling constant,  
as all others, depends on the energy scale, for example, the energy 
transfer, of the process. At 100 GeV energy $\alpha_{em}$ is $\sim 1/128$.} 
(Usually, it is denoted just $\alpha$ without the subscript.) 
The weak coupling constant is of similar magnitude: 
\beq 
g_{w}={e\over \sin\theta_{W}}\label{eq:gweak} 
\eeq 
with $\theta_{W}$ the weak interaction (or Weinberg) angle, which has  
the numerical value $\sin^2\theta_{W}\sim 0.23$. The fact that the weak  
and electromagnetic coupling constants are of the same order of  
magnitude is of course related to the fact that they are unified in  
the Standard Model to the `electroweak' interaction.
 
The coupling constant of the strong interaction, $g_{s}$, is somewhat  
higher. Also, it runs faster (it decreases) with energy  than the  
electromagnetic coupling. At energies of a few GeV,
\beq 
\alpha_{s}\equiv {g_{s}^2\over 4 \pi}\sim 0.3 
\eeq 
 
\begin{figure}[!htb] 
\begin{center} 
\includegraphics[width=8cm]{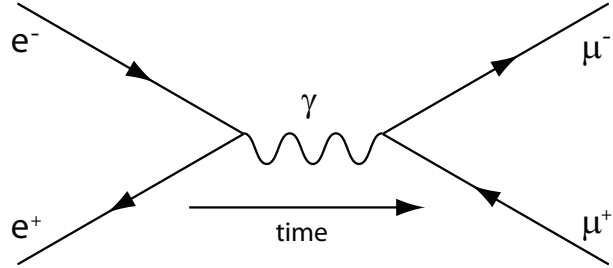} 
\end{center} 
        \caption{\small A Feynman diagram representing the annihilation of an 
        electron and a positron to a muon pair.\label{fig:eetomumu}} 
 
\end{figure}

Let us look at the Feynman diagram for a simple process like  
$e^+e^-\to\mu^+\mu^-$ (Fig.\,~(\ref{fig:eetomumu})). The amplitude  
will be proportional to the coupling constants at both vertices,  
which in this case are both equal to $e$. The cross section, being  
proportional to the square of the amplitude, is thus proportional to 
$e^4\propto (\alpha/4\pi)^2$. 
 
The total energy of the $e^+e^-$ pair in the centre of  
momentum frame is 
$E_{cm}(e^+)+E_{cm}(e^-)=\sqrt{s}$. Since the total momentum in this  
frame is zero, the four-momentum $p^\mu=(\sqrt{s},0,0,0)$ is identical  
to that of a massive particle of mass $M=\sqrt{s}$ which is at rest. 
Energy and momentum conservation then tells us that the photon in the  
intermediate state has this four-momentum. However, a freely propagating  
photon is massless, which means that the intermediate photon is  
virtual by a large amount. In quantum field theory one can show that  
the appearance of an intermediate state of virtual mass $\sqrt{s}$ for 
 a particle with real rest 
mass $M_{i}$ is suppressed in amplitude by a factor (called the  
propagator factor)  
\beq 
P(s)=1/(s-m_{i}^2)\label{eq:prop} 
\eeq 
In this case ($m_{\gamma}=0$), we have  a  
factor of $1/s$. (If one does this rigorously, one should insert a small imaginary part in the denominator, which defines how the singularity on the mass shell
is treated.)
The outgoing particles (in this case the muons) have a large number of  
possible final states to enter (for example, all different scattering angles 
in the centre of momentum frame). This is accounted for by the  
so-called phase space factor $\phi$, which generally grows as $s$ for large  
energies.  
For the cross section $\sigma$ 
\beq 
\sigma(e^+e^-\to\mu^+\mu^-)\propto\phi\left( {\alpha^2\over  
s^2}\right)\label{eq:approxsig} 
\eeq 
with $\phi$ the phase space factor. If $s$ is large compared to  
$m_{e}^2$ and $m_{\mu}^2$, $\phi\propto s$, and 
\beq 
\sigma(e^+e^-\to\mu^+\mu^-)\sim {\alpha^2\over s}\label{eq:ecross} 
\eeq 
This is not an exact ex\-pression. A careful calculation  (see next section) gives $4\pi\alpha^2/(3s)$),  
but it is surprisingly accurate and often  
accurate enough for the estimates we need in big bang cosmology. 
  
\begin{figure}[!htb] 
\begin{center} 
\includegraphics[width=8cm]{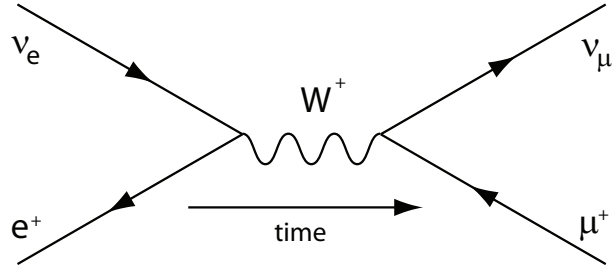} 
\end{center}  
        \caption{\small A Feynman diagram representing the annihilation of an 
        electron neutrino and a positron to a muon neutrino and a muon.\label{fig:w}} 
 
\end{figure}

Since the weak coupling strength is similar to the electromagnetic strength,  
the same formula is valid for, for example, $\nu_{e}+e\to\nu_{\mu}+\mu$  
which goes through $W$ exchange (see Fig.\,~\ref{fig:w}). The only  
replacement we need is $1/s\to 1/(s-m_{W}^2)$ for the propagator, 
thus 
\beq 
\sigma(\bar\nu_{e}+e^-\to\bar\nu_{\mu}+\mu^-)\sim {\alpha^2s\over  
(s-m_{W}^2)^2}\label{eq:wcross} 
\eeq 
When $s\ll m_{W}^2$, this gives $\sigma_{w}\sim \alpha^2s/m_{W}^4$, 
which is a very small cross section, e.g., for  
MeV energies. One should  notice, however, 
the fast rise with energy due to the factor $s$. This is  
the historical reason for the name `weak interaction', which 
is really not appropriate at high energies (much larger than  
$m_{W}$), where the two types of cross sections become of similar size. 
 
Note that once one remembers the factors of coupling constants and  
the propagators, the magnitude of cross sections can often be  
estimated by simple dimensional analysis. A cross section has the dimension  
of area, which in our units means (mass)$^{-2}$. It is very 
useful to check  
that the expressions (\ref{eq:ecross}) and (\ref{eq:wcross}) have the  
correct dimensions. 
 
A decay rate $\Gamma$ can be estimated in a similar way. If a  
collection of identical unstable particles decay, their number density  
decreases as $e^{-\Gamma t}$ which means that $\Gamma$ has the  
dimensions of mass.

\begin{figure}[!htb] 
\begin{center} 
\includegraphics[width=8cm]{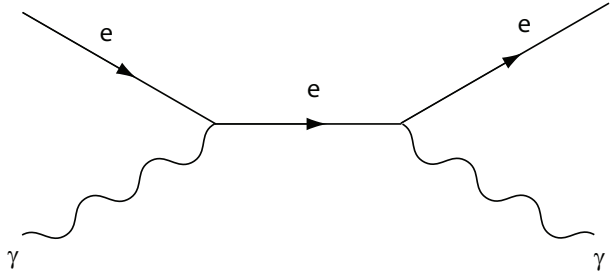} 
\end{center} 
        \caption{\small A Feynman diagram representing the $\gamma e\to\gamma e$ 
        process. In the classical limit, this is called Thomson scattering. The quantum version is called Compton scattering, and in the relativistic regime, the result is given by the Klein-Nishina formula.\label{fig:thomson}} 
 
\end{figure}

A fermion has a propagator that behaves as $1/m$ (instead of $1/m^2$) at 
low energies. This means that the Thomson cross section 
$\sigma(\gamma e\to\gamma e)$ at low energies $E_{\gamma}\ll m_{e}$  
can be estimated to be (see Fig.\,~\ref{fig:thomson}) 
\beq 
\sigma_{T}\equiv\sigma(\gamma e\to\gamma e)\sim {\alpha^2\over m_{e}^2} 
\label{eq:thomson} 
\eeq 
 
\subsection{{Examples of Cross Section Calculations}} 
The estimates we have given are in many cases sufficient for 
cosmological and astrophysical applications. However, there  
are cases when one would like to have a more accurate formula. We now provide only a couple of examples and 
summarize the general framework for calculation and the main results. 
 
\subsection{Definition of the Cross Section} 
 
The differential cross section 
$d\sigma/dt$ for $2\to 2$ scattering $a+b\to c+d$ 
is given by the expression 
\beq 
{d\sigma\over dt}= 
{|\widetilde T|^2\over  
16\pi\lambda\left(s,m_{a}^2,m_{b}^2\right)}\label{eq:dsigmadt1} 
\eeq 
where  the Lorentz invariant Mandelstam variables are $t=(p_a-p_c)^2$ and 
$s=(p_a+p_b)^2$. $|\widetilde T|^2$ is the polarization-summed and  
squared quantum mechanical transition amplitude. 
For a $2\to 2$ process, the kinematically allowed region in $s$ is 
\beq 
s>(m_{3}+m_{4})^2\label{eq:smin} 
\eeq 
which can be understood from energy conservation: In the centre of momentum system,  
where  $\sqrt{s}$ corresponds to the total  
energy, at least the rest mass energy $m_{3}+m_{4}$ has to be provided. 
 
The kinematical limits for $t$ can be  
obtained from the condition $|\cos\theta_{13}^{\rm cms}|\leq  
1$, with 
\beq 
\cos\theta_{13}^{\rm  
cm}={s(t-u)+(m_{1}^2-m_{2}^2)(m_{3}^2-m_{4}^2)\over 
\sqrt{\lambda(s,m_{1}^2,m_{2}^2)}\sqrt{\lambda(s,m_{3}^2,m_{4}^2)}}. 
\label{eq:tlimits} 
\eeq 
 
 A  typical calculation involves (here we follow the treatment of \cite{BGbook}) 
computing the matrix element in terms of $s$ and $t$ and carrying out 
the $t$ integration to obtain the total cross section. 
 
In the one-photon exchange approximation, the cross section for the annihilation process 
$e^+e^-\to\mu^+\mu^-$ is  
\beq 
\sigma(e^+e^-\to \mu^+\mu^-)={2\pi\alpha^2\over s}\beta\left(1-{\beta^2\over 3}\right) 
\label{eq:eemumu1} 
\eeq 
where the only app\-roxi\-mation made is to ne\-glect $m_e$ (this is allowed, since 
$m_e^2/m_\mu^2\ll 1$). Here $\beta$ is the velocity of one of the outgoing muons  
in the 
centre of momentum system, $\beta=\sqrt{1-4m_\mu^2/s}$. 
In the relativistic limit of $s\gg  m_{\mu}^2$, ($\beta\to 1$), this becomes 
\beq 
\sigma\left(e^+e^-\to\mu^+\mu^-\right)_{{\rm large}\  
s}={4\pi\alpha^2\over 3s}\label{eq:deepin} 
\eeq 
in agreement with our simple estimate (\ref{eq:ecross}). 
 
\subsection{The $\gamma\gamma e e$ System}\label{subs:ggeesystem} 
 
\begin{figure}[!htb] 
\begin{center} 
\includegraphics[width=10cm]{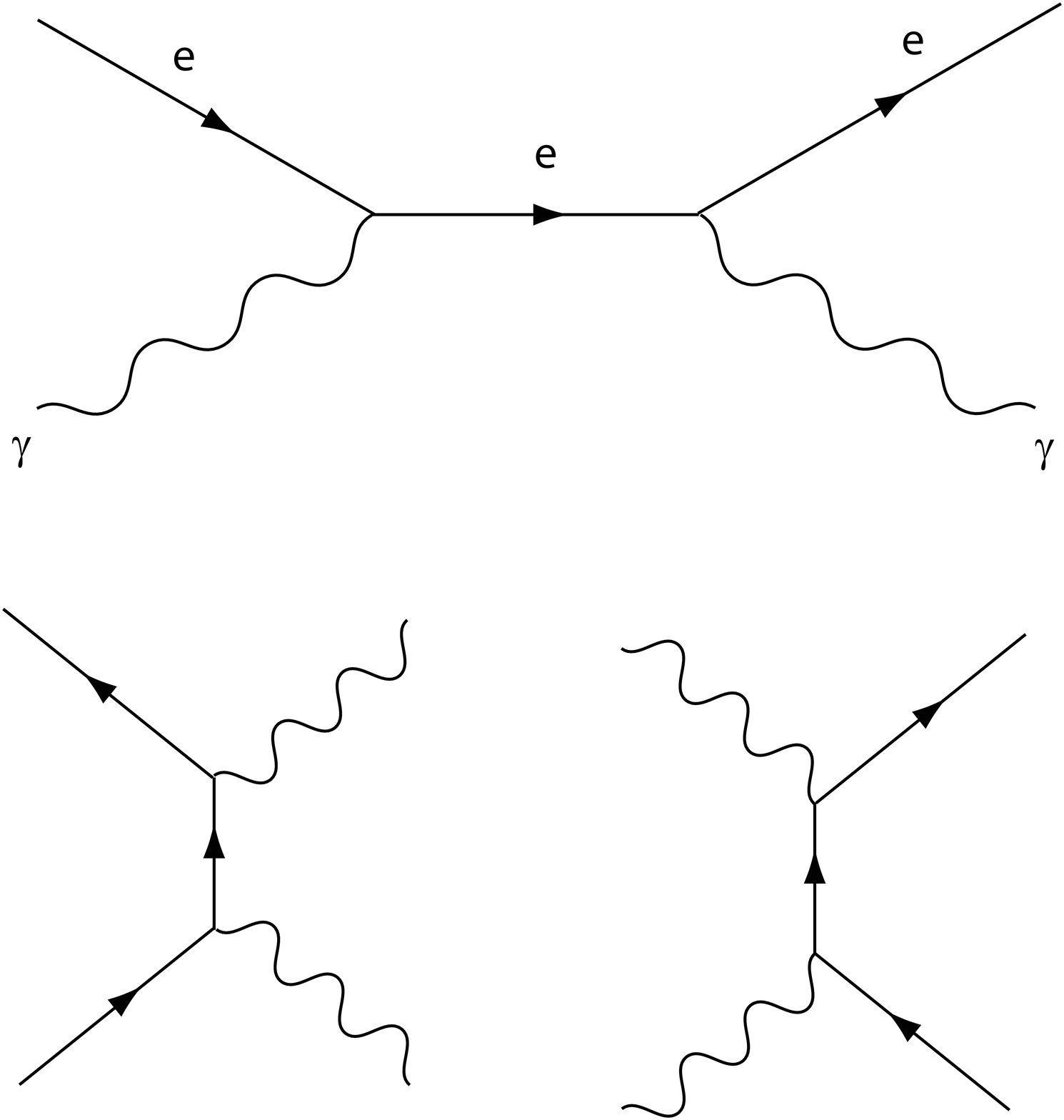} 
\end{center} 
        \caption{\small The same Feynman diagram can, after rotation of the external legs, describe both $e\gamma\to e\gamma$, $\gamma\gamma\to e^+e^-$, and  
$e^+e^-\to \gamma\gamma$. Here time is as usual flowing from left to right.\label{fig:ggee}} 
 
\end{figure}

By different permutations of the incoming and outgoing particles, 
the basic $\gamma\gamma e e$ interaction (shown in  
Fig.~\ref{fig:thomson}) can describe all of the astrophysically important processes (see the contributions by F. Aharonian and C. Dermer in this volume) $\gamma\gamma\to e^+e^-$,  
$e^+e^-\to \gamma\gamma$,  and $\gamma e^\pm\to\gamma e^\pm$, see Fig.~\ref{fig:ggee}. 
 
For $\gamma\gamma\to e^+e^-$ the result is 
\beq 
\sigma\left(\gamma\gamma\to e^+e^-\right)={\pi\alpha^2\over 2m_{e}^2} 
\left(1-\beta^2\right)\left[\left(3-\beta^4\right)\ln\left({1+\beta\over  
1-\beta}\right)+2\beta\left(\beta^2 -2\right)\right]\label{eq:ggee1} 
\eeq 
where $\beta$ now is the velocity of one of the produced electrons in the  
centre-of-momentum frame, $\beta=\sqrt{1-4m_e^2/s}$. 
Near threshold, i.e. 
for small $\beta$, the expression in square brackets can be series expanded 
to $2\beta +{\cal O}(\beta^2 )$, and thus 
\beq 
\sigma\left(\gamma\gamma\to e^+e^-\right)_{{\rm small}\ \beta}\simeq  
{\pi\alpha^2 \over m_{e}^2} 
\eeq 
In the other extreme, $\beta\to 1$,  
\beq 
\sigma\left(\gamma\gamma\to e^+e^-\right)_{s\gg 4m_{e}^2}\simeq 
{4\pi\alpha^2\over s}\left[\ln\left({\sqrt{s}\over  
m_{e}}\right)-1\right]\label{eq:larges} 
\eeq 
 
One could in fact have guessed most  
of this to a fair amount of  accuracy by the simple dimensional and  
vertex-counting rules. At low  
energy, the only available mass scale is $m_{e}$, so the factor $\alpha^2  
/m_{e}^2$ could have been guessed for that reason. The  
factor $\beta$ could also have been inferred with some more knowledge of   
non-relativistic partial wave  amplitudes. At low energy, the  
$\ell=0$ ($S$-wave) 
amplitude should dominate, and this  contributes to the cross section  
proportionally to $\beta$. A partial wave $\ell$ contributes 
to the total cross section with a term proportional to $\beta^{2\ell +  
1}$. We see from (\ref{eq:eemumu1}) that in the case of  
$e^+e^-\to \mu^+\mu^-$ the $S$-wave dominates at low energy, but 
when $\beta\to 1$, the $P$-wave contribution is $1/3$. 
At high energy, when $m_{e}$ can be neglected,  
the dimensions have to be carried by $s$. Only the logarithmic  
correction factor in (\ref{eq:larges}) could not have been easily 
guessed.  
 
\begin{figure}[!htb] 
\begin{center} 
\includegraphics[width=\textwidth]{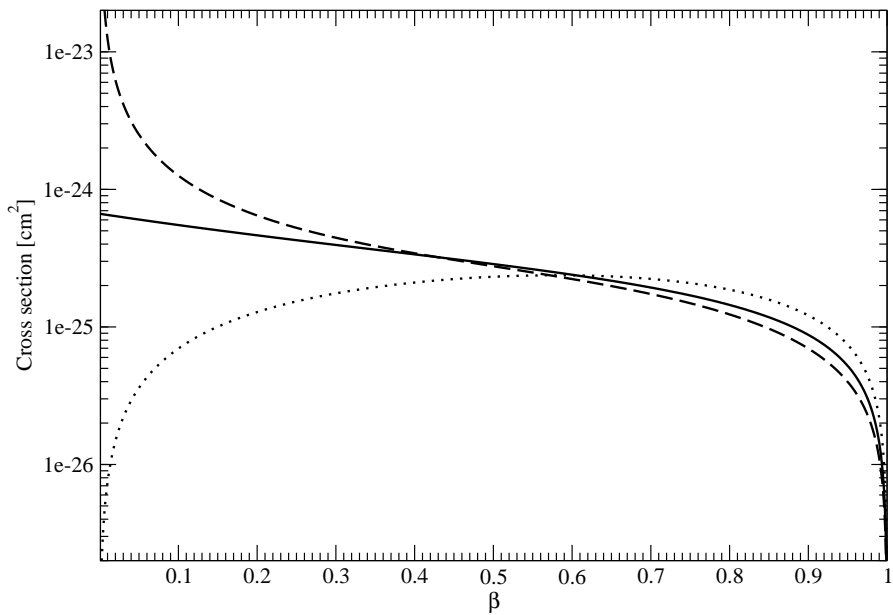} 
\end{center} 
\caption{\small The cross sections (in cm$^2$)  
for photon photon annihilation, $e^+e^-\to \gamma\gamma$ and 
Compton scattering as a function of the  
cms velocity 
$\beta$ of the electron.\label{fig:eeggres}} 
 
\end{figure}

These formulas show that the $\gamma\gamma\to e^+e^-$ cross section  
rises from threshold to a maximum at intermediate energies and  
then drops roughly as $1/s$ at higher $\beta$, i.e., higher cms energy in the process (see Fig.~\ref{fig:eeggres}). 
 
The results for the reverse process $e^+e^-\to \gamma\gamma$ are of 
course extremely similar. Now, the process is automatically always  
above threshold. For $\beta\to 0$ (with $\beta$ now the velocity of 
one of the incoming particles in the cm-system, still given 
by the formula $\beta=\sqrt{1-4m_{e}^2/s}$), the flux factor $\sim  
1/\beta$ in (\ref{eq:master}) diverges. Since the outgoing photons move  
away with $\beta=c=1$ there is no partial-wave suppression factor, and we  
can thus expect the cross section at low energy to behave as 
\beq 
\sigma\left(e^+e^- \to\gamma\gamma\right)_{{\rm low\ energy}}\sim  
{\alpha^2 \over \beta m_{e}^2} 
\eeq 
and the high-energy behaviour by the same formula, with $m_{e}^2$  
replaced by $s$ (and possibly a logarithmic factor). These  
expectations are borne out by the actual calculation, which gives 
\beq 
\sigma\left(e^+e^- \to\gamma\gamma\right)= 
{\pi \alpha^2\left(1-\beta^2\right)\over 2\beta m_{e}^2} 
\left[ {3-\beta^4\over 2\beta}\ln\left({1+\beta\over 1-\beta}\right)-2+\beta^2\right] 
\eeq 
Note the similarity with (\ref{eq:ggee1}). The $1/\beta$ behaviour of the cross section (see the dashed curve in Fig.~\ref{fig:eeggres}) was noted by Arnold Sommerfeld in the 1930's, and he showed how one can make an improved calculation valid at very small velocities by not only treating the annihilating particles as plane waves, but using wave functions appropriate for the attractive Coulomb interaction between the electron and positron. He thereby described a generic mechanism, so so-called Sommerfeld enhancement mechanism, which recently has played an important role for dark matter calculations, as we will see later.  
 
\subsubsection{Compton and Inverse Compton Scattering} 
As the final example, we consider Compton scattering $\gamma + e^-\to 
\gamma + e^-$. Historically, this was first computed for  
an incoming beam of photons of energy $\omega$  
which hit electrons at rest. Later on, the related process of a very high energy electron or positron colliding with a low-energy photon (such as coming from the cosmic microwave background, or from infrared or optical radiation created in stellar processes) and upscattering that photon to high, maybe GeV energy or higher, has been found to be very important in astrophysics. Despite being really one and the same process, the latter situation is often referred to as the inverse Compton or IC process. In fact, the inverse Compton process is one purely leptonic process of creating high-energy $\gamma$-rays, and could  be important for the emission $\gamma$-rays in several cases, such as AGNs, GRBs and various supernova remnants. However, to completely prove such a scenario, it is important to search for, or give upper limits on, neutrino emission. In competing hadronic models of emission, where  $\gamma$-rays mainly come from  $\pi^0$ decays, one should also have essentially the same amount of charged pions which decay into a leptons and neutrinos. Also for some ``leptophilic'' models of dark matter, where electrons and muons are main annihilation products, inverse Compton processes may be quite important, e.g., near the galactic centre where radiation fields are large.  
 
For  
scattering of a photon by an angle $\theta$ with respect to the incident photon direction, 
the outgoing photon energy $\omega'$ is given by energy-momentum 
conservation 
\beq 
{\omega'}={m_{e}\omega\over m_{e}+\omega\left(1-\cos\theta\right)} 
\eeq 
In this frame, the unpolarized differential cross section,  
the Klein-Nishina formula as it was first computed by Klein  and  
Nishina shortly after Dirac had presented his equation describing relativistic electrons (and positrons), is  
\beq 
{d\sigma\over d\Omega}={\alpha^2\over 2 m_{e}^2} 
\left({\omega'\over \omega}\right)^2 
\left[{\omega'\over \omega}+{\omega\over \omega'}-\sin^2\theta\right] 
\eeq 
Integrated  over all possible scattering angles this gives the 
total cross section 
$$ 
\sigma(\gamma +e\to \gamma + e)= 
{\pi\alpha^2\left(1-\beta\right)\over m_e^2\beta^3}\times 
$$ 
\beq\left[{4\beta\over 1+\beta}+\left(\beta^2+2\beta-2\right)\ln\left({1+\beta\over 1-\beta}\right) 
-{2\beta^3\left(1+2\beta\right)\over (1+\beta)^2}\right] 
\eeq 
where $\beta$ is now the incoming electron velocity in the centre of  
momentum frame, $\beta=(s-m_{e}^2)/(s+m_{e}^2)$. If one expands this  
result around $\beta=0$, one recovers the Thomson scattering result 
\beq 
\sigma_{{\rm Thomson}}={8\pi\alpha^2\over 3 m_{e}^2}\sim 6.65\cdot 10^{-25}\ {\mathrm cm}^2 
\eeq 
and the large-$s$, so-called Klein Nishina regime gives 
\beq 
\sigma_{{\rm KN}}={2\pi\alpha^2\over s}\left[\ln\left({s\over m_{e}^2}\right) 
+{1\over 2}\right] 
\eeq 
We see that for photon energies much larger than $m_{e}$ -- that is, in  
the Klein-Nishina regime -- the Compton cross section falls quite  
rapidly.

In the classical Compton  
scattering situation, the outgoing photon energy is always less 
than the incoming one. Thus, energetic photons travelling through a 
gas of cold electrons will be `cooled' by Compton scattering. In the IC case (for example for the cosmic microwave background radiation passing  
through a galaxy cluster with hot gas) energetic electrons may instead  
transfer energy to photons, thereby `heating' them. For CMBR this is called
the Sunyaev-Z'eldovich effect, and has a large range of applicability (for instance, it has recently been used to find galaxy clusters).   
 
When computing actual numbers for the cross sections (which should have 
the dimensions of area) in our units, a useful conversion factor 
is 
\beq 
1\ {\rm GeV}^{-2}=0.389\cdot 10^{-27}\ {\rm cm}^2 
\eeq 
In Fig.~\ref{fig:eeggres} the numerical results are summarized.  
The cross sections are shown (in cm$^2$)  
for $\gamma\gamma\to ee$, $ee\to\gamma\gamma$ and 
$\gamma e\to \gamma e$ as a 
a function of the  
cms velocity 
$v$ of the electron. 
 We see in the figure the different 
behaviour at low cms velocity (small $\beta$) already discussed, but that they 
show a similar decrease at high energy.

Another process of great astrophysical importance is that of  
 bremsstrahlung. By this is meant the emission of photons from 
charged particles which are accelerated or decelerated. If this acceleration 
is due to circular motion in a magnetic field, the term  synchrotron 
radiation is used. Through these processes (unlike Compton scattering) 
 the number of photons can change. This is needed, for instance in the early universe, if thermal 
equilibrium is to be maintained, since  
the number density of photons has to vary, as it depends strongly on temperature. 
 Most of the produced 
photons have very low energy (long wavelength). If fast electrons pass  
through a region where synchrotron radiation and bremsstrahlung 
occur, these low-energy photons may  be upscattered in energy through 
the inverse Compton process. This may for example explain the observations of 
very high-energy photons in active galactic nuclei (see the contributions by Aharonian and Dermer in this volume, where many other applications are discussed). 
 
For a detailed discussion of these and other QED processes, see  
standard textbooks in quantum field theory, for example, \cite{itzykson}, or a simplified treatment along the lines given here, in \cite{BGbook}. And, of course, for good examples of the use of these processes in astrophysics, see the accompanying lectures by F. Aharonian and C. Dermer in this volume.  
 
\subsection{{Processes involving Hadrons}} 
Since protons and neutrons belong to the most common particles in the universe, 
it is of course of great interest to compute processes where these and other 
hadrons (such as pions) are involved. This is, however, not easy to do 
from first principles. The reason that in the previous section we could 
compute so accurately weak and electromagnetic processes is that 
we could use perturbation theory (as summarized, for example, in 
Feynman diagrams). The expansion parameter, the electroweak  
gauge coupling constant $g$ 
or rather $\alpha_{ew}=g^2/(4\pi)\sim 10^{-2}$, is  
small enough that a lowest-order 
calculation is enough to obtain very accurate results.

In QCD, which also is a gauge theory just as QED, we also have a coupling constant $\alpha_s$. Due to the fact that the gauge group of QCD is SU(3), which involves self-interactions of the 8 spin-1 gluons, there are important differences. We say that QCD is  a non-abelian gauge theory whereas QED is based on the abelian group U(1) with only one spin-a gauge field, the photon.  One consequence of this difference is that QCD has what is called asymptotic freedom meaning that the coupling strength which is of order unity at a few hundred MeV, ``runs'' to smaller values for large energies. The energy scale is set, for example, 
by the energy or momentum transfer $Q$ ($Q^2\equiv -t$ with $t$ in the process.  
Thus, for processes with  large $Q^2$, we should be able to use low-order perturbative QCD, 
although with lower accuracy than for QED due to the 
 possible importance of higher-order corrections. At low energies when the QCD coupling becomes of the order unity  perturbation theory  
breaks down. In the nonperturbative regime we have to rely on empirical 
methods, such as ``QCD sum rules'' \cite{qcd_sum_rules} or large computer simulations, where one tries to solve QCD is solved 
by formulating it as a field theory on a lattice. Although the problem is notoriously difficult, the agreement between the spectrum of hadrons, i.e., the masses and quantum numbers of the lowest-lying states with experimentally measured quantities, is quite satisfactory for the most recent numerical simulations \cite{lattice_qcd}. 
 
For processes like proton proton scattering at low energies, the picture 
of strong interactions being due to the exchange of gluons breaks down. 
Instead one may approximate the exchange force as being due to pions and 
other low-mass mesons with surprisingly good results (this is in fact what motivated 
Yukawa to predict the existence of pions). If one wants to make crude  
approximations of the strong interaction cross section in this regime, 
$\sigma_{strong}\sim 1/m_\pi^2$ is a good estimate.

In the perturbative regime at high $Q^2$, the scattering, for example,  
of an electron 
off a proton (`deep inelastic scattering') can be treated by the successful  parton model. Here, the momentum of 
a hadron at high energy is shared between its different constituents. Among the constituents 
are of course the quarks that make up the hadron (two $u$ and one $d$ 
quarks in the case of the proton), i.e., the valence quarks. 
In addition, there may be pairs of quarks and antiquarks produced through 
quantum fluctuations at any given ``snapshot'' of the hadron. The incoming exchange photon sent 
out from an electron in $ep$ scattering may hit these ``sea quarks'', which will 
therefore contribute to the scattering process. 
 
Since the partons interact with each other, they can share the momentum 
of the proton in many ways. Thus, there will be a probability distribution, 
$f_i(x,Q^2)$, for a parton of type $i$ (where $i$ denotes  
any quark, antiquark or gluon) to carry  a fraction $x$ of the proton 
momentum. These functions cannot be calculated from first principles. 
However, once determined (by guess or by experimental information 
from various scattering and annihilation processes) at a particular value of $Q_0^2$, then the evolution 
of the structure functions with $Q^2$ can be predicted. This analysis, 
as first convincingly performed by Altarelli and Parisi, gives rise to a predicted 
variation of the deep inelastic scattering probability with $Q^2$ 
(so-called scaling violations) which has been successfully compared  
with experimental data. The success of the perturbative QCD program, including the running of $\alpha_s$ in agreement with the asymptotic freedom prediction, and the agreement of scaling violations in several processes with data, resulted in a Nobel Prize for Gross, Wilczek and Politzer in 2004.

With the successful QCD parton model, we can now compute many electromagnetic and weak 
processes, including those 
 when hadrons are involved. For instance, the neutrino proton 
scattering cross section is be given by the scattering of a neutrino on 
a single quark or antiquark. This calculation is easily done in a way similar  
to how we computed the $\bar\nu_{e}+e^-\to  
\bar\nu_{\mu}+\mu^-$ cross section. The only change is that the contributions 
from all partons have to be summed over, and an integral of $x$ performed.

As an example, we give the expression for the electromagnetic cross section  
$p+p\to \mu^++\mu^-$, which is called the Drell-Yan process, in the QCD 
parton model. 
The fundamental process must involve charged partons, i.e., quarks (since we assume that strong interactions dominate and thus neglect 
the weak contribution), $q+\bar q\to \gamma^*\to \mu^++\mu^-$, with 
the (valence) quark taken from one of the protons and the (sea) antiquark from the other. 
The momentum transfer in the process is $Q^2=\hat s$, 
where $\hat s =(p_{\mu^+}+p_{\mu^-})^2$. We know from  
(\ref{eq:deepin}) 
that the parton level cross section is $4\pi\alpha e_q^2/3\hat s$ (where 
we have to take into account that the quark charge $e_q$ is not the unit  
charge). Since the parton from proton~1 carries the fraction $x_1$ 
and that from proton~2  $x_2$ of the respective parent proton, 
$\hat s=x_1 x_2 s$, with $s=(p_1+p_2)^2$. The total cross section for producing 
a muon pair of momentum transfer $\hat s$ is thus 
\beqa 
{d\sigma\over d\hat s}={4\pi\alpha^2\over 3\hat s}\times\  
\ \ \ \ \ \ \ \ \ \ \ \ \ \ \ \ \ \ \ \ \ \ \ \ \ \ \ \ \ \  
\ \ \ \ \ \ \ \ \ \ \ \ \ \ \ \ \ \ \ \ \ \ \ \ \ \ \ \ \ \  
 & &\nonumber \\ 
k_c\sum_qe_i^2\int_0^1dx_1\int_0^1dx_2\left[f_q(x_1,\hat s)f_{\bar q}(x_2,\hat s)+ 
f_{\bar q}(x_1,\hat s)f_{q}(x_2,\hat s) 
\right]\delta(\hat s -x_1x_2s) & &\nonumber  
\eeqa 
Here $k_c$ is a colour factor, which takes into account that for a given 
quark of a given colour, the probability to find in the other proton an 
antiquark with a matching (anti-) colour is $1/3$. Thus, in this case 
$k_c=1/3$. In the reverse process, $\mu^+ + \mu^-\to q+\bar q$, all the 
quark colours in the final state have to be summed over (each contributes 
to the cross section), so in that case $k_c=3$. 
 
\subsection{Neutrinos} 
Neutrinos would  
provide an important contribution to the total energy density of the  
universe if they had a mass in the eV range. Present-day analyses of the microwave background and 
 the matter distribution favour as we mentioned, cold dark matter over hot dark matter which eV-scale Standard Model neutrinos would constitute.  
However, neutrinos are plausibly part of dark matter, as one of the most important discoveries the last 15 years has been the definitive demonstration that 
neutrinos are not massless. It has been shown that neutrinos species can transform into each other through quantum mechanical mixing, mean that in principle they could also decay into each other, e.g., by emitting a photon. The huge mass scale of the $W$ and $Z$ bosons means, however, 
that the lifetime also for the heaviest neutrino is much longer than the  
age of the universe, so they are effectively stable.  
Their production through thermal processes in the early 
universe would then mean that their number density is of the same order of magnitude today as that of the microwave background photons, and since they are electrically neutral they qualify as dark matter. This was in fact one of the most studied 
candidates for dark matter in the 1970's, when one did not know how many different neutrino types there were (this was fixed in the 1990's to 3 standard species, thanks to the CERN LEP experiments), neither very much about their mass. Today's knowledge give estimates of far less than a percent for their contribution to 
$\Omega$, however. 
 
There is of course a possibility that there may exist neutrinos with weaker 
couplings than standard model ones, and if they have mass in the keV range they 
would qualify as ``warm'' dark matter, that would still be allowed from 
structure formation. However, this involves some fine-tuning to get the correct 
relic density and to avoid other experimental bounds \cite{shaposhnikov}. 
   
Neutrinos  
are also very important information carriers from violent astrophysical   
processes, which is why neutrino astrophysics is a very active field of research 
at the present time. An attractive property of neutrinos is in fact  
their feeble interactions at low energies, which means that they may 
penetrate regions with dense matter, e.g., the solar interior, or near remnants of supernovae, without being absorbed.   
Where other particles become trapped or can only propagate through very slow  
diffusive processes (for instance, it takes on the order of a million years for a photon created near the centre of the sun to diffuse out to the solar surface), neutrinos are able to escape. 
Neutrinos can thus connect regions of matter that would otherwise be isolated 
from each other.  
Because they almost massless, they move effectively with the speed of light,  
which makes energy transfer (i.e., radiative heat conduction)  very efficient, e.g, from the interior of the sun.   Unfortunately, the fact that neutrinos are so weakly interacting, also means that they are extremely difficult to detect. As of today, the only neutrinos of astrophysical interest that have been detected are those created in the fusion processes in the solar interior, and the exceptional supernova in 1987, where a handful of neutrinos was detected a few hours before it was spotted optically in the Southern sky.

Neutrinos are the  
neutral counterparts of the charged leptons:  
$e, \mu$ and $\tau$. There are therefore three types of ``active'' neutrinos in  
the Standard Model of particle physics: $\nu_e$, $\nu_\mu $ and 
$\nu_\tau$. Neutrinos are fermions, i.e., spin-$\frac{1}{2}$  
particles. 
Apart from their possible gravitational interactions,  
$\nu$s interact  with matter  only through  the exchange of the mediators  
of the weak force, the  $W$ and $Z$  
bosons. They are fundamental particles 
without constituents, as far as is known,  
have extremely small masses 
and lack electric charge.  
 Among the most spectacular events in astrophysics are supernova explosions. 
In a few seconds, more energy is released in neutrinos from the forming  
neutron star than all the electromagnetic emission from an entire galaxy 
over a decade. 
 
Neutrino interactions with matter are divided into two kinds, neutral  
current (NC) interactions  mediated by the neutral Z bosons, and  charged 
current (CC) interactions involving the exchange of  
W$^+$ and W$^-$ bosons. 
 NC interactions are responsible for annihilation reactions involving 
neutrinos, \[e^{+}+ e^{-} \rightarrow \nu_{\mu}+ \bar 
\nu_{\mu} \]  
for example, and elastic scattering interactions such as  
\[ \nu_\mu+ e^- \rightarrow \nu_\mu+ e^-. \] 
 
In CC interactions there is a change of fermion type, of ``flavour''. For example, an  
antineutrino can be absorbed by a proton, producing a neutron and a positron 
in the final state. This comes about because at the parton level a $u$-quark
in the proton is changed into a $d$-quark, which means it is transformed to a neutron. In this process charge is transferred, both for the leptons as the neutrino becomes a charged lepton, and  for the hadrons as the positively charged proton becomes a neutron.

\subsection{Neutrino Interactions} 
 
For the neutrino process (the flavour-changing charged current interaction) 
$\bar\nu_{e}e^-\to  
\bar\nu_{\mu}\mu^-$  
the cross section at low energies (but still high enough to produce the heavier muon) is 
\beq 
\sigma\left(\bar\nu_{e}e^-\to  
\bar\nu_{\mu}\mu^-\right)\sim {g_{w}^4 s\over  
96 \pi 
m_{W}^4} 
\eeq 
 
Before it was known that $W$ bosons existed, Enrico Fermi had written a  
phenomenological theory for weak interactions with a dimensionful  
constant (the Fermi constant) $G_{F}$. (Enrico Fermi is of course also well-known today as the one who has been honoured by giving the $\gamma$-ray satellite, FERMI, its name. This has to do with his description of acceleration processes in astrophysics.) The relation between Fermi's constant and the gauge theory quantities is 
\beq 
{G_{F}\over \sqrt{2}}={g_{w}^2\over 8 m_{W}^2}\simeq  
1.166 \cdot 10^{-5} \  \mbox{GeV}^{-2} 
\eeq 
Using the Fermi  constant, the cross section can now be written  
\beq 
\sigma\left(\bar\nu_{e}e^-\to  
\bar\nu_{\mu}\mu^-\right)={G_{F}^2 s\over 3\pi}. 
\eeq 
 
We note that the cross section rises with $s\simeq 2E_{\nu}m_{e}$ and  
thus 
  linearly with neutrino energy. When $s$ starts to approach $m_{W}^2$, 
  the $W$ propagator $1/(s-m_W^2)$ has to be treated more carefully.  
It can be improved by writing it in the so-called Breit-Wigner form 
\beq 
{1\over s-m_W^2}\to {1\over s-m_w^2 +i\Gamma m_W} 
\eeq 
where $\Gamma$ is the total decay width (around 2 GeV) of the $W$. We 
 see from this that a substantial enhancement of the cross section is 
 possible for $s\simeq m_{W}^2$. This is an example of a resonant 
 enhancement in the $s$-channel. For a target electron at rest, 
 this resonance occurs at around $6.3$ PeV and is sometimes called the Glashow 
 resonance. If astrophysical sources exist which produce  
 electron antineutrinos 
 with  such high energies, the prospects of detecting  
 them would be correspondingly enhanced. However, well above the resonance, 
 the cross section will again start to decrease like  
 $1/s$, just as in the electromagnetic case, $e^+e^-\to\mu^+\mu^-$. 
  
 It should be noted that the latter process, $e^+e^-\to\mu^+\mu^-$, 
 also receives a contribution from an intermediate $Z$ boson. At low 
 energies this is negligible, but due to the resonant  
 enhancement it will dominate near $s\simeq m_{Z}^2$. This is the  
 principle behind the $Z$ studies performed at the LEP accelerator at  
 CERN (where all other fermion-antifermion pairs of the Standard  
 Model were also 
 produced except for $t\bar t$, which was not kinematically allowed). In a full  
 calculation, the two contributions have to be added coherently and may 
 in fact interfere in interesting ways, producing for example, a  
 backward-forward asymmetry between the outgoing muons. 
 
A detailed calculation for  
neutrino energies above around 5 MeV  shows that the total cross section for the  
reaction $\nu_X e^- \rightarrow \nu_X e^- $ is well approximated  
by \cite{Raffelt}: 
 
\begin{equation} 
 \sigma_{\nu e} = C_X \cdot 9.5 \cdot 10^{-45}  \cdot 
       \left(\frac{E_\nu}{\mbox{1\ MeV}} \right) \ {\rm cm}^2  
 \label{sigma_nue} 
\end{equation}  
where the flavour-dependent constants $C_X $ are 
 
\beq 
 C_{e}       = 1  
 \eeq 
 and 
\beq  
 C_{\mu}    =  C_{\tau}    = \frac{1}{6.2}  
\eeq 
 
The cross section is larger for electron neutrinos as they can, unlike 
the other neutrino species,  
couple to the electrons in the target through both NC and CC interactions.

Laboratory experiments have, so far, not succeeded in  directly measuring the mass of 
any neutrino. Instead, the negative results have been expressed in the form 
of upper limits, due to the finite resolution of the experiments. The best 
(lowest) upper limits on the mass of the electron neutrino come from the studies 
of the electron energy spectrum in tritium decay: 
 
\[ ^3{\rm H} \rightarrow ^3{\rm He} + e^- + \bar \nu_e \] 
 
As the minimum amount of energy taken by the $\nu_e $ is its mass,  
the end-point energy of the emitted electron is a measurement of 
$m_{\nu_e}$. According to these experiments the mass of the electron neutrino 
is lower than 3 eV at the 95 per cent confidence level \cite{PDG}. With 
the KATRIN experiment being constructed in Karlsruhe, one hope to decrease this  upper limit (or find a non-zero value) by an order of magnitude \cite{katrin}.  
 
The discovery of the tau neutrino was announced in  2000 by the DONUT collaboration at Fermilab, through appearance in charm meson decays in photographic emulsion.

Mixing of neutrino species is a very interesting quantum mechanical effect which may occur if the weak-interaction eigenstates  
$\nu_e$, $\nu_\mu$ and $\nu_\tau$  are not  the mass eigenstates that propagate in vacuum.  We can then  express a flavour or weak-interaction neutrino eigenstate, 
 $\nu_f$,  as a linear superposition of orthogonal mass eigenstates, $\nu_m$: 
 
\[ |\nu_f>  = \sum_{m}^{} {c_{fm} |\nu_m> }. \] 
 
Of course, all three neutrinos may mix, but it is instructive to see what 
happens if just two of the neutrinos mix, e.g, $\nu_\mu\rightleftarrows \nu_e$ 
mixing with mixing angle $\theta$. 
The time evolution of a muon neutrino wave function, produced e.g. in pion 
decays,  with  
momentum $p$ is then 
 
\begin{equation} 
|\nu_e(t)> = -\sin{\theta} e^{-iE_1 t} |\nu_1> +  
               \cos{\theta} e^{-iE_2 t}|\nu_2>  
\label{oscillation} 
\end{equation}  
with $E_1$ and $E_2$ are the energies of the two mass eigenstates.  
Two energy levels arise if $\nu_1$ and $\nu_2$ have different masses, for the same momentum, $p$. Then, for  small 
neutrino masses $m_i \ll E_i$,  
\begin{equation} 
 E_i = p + \frac{m^2_i}{2 p} 
\label{smallm} 
\end{equation} 
 
The probability $P(\nu_e \rightarrow \nu_e) = {| <\nu_e|\nu_e>| }^2$, 
that an electron neutrino remains a $\nu_e$ 
after a time $t$ then becomes 
 
\begin{equation} 
P(\nu_e \rightarrow \nu_e) 
  = 1 - \sin^2{(2\theta)}  
 \sin^2{[\frac{1}{2}(E_2 - E_1)t]}\label{eq:nuenue} 
\end{equation} 
 
For very small neutrino masses, using (\ref{smallm}), 
\begin{equation} 
P(\nu_e \rightarrow \nu_e) 
  = 1 - \sin^2{(2\theta)} \sin^2{\left[\left(\frac{m^2_2 - m^2_1}{4E}\right)t 
\right]} 
\label{Pe_e} 
\end{equation} 
where $E$ is the energy of $\nu_e$. 
 
Thus the probability  the electron neutrino transforming to 
 a muon neutrino at a time $t$ is 
 
\begin{equation} 
P({\nu_e \rightarrow \nu_\mu }) =  
 \sin^2{(2\theta)} \sin^2{\left[\frac{\Delta m^2}{4E} t \right]} 
\label{pe_mu} 
\end{equation} 
where $\Delta m^2 = |m^2_2 - m^2_1|$.

From (\ref{pe_mu}) it is seen that 
the probability function for flavour change oscillates, 
with an amplitude given by $\sin^2(2\theta)$ and oscillation 
frequency $\sim \Delta m^2/E$. This is now the generally accepted reason 
for the  deficit of solar 
electron neutrinos, as deduced by combining data from the Super-Kamiokande 
experiment in Japan (most recently \cite{super-k}), which sees the deficit of electron neutrinos, with SNO in Canada, which has measured the neutral current cross section, which shows no deficit \cite{sno}. As the neutral current has the same strength for all three neutrinos, this is strong evidence that the total flux is unchanged, 
but the flux of electron neutrinos has decreased due to mixing. 
 
Numerically, the oscillation length becomes 
\begin{equation} 
 L_\nu = 1.27 \left( \frac{E}{1\, {\rm MeV}}\right)  
 \left(\frac{1\ {\rm eV}^2}{\Delta m^2 }\right) \ {\rm metres.} 
 \label{losc} 
\end{equation} 
In fact, a direct proof that oscillations occur in the (anti-) neutrino sector is given by recent results from the KamLAND experiment \cite{kamland}, where reactor antineutrinos have been shown to oscillate over more than one period of oscillation in $L/E$. 
\subsection{Atmospheric Neutrinos} 
Neutrinos are copiously produced in the atmosphere by hadronic and muonic decays following the  
interaction of cosmic rays, 
\begin{eqnarray} 
\left\{\begin{array}{cccc} 
 p/n + N  \rightarrow &  \pi^+/K^+ + ...  &   &    \\ 
      &  &  \pi^+/K^+  \rightarrow  \mu^+ + \nu_\mu &  \\ 
      &  &  &  \mu^+ \rightarrow e^+ + \bar \nu_\mu + \nu_e, \\ 
 \end{array} \right. \nonumber  \\    
\left\{\begin{array}{cccc} 
 p/n + N  \rightarrow & \pi^-/K^-  + ...  &  &  \\ 
         &  &  \pi^-/K^-  \rightarrow  \mu^- + \bar \nu_\mu  &  \\ 
         &  &  &   \mu^- \rightarrow e^- + \nu_\mu + \bar \nu_e  \\ 
 \end{array} \right. 
\end{eqnarray}  
 
Studying the end result of these reactions   
one expects that there are about twice as many muon neutrinos  
than electron neutrinos produced in the atmosphere: 
\begin{equation} 
 \frac{\phi_{\nu_\mu} + \phi_{\bar \nu_\mu}} 
 {\phi_{\nu_e }+ \phi_{\bar \nu_e}} = 2   
 \label{atmosratio}  
\end{equation} 
 
This expectation holds at low energies. At higher energies, additional  
effects have to be taken into account: for example, the competition between  
scattering and decay of the produced pions, and also time dilation. 
 As the energy spectrum of the primary nuclei reaches out to $\sim$10$^{20}$  
eV,  
one expects neutrinos to be produced up to comparable energies.  
 
Due to  the complicated chain of reactions in  
the cascade computer simulations are needed 
to find the differential spectrum of atmospheric neutrinos.  
One finds that there is a broad peak around 0.1 GeV  
($\sim$1 cm$^{-2}$ s$^{-1}$)  and at very 
high energies, $E_\nu$ much larger than TeV, a falling flux $\sim E^{-3.7}$.  
 
The cross section for  neutrino-nucleon interactions in a target can be  
calculated by inserting the nucleon mass instead of $m_e$  in our previous example. In the region of the maximum flux of 
atmospheric neutrinos the cross section is  
$\sigma_{\nu N}\sim$10$^{-39}$ cm$^2$. The Super-K experiment showed that also atmospheric neutrinos oscillate, and that part of the 
muon neutrinos disappear due to $\nu_\mu\to\nu_\tau$ mixing taking place.  
(Due to the 
high $\tau^\pm$ lepton mass (1.8 GeV) the $\nu_{\tau}$s generated by  
mixing will not have enough energy on average to make the charged  
current interaction $\nu_{\tau}+N\to \tau + X$ kinematically possible. 
Their contribution to the neutral current  events is too small  
to be easily detected.)  The Super-K data on the angular and energy dependence of the  
oscillation is  
in excellent agreement with the $E/L$ ratio given by (\ref{losc}). 
 
\subsection{{Neutrinos as Tracers of Particle Acce\-le\-ra\-tion}} 
 
A kiloton-size detector is necessary 
to observe neutrinos from sources as close as  the Earth's atmosphere,  
or the Sun (this actually gave a Nobel Prize to Davies and Koshiba, in 2002).  
To  
be able to search other astrophysical objects, the required 
detector mass becomes very large, megaton or even gigaton (or volume of order km$^3$). 
 
Consider the $\nu_\mu \to \mu$ charged current weak interaction in a medium,  
\begin{center} 
$  \nu_\mu $ + N $\rightarrow$ $ \mu $ + ..., 
\end{center} 
where N is a nucleon in the medium in or  surrounding the detector. 
The muon range rises with energy, and around 1 TeV ($10^{12}$ eV) it  
is more than 
one kilometre. The detection area is therefore greatly enhanced at high  
energies. In water or ice, a good approximation of the muon range as a function of  
energy is given by 
\begin{equation} 
 R_\mu \approx  
 2.5 \ln{ \left(2 \cdot \frac{E_\mu}{\mbox{1 TeV}} + 1\right)} \ \mbox{km} 
 \label{muonrange} 
\end{equation} 
T 
he muon produced conserves, on average, the direction of the 
incoming neutrino. The average of the square of the $\nu_\mu$-$\mu$ angle 
is approximately (see \cite{BGbook}) 
 
\begin{equation} 
\sqrt{<\theta^2>} \approx 2   
 \left(\frac{\mbox{1 TeV}}{E_\nu} \right)^\frac{1}{2}   \ \mbox{deg}. 
 \label{angneu} 
\end{equation} 
 
The cross section for neutrino interaction with 
a fixed target rises linearly with energy. Neutrino telescopes for very high energies become 
efficient at a few GeV, where the product of the neutrino-matter cross section 
and the muon range rises approximately as $E^2_\nu$.  
Above 1 GeV, the induced flux 
of muons from atmospheric neutrinos, for example, is  
about 1 m$^{-2}$ year$^{-1}$.

This detection scheme does not work as well for other types of 
neutrinos. Electrons (from $\nu_e $ + N $\rightarrow$ e + \ldots) have a very short range as they lose energy through radiative processes, due to their small mass. On the other hand $\tau$ leptons, the heaviest known charged 
leptons with $m_\tau$ = 1.78 GeV, are 
produced in charged current interactions  of 
$\nu_\tau$, but they are very short lived ($t_\tau\sim 10^{-13}$ s). 
Therefore they are 
not suitable for detection, except for the fraction of times where the $\tau$ 
decays into $\mu \bar \nu_\mu \nu_\tau$, which happens in roughly 20 percent of the  
cases. However, in large neutrino detectors such as the IceCube, one may perhaps 
 detect ultra-high-energy electron and $\tau$ neutrino events by 
the intense cascade of light that is produced by secondary electrons, positrons 
and photons. In the case of $\tau$ neutrinos, special relativity may help to 
produce a good signature. If sources of PeV ($10^{15}$ eV) $\tau$ neutrinos 
exist, the produced charged $\tau$ lepton would have a relativistic $\gamma$ 
factor as large as 
\beq 
\gamma\sim{E_\nu\over m_\tau}\sim 10^6 
\eeq 
which means, thanks to time dilation,  that in the detector  
reference frame the $\tau$ lepton will travel 
a distance 
$\ell \gamma c t_\tau\sim 100\ {\rm m}.$ 
 
The ``double bang'' created by the charged current interaction  and the 
subsequent decay of the $\tau$ lepton, separated by 100 m, would be the 
signature of PeV $\tau$ neutrinos. 
 
Since neutrinos oscillate, very energetic $\tau$ neutrinos could be produced 
by mixing with muon neutrinos created in high-energy pion decays in  
cosmic accelerators. This is, e.g., the basis for the experiments OPERA \cite{opera} at Gran Sasso 
and MINOS \cite{minos} at Fermilab, where a few events of produced $\tau$ leptons have in 
fact been reported. 
 
In present detectors, only  neutrino-induced muons moving upwards in the  
detectors (or downwards but near the horizon) 
are safe tracers of neutrino interactions. Most muons moving  
downwards have 
their origin in cosmic-ray nuclei interacting with the Earth's atmosphere and produce a very difficult background.  
At the surface of the Earth, the flux of  
downward-going muons produced in the atmosphere 
is about 10$^6$ times larger than the flux of neutrino-induced  
 upward-moving muons.  
 
By going underground, the 
material (rock, water, ice, etc.)  above the detector attenuates the  
flux of atmospheric muons considerably. In addition, if it is experimentally  
possible to select events where a muon is moving upwards the Earth  
itself acts as a filter since only neutrino-induced muons 
can be produced upward-going close to the detector.  

\subsection{{AMANDA, IceCube and Direct Detection of WIMPs}} 
 
Neutrinos may give clues to the dark matter problem in another way  
than just being a small part of the dark matter due to their tiny mass. 
If the dark matter has a component that is massive and weakly  
coupled (electrically neutral) it will be non-relativistic at  
freeze-out, which is of course the WIMP paradigm of  cold dark matter. 
A good template for a dark  
matter WIMP candidate is as we mentioned the lightest supersymmetric particle -- plausibly  
the  
neutralino $\chi$ (see Section~\ref{ch:susy} for more details). 
 
 Neutralinos (or other WIMPs) have interactions with ordinary matter  
which are equally as small as those of neutrinos. However, since they  
move with non-relativistic velocity there is a chance that they become  
gravitationally trapped inside, for example, the Sun or the Earth. A  
neutralino scattering e.g., in the Sun will lose energy and fall further inside  
the solar volume, and successive scatterings in the solar medium will soon  
make it lose more and more energy. In the end, neutralinos will assemble near the centre. As they are their own antiparticles (they are Majorana fermions), they  
can annihilate with each other, resulting in ordinary particles  
(quarks, leptons, gauge particles).  
 
As the annihilation rate is proportional to the scattering rate, and the interior of the Earth is almost entirely spin-0 nuclei, constraints on the spin-independent scattering rate from experiments described in section
\ref{ch:4} mean that  
neutrinos from the center of the  Earth are not a very promising signal for canonical WIMPs. However, as the Sun  
consists to some 75 \% of single protons (i.e., hydrogen nuclei) with spin-1/2, spin-dependent scattering is important and searching for neutrinos from the Sun stands well in competition with other experiments. We will return also to this later.

Most of the annihilation  
products in the Sun create no measurable effect; they are just stopped and  
contribute somewhat to the energy generation. 
However, neutrinos have the unique property that  
they can penetrate the whole Sun without being much absorbed, at least for 
WIMPs less massive than a few hundred GeV. An annihilating    
neutralino pair  of mass $m_{\chi}$ would  
thus give rise to 
high-energy neutrinos of energy around $m_{\chi}/3$ or so (the  
reason that $E_{\nu}\ne m_{\chi}$ is that other particles created in  
the annihilation process share the energy). The signal of high-energy  
 neutrinos (tens to hundreds of GeV -- to be compared with the `ordinary' MeV 
solar neutrinos) from the centre of the Sun would be an  
unmistakable signature of WIMP annihilation. 
 
The detection of muons in IceCube, for instance, relies on the Cherenkov effect. 
This coherent emission of light follows a characteristic angle given by the 
Mach relation 
$$ 
        \cos\theta = \frac{1}{\beta n} 
$$ 
where $\beta$ is the speed of the particle traversing the medium in  
units of the speed of light. The Cherenkov effect takes place when 
$$ 
        \beta > \frac{1}{n}. 
$$ 
 
Cherenkov radiation constitutes a very small 
fraction of the total energy loss of a charged particle as it crosses 
a medium. The superluminal condition is fulfilled only  
between the UV and near-infrared    
region of the electromagnetic spectrum. In water or ice, for example, where 
the index of refraction for UV and optical wavelengths  
averages around 1.3, the Cherenkov radiation 
cut-off in the UV region is around 70 nm.  
The differential energy loss into Cherenkov photons in water or ice is  
 a few per cent of the total differential 
energy loss of a charged track moving with a speed very close to $c$. 
  
\subsection{Water and Ice Cherenkov Telescopes} 
 
Neutrinos can thus be detected indirectly by the Cherenkov radiation from  
charged leptons and hadrons produced in neutrino  
interactions with  
matter. The  extremely large detector volumes needed to detect  
neutrinos from distances beyond our Sun makes the use of any other material 
than water or ice very difficult.
 
A typical detector consists of an array of light sensors (photomultipliers, PM)  with good  
time resolution ($\sim 1 $ ns) distributed in the medium. The pattern of the  
hit PMs, and relative arrival times, are then used to fit the direction of the 
particle that generated the Cherenkov radiation. 
The correlation between the original direction of  
the neutrino and the produced charged lepton means that one may reconstruct the direction of the incoming neutrino. 
 
 Antares is a good prototype, for a larger detector being planned with the working name KM3NET, near Toulon in the Mediterranean.  
The AMANDA  experiment  
at the South Pole was similarly an excellent working prototype, where 
the disadvantages related to the remote location of the telescope were  
compensated by the virtues of the glacier ice, found to be the clearest  
natural solid on Earth. The Cherenkov photons emitted along the path 
of a muon at some wavelengths be selected hundreds of metres away from  
the muon track.  
 
The AMANDA detector was a great success, but was too small and has recently been
abandoned, replaced by a much larger detector, 
the IceCube, with 80 strings  encompassing roughly a cubic kilometer 
of ice. Construction was finished in 2010, and at that time also a smaller and 
denser inset, the DeepCore detector, was completed. This allows a lower  
detection energy threshold which is particularly beneficial for the WIMP  
search. Unfortunately, despite the heroic effort to build the first large neutrino detector in this remote location, no astrophysical neutrino source including WIMPs has yet been detected, 
but it is only a year that data have been collected (for a recent review of 
dark matter detection in neutrino telescopes, 
see \cite{icecube}).

\section{Supersymmetric Dark Matter}\label{ch:susy} 
As we have mentioned several time already, one of the prime 
candidates for the non-baryonic cold dark matter particle is provided  
by the lightest 
supersymmetric particle, most likely the lightest neutralino $\chi$. Even
it would be that supersymmetry were not realized in nature, the neutralino is still important as a nice, calculable template for a generic WIMP. 
 
In most versions of 
the low-energy theory  which results from the largely unknown mechanism of 
supersymmetry breaking, there is a conserved multiplicative quantum 
number, R-parity: 
\beq 
R=\left(-1\right)^{3(B-L)+2S}, 
\eeq 
where $B$ is the baryon number, $L$ the lepton 
number and $S$ the spin of the particle. This implies that 
$R=+1$ for ordinary particles 
and $R=-1$ for supersymmetric particles. In fact, for phenomenological reasons, this symmetry is required, as its conservation protects the proton from decaying 
through the mediation of $R$-symmetry breaking interactions. The $R$-symmetry  means that supersymmetric 
particles can only be created or annihilated in pairs in reactions 
of ordinary particles. It also means that a single supersymmetric particle 
can only decay into final states containing an odd number of 
supersymmetric particles. In particular, this 
makes the lightest supersymmetric particle 
stable, since there is no kinematically allowed state with negative 
$R$-parity which it can decay to. This is of course of utmost importance for the dark matter problem. Also other WIMP models of dark matter needs  
some mechanism to prevent decay, and the simplest mechanism is a discrete 
symmetry like the double-valued ($Z_2$) $R$-symmetry. (Another reason for stability could be the quantum numbers of the particles in the theory. There are in fact models with high  spin or isospin multiplets which also have a stable particle which could act as dark matter \cite{Cirelli:2007xd}.) 
 
Pair-produced neutralinos in the early universe which 
left thermal equilibrium as the universe kept expanding should, due to their stability, have a non-zero 
relic abundance today. If the scale of supersymmetry breaking is related to 
that of electroweak breaking, the neutralino will act as a WIMP and therefore a dark matter candidate with a relic density of the same order of magnitude as the value implied by the  WMAP measurements. 
This is a very elegant and economical method 
to solve two of the  most outstanding problems in fundamental 
science,  dark matter and the unification of the basic 
forces, if they have a common element of solution - supersymmetry. 
 
\subsection{Supersymmetric Dark Matter Particles}\label{sec:mssm} 
 
If R-parity is conserved, the lightest supersymmetric particle should be stable. 
The most plausible candidate is 
the lightest neutralino $\chi$. As we saw in section~\ref{ch:2} it is a mixture 
of the supersymmetric partners of the photon, the $Z$ and the two neutral 
$CP$-even Higgs bosons present in the minimal extension of the 
supersymmetric standard model. It is electrically 
neutral and thus neither absorbs nor emits light, and is  stable, 
surviving since earliest epoch after the big bang. 
Its gauge couplings and mass means that  
for a large range of parameters in the supersymmetric sector 
a relic density is predicted in the required range to explain 
the observed $\Omega_\chi h^2\sim 0.11$. Its electroweak  couplings to ordinary 
matter also means that its existence as dark matter in our galaxy's halo 
may be experimentally tested. 
 
Unfortunately, very little is known about how  
supersymmetry is broken (for a discussion, see \cite{Intriligator:2007cp}), and therefore any given supersymmetric 
model contains a large number of unknown parameters (of the order of 100). 
Such a large parameter space is virtually impossible to explore by 
present-day numerical methods, and therefore simplifying assumptions 
are needed. Fortunately, most of the unknown parameters 
such as CP violating phases influence the properties relevant 
for cosmology, and for detection, very little.  
 
Usually, when scanning the large space 
of a priori unknown parameters in supersymmetry, one thus makes 
reasonable simplifying assumptions and accepts solutions 
as cosmologically appropriate if they give a neutralino relic 
density in the range 
\beq 
0.09 \lsim \Omega_\chi h^2 \lsim 0.12\label{eq:bound} 
\eeq 
Recently, there has been a number of analyses where the relic density, and other parameters or experimental quantities known within some error bounds are allowed to vary. By using  
so-called Multi-Chain Monte Carlo methods (MCMC), one can get a ``global fit'' of the best-fit models using statistical methods \cite{trotta,scottetal}.  
Usually, one employs what is called a Bayesian method which need some assumption about the prior distribution of probabilities. In the case of mass parameters one may, for instance, choose linear or logarithmic scans. If experimental data are good enough, it can be shown that the choice of priors is not crucial. However, so far there has been a lack of experimental information, meaning that the predicted most likely regions in parameter space may depend quite sensitively 
on priors (see, e.g., \cite{gianfranco_11}). Hopefully, the situation may soon  
change with new results from the LHC. A drawback of the method of global fits is that it is very computer intensive, meaning that only very simplified models of supersymmetry have been fully investigated so far.

Besides its interesting implications for cosmology, the motivation 
from particle physics for supersymmetric particles 
at the electroweak mass scale 
has  become stronger due to 
the apparent need for 100 GeV - 10 TeV scale supersymmetry to achieve 
unification of the gauge couplings in view of LEP results. (For an extensive review of the literature on 
supersymmetric dark 
matter up to mid-1995, see Ref.\,~\cite{jkg}. More recent reviews are \cite{hoopersilk} and \cite{lb_new}). 
 
A great virtue of supersymmetry 
at the phenomenological level is that it  gives an attractive solution to the 
so-called hierarchy problem, which is to understand why the 
electroweak scale at a few hundred GeV is so much smaller than the Planck scale 
$\sim 10^{19}$ GeV despite the 
fact that there is nothing in non-supersymmetric theories to 
cancel the severe quadratic divergences of loop-induced mass terms. 
In supersymmetric theories, the partners of differing spin would 
exactly cancel those divergencies (if supersymmetry were unbroken). 
Of course, supersymmetric models are not guaranteed to contain good 
dark matter candidates, but in the simplest models $R$-parity is conserved 
and the neutralino naturally appears as a good candidate.

\subsubsection{The MSSM}

The minimal supersymmetric extension 
of the standard model is defined by the particle content and 
gauge couplings required by supersymmetry and a gauge-invariant 
  superpotential. Thus, to each particle degree of freedom in the 
non-supersymmetric Standard Model, there appears a supersymmetric partner 
with the same charge, colour etc, but with the spin differing by half 
a unit. The only addition to this  doubling of the 
 particle spectrum of the Standard Model concerns the Higgs sector. It 
 turns out that the single scalar Higgs doublet is not enough to give 
 masses to both the $u$- and $d$-like quarks and their 
 superpartners (since supersymmetry forbids using both 
a complex Higgs field and its complex conjugate at the same time, which 
one does in the non-supersymmetric Standard Model). 
Thus, two complex Higgs doublets have to be 
 introduced. After the usual Higgs mechanism, three of these states 
 disappear as the longitudinal components of the weak gauge bosons 
 leaving five physical states: two neutral scalar Higgs particles $H_{1}$ and 
 $H_{2}$ (where by convention $H_{2}$ is the lighter state), one 
 neutral pseudoscalar state $A$, and two charged scalars $H^{\pm}$. 

 The $Z$ boson mass gets a contribution from the vacuum expectation 
 values $v_i$ of both of the doublets, 
\begin{equation} 
  \langle H^1_1\rangle = v_1 , \qquad \langle H^2_2\rangle = v_2, 
\end{equation} 
with $g^2(v_1^2+v_2^2) = 2 m_W^2$. One assumes that 
vacuum expectation values of all other scalar fields (in particular, 
squark and sleptons) vanish, as this avoids color and/or charge breaking 
vacua. 

  The supersymmetric theory  also  contains the 
 supersymmetric partners of the spin-0 Higgs doublets. In particular, 
 two Majorana fermion states, higgsinos,  appear as the supersymmetric 
 partners of the electrically neutral parts of the 
$H_{1}$ and $H_{2}$ doublets. These can mix 
quantum mechanically with each other and 
 with two other neutral 
 Majorana states, the supersymmetric partners of the photon (the photino) 
 and the $Z$ (the zino). When diagonalizing the mass matrix of these 
 four neutral Majorana spinor fields (neutralinos), the lightest physical state 
 becomes an excellent candidate for cold dark matter, CDM. 

The one-loop effective potential for the Higgs fields 
 has to be used used to obtain  realistic Higgs mass estimates. 
 The minimization conditions of the potential allow one to 
trade two of the Higgs potential parameters 
for the $Z$ boson mass $m_Z^2 = {1\over2} (g^2+g'^2) 
(v_1^2+v_2^2)$ (where $g=e/\sin\theta_W$, $g'=e/\cos\theta_W$) and the ratio of VEVs, $\tan\beta$.
This ratio of VEVs 
 \beq 
 \tan\beta\equiv {v_{2}\over v_{1}} 
 \eeq 
 always enters as a free parameter in the MSSM, although it seems 
 unlikely to be outside the range between around 1.1 and 60, with some preference for the higher values. 
The third parameter 
can further be re-expressed in terms of the mass of one of the physical 
Higgs bosons, for example $m_{A}$. 

\subsection{Higgs and Supersymmetry}
At the ATLAS and CMS experiments at the CERN Large Hadron Collider, a discovery of the Higgs particle has not yet been claimed (by the end of 2011), as the statistical significance is still below the wanted 5 $\sigma$ (standard deviations). However, there are intriguing indications showing up at more than 3$\sigma$ at a mass value around 125 GeV. If this would stand when more statistics is gathered in 2012, it could means that the Standard Model of particles and fields would be completed with a most wanted spin-0 boson, the Higgs particle. Moreover, a mass below 130 GeV is a firm prediction of supersymmetry, so it may also show the way to a whole new phenomenology, 
including a very interesting dark matter candidate - the lightest supersymmetric particle, generally thought to be the neutralino. This is a quantum mechanical mixture of the supersymmetric partner of the photon, the neutral weak gauge boson and the neutral spin-1/2 partners of each of the two Higgs doublets which are needed by supersymmetry
.  
In  
supersymmetric theories, 
the most likely dark matter candidate is a quantum mechanical superposition, called the neutralino $\chi$ 
 of electrically neutral supersymmetric fermions.
 
Of course, if the 125 GeV Higgs also signals the presence of supersymmetry, then a rich spectrum of particles, several of which may be in reach kinematically at the LHC, is expected. Even if supersymmetry is not realized in nature, it will continue to play a role as an important template for dark matter, as the neutralino is a very attractive, calculable candidate for a generic WIMP. We will return to this later. 
 
\subsection{The Neutralino Sector}

The neutralinos $ \tilde{\chi}^0_i$, of which the lightest is the dark matter candidate, are linear combination of the 
neutral gauge bosons ${\tilde B}$, ${\tilde W_3}$ (or equivalently 
$\tilde\gamma$, $\tilde Z$) and of the neutral 
higgsinos ${\tilde H_1^0}$, ${\tilde H_2^0}$.  In this basis, their 
mass matrix 
\begin{eqnarray} 
  {\cal M} = 
  \left( \matrix{ 
  {M_1} & 0 & -{g'v_1\over\sqrt{2}} & +{g'v_2\over\sqrt{2}} \cr 
  0 & {M_2} & +{gv_1\over\sqrt{2}} & -{gv_2\over\sqrt{2}} \cr 
  -{g'v_1\over\sqrt{2}} & +{gv_1\over\sqrt{2}} & 0 & -\mu \cr 
  +{g'v_2\over\sqrt{2}} & -{gv_2\over\sqrt{2}} & -\mu & 0 \cr 
  } \right) 
\end{eqnarray} 
can be diagonalized  to give four neutral Majorana states, 
\begin{equation} 
  \tilde{\chi}^0_i = 
  a_{i1} \tilde{B} + a_{i2} \tilde{W}^3 + 
  a_{i3} \tilde{H}^0_1 + a_{i4} \tilde{H}^0_2\label{eq:mix} 
\end{equation} 
($i=1,2,3,4$) the lightest of which,  $\chi_1^0$ or simply $\chi$, is then the candidate for 
the particle making up the dark matter in the universe (in section~\ref{ch:2} we called the coefficients $a_{1j}$ $N_j$.). 
 
The coefficients in (\ref{eq:mix}) are conveniently normalized such that 
for the neutralino 
\beq 
\sum_{j=1}^4|a_{1j}|^2=1.
\eeq 
The properties of the neutralino are quite different depending 
on whether is consists mainly of gaugino ($j=1,2$) or higgsino ($j=3,4$) 
components. We therefore define a parameter, $Z_g$, 
which tells the size of the gaugino fraction: 
\beq 
Z_g=\sum_{j=1}^2|a_{1j}|^2. 
\eeq 
A neutralino is often said to be gaugino-like if $Z_g\gsim 0.99$, 
higgsino-like if $Z_g\lsim 0.01$, and mixed otherwise. 
 
For simplicity, one often makes a 
diagonal ansatz for the 
 soft supersymmetry-breaking parameters in the sfermion sector. 
This allows the squark mass matrices to be diagonalized analytically. 
Such an ansatz implies the absence of 
tree-level flavor changing neutral currents (FCNC) in all sectors of the 
model. In models inspired by low-energy supergravity with 
a universal scalar mass at the grand-unification (or Planck) scale 
 the running of the scalar masses down to the electroweak scale 
generates off-diagonal terms and tree-level FCNC's in the squark 
sector.  
 
In the estimates of detection rates here, we will adhere to a purely phenomenological approach, where the 
simplest unification and scalar sector constraints are assumed, and 
no CP violating phases outside those of the Standard Model, but no 
supergravity relations are used. This reduces the number of free parameters 
to be scanned over in numerical calculations to seven: $\tan\beta$, 
$M_1$, $\mu$, $m_A$, and three parameters related to the sfermion 
sector (the exact values of the latter are usually not very important). 
In fact, on can reduce the number of parameters further by choosing, 
e.g., explicit supergravity models, but this only corresponds to 
a restriction to a subspace of our larger scan of parameter space. In fact,
data from the LHC have already excluded large sectors of the simplified models.

 The non-minimal character of the Higgs sector may well be the first 
 experimental hint at accelerators of supersymmetry. At tree level, 
 the $H^ 0_{2}$ mass is smaller than $m_{Z}$, but radiative (loop) 
corrections are important and shift this bound by a considerable amount. 
However, even after allowing for such 
 radiative corrections it can hardly be larger than around 130 GeV. When  
there were some weak indications of a Higgs signature at 140 GeV in LHC data 
reported in mid-2011, this looked like bad news for the MSSM. However, 
with further data, the preferred mass is now around 125 GeV, which is easily accomodated. 
 
\subsection{Experimental Limits}

The successful operation of the CERN accelerator 
LHC at centre of mass energies above 7 TeV without observing any supersymmetric 
particles, in particular squarks of gluinos, puts important constraints on the parameters of the MSSM. However, it may be that the mass scale of neutralinos  
is decoupled from the other supersymmetric particle masses (e.g, in ``split susy'' models \cite{split}).   
 
It has proven to be 
very difficult, however, to put very tight lower limits on the 
mass of the lightest neutralino, because of the multitude of 
couplings and decay modes of the next-to-lightest supersymmetric 
particle. The lightest neutralino can in general only be 
detected indirectly in accelerator experiments 
through the missing energy and momentum it 
would carry away from the interaction region. 
 
The upper limit of dark matter neutralino masses in the MSSM  
of the order of 7 TeV \cite{coann}. Above that mass, which is still 
far from the unitarity bound of  340 TeV \cite{unitarity}, the relic density becomes 
larger than the allowed WMAP upper limit. 
To get values for the lightest neutralino mass larger than a few 
hundred GeV, however, some degree of ``finetuning'' is necessary. (On the other hand, we have seen that for the other important unknown part of the energy density of the universe, the cosmological constant $\Lambda$, a ``finetuning'' of many many orders of magnitude seems also to be necessary.) 
 
By making additional well-motivated but not mandatory 
 restrictions on the parameter space, such as 
in supergravity-inspired models or in simplified constrained MSSM models (CMSSM), one gets in general masses 
below 600 GeV \cite{ellisco,cmssm} for the lightest neutralino, but as mentioned these models are feeling some tension from early LHC data. 

\subsection{Supersymmetry Breaking}

Supersymmetry is a mathematically beautiful theory, and would give 
rise to a very predictive scenario, if it were not broken in an 
unknown way which unfortunately introduces a large number of unknown 
parameters. 
 
Breaking of supersymmetry has  to be present 
since no supersymmetric particle has as yet been 
detected, and unbroken supersymmetry requires 
particles and sparticles to have the same mass. This breaking can be 
 achieved in the MSSM by a soft 
 potential which does not reintroduce large 
 radiative mass shifts (and which indicates that the 
 lightest supersymmetric particles should perhaps not be too much heavier than the 250 
 GeV electroweak breaking scale). The origin of the effective 
 low-energy  potential need not be specified, 
 but it is natural to believe that it is induced through explicit 
 breaking in a hidden sector of the theory at a high mass scale. The 
supersymmetry breaking terms could then transmitted to the visible sector 
 through gravitational interactions. 
 
 Another possibility is that 
 supersymmetry breaking is achieved through gauge interactions at 
 relatively low energy in the hidden sector. This is then 
 transferred to the visible sector through some messenger fields which 
 transform non-trivially under the Standard Model gauge group. 
 However, we shall assume the  ``canonical'' 
 scenario in most of the following. 
 
 Since one of the virtues of supersymmetry is that it may establish grand unification of the gauge interactions at a common mass scale, 
  a simplifying assumption is often used
for the gaugino mass parameters, 
\beq
  M_1 = {5\over 3}\tan^2\theta_wM_2\esim 0.5 M_2,
 \eeq
and
\beq  
M_2 = { \alpha_{\rm em} \over \sin^2\theta_w \alpha_s } M_3 \esim 0.3 M_3, 
  \label{gauginounif} 
\eeq
where $\theta_W$ is the weak mixing angle, $\sin^2\theta_W\approx 0.22$.

When using the minimal supersymmetric standard model 
in calculations 
 of relic dark matter density, one should make sure that all 
accelerator  constraints on supersymmetric particles and couplings are 
imposed. In addition to the significant restrictions on parameters given by 
LEP and LHC, the measurement of the \bsg\ process 
is providing important bounds, since 
supersymmetric virtual particles 
may contribute  significantly to this loop-induced decay. There are also constraints arising if one wants to attribute the slightly abnormal value of $g-2$ for the muon \cite{gm2} to supersymmetric contributions from virtual particles. 
 The relic density calculation in the MSSM for a given set of 
parameters is nowadays accurate to a few percent or so \cite{ds}.

\subsection{Other Supersymmetric Candidates} 
 
Although the neutralino is considered by most workers in the 
field to be the preferred supersymmetric dark matter candidate, 
we mention briefly here also some other options. 
 
If the axion, the spin-0 pseudoscalar field which solves the strong CP problem exists, and if the underlying theory is supersymmetric, 
there should also exist a spin-1/2 partner, the axino. If this 
is the lightest supersymmetric particle and is in the multi-GeV 
mass range, it could compose the cold dark matter of the universe (for a review, see \cite{axinos}). 
 
A completely different type of supersymmetric dark matter candidate 
is provided by so-called Q-balls \cite{kusenko}, non-topological 
solitons predicted to be present in many versions of the 
theory. These are produced in a non-thermal way and may have large 
lepton or baryon number. They could produce unusual ionization signals 
in neutrino telescopes, for example. However, the unknown properties 
of their precise 
formation mechanism means that their relic density may be far below 
the level of observability, and a value around the observationally 
favoured $\Omega_Q\sim 0.22$ may seem fortuitous (for a recent review of the physics of Q-balls, see \cite{qballs}. 
 
Of course, there remains the possibility of dark matter being 
non-supersymm\-etric WIMPs. However,  the interaction cross sections should then 
be quite similar as for supersymmetric particles. Since, the rates 
in the MSSM are completely calculable once the supersymmetry parameters 
are fixed, these particles, in particular neutralinos, serve 
as important templates for reasonable dark matter candidates when 
it comes to designing experiments with the purpose of  detecting 
dark matter WIMPs.

\section{Detection Methods for Neutralino Dark Matter}\label{sec:detection} 
 
The ideal situation would appear if supersymmetry were 
discovered at accelerators, so that direct measurements of the 
mass of the lightest supersymmetric particle, its couplings and 
other properties could be performed. This would give a way 
to check from very basic principle if this particle is a good 
dark matter candidate - if it is electrically neutral and has the 
appropriate mass and couplings to give the required relic density 
to provide $\Omega_\chi h^2\sim 0.11$. So far, no signal of supersymmetry 
has been found at either LEP, Fermilab, or LHC. 
An indirect piece of evidence for supersymmetry would be the discovery 
of a Higgs particle below around 130 GeV, since this is the maximal value
of the lightest Higgs mass after radiative corrections, in the MSSM. In the non-supersymmetric Standard 
Model the Higgs could be much heavier. It is indeed encouraging that  
the first signs of the Higgs at LHC seems to correspond to a mass of 125 GeV. 
 
If we assume a local neutralino halo 
density of $\rho_\chi= 
\rho_\odot\sim 0.4$ GeV/cm$^{3}$ \cite{catena}, and a typical galactic velocity 
of neutralinos of $v/c\sim 10^{-3}$, the flux of particles of mass 100 GeV at 
the location of a detector at the Earth is roughly $10^{9}$ 
m$^{-2}$\,s$^{-1}$. Although this may seem as a high flux, the 
interaction rate has to be quite small, since the correct magnitude 
of $\Omega_\chi h^2\sim 0.11$ is only achieved if the annihilation 
cross section, and therefore by expected crossing symmetry also 
the scattering cross section, is of weak interaction strength.

The rate for direct detection of galactic neutralinos, integrated over 
deposited energy assuming no energy threshold, is 
\begin{equation} 
  R = \sum_i N_i n_\chi \langle \sigma_{i\chi} v \rangle , 
\end{equation} 
where $ N_i $ is the number of nuclei of species $i$ in the detector, 
$n_\chi$ is the local galactic neutralino number density, $ 
\sigma_{i\chi} $ is the neutralino-nucleus elastic cross section, and 
the angular brackets denote an average over $ v $, the neutralino 
speed relative to the detector. 
 
The most important non-vanishing 
contributions for neutralino-nucleon scattering are 
 the scalar-scalar coupling giving a spin-independent 
effective interaction, and the spin-dependent axial-axial interaction,
\beq 
{\cal L}_{\rm eff}=f_{SI}\left(\bar\chi\chi\right)\left(\bar NN\right) 
+f_{SD}\left(\bar\chi\gamma^\mu 
\gamma^5\chi\right)\left(\bar N\gamma_\mu\gamma^5 N\right). 
\eeq 
Usually, it is the spin-independent interaction that gives the 
most 
important contribution in realistic target materials (such as Na, Cs, 
Ge, I, or Xe), due to the enhancement caused by the coherence of 
all nucleons in the target nucleus.

The neutralino-nucleus elastic cross section can be written as 
\begin{equation} 
  \sigma_{i\chi} = {1 \over 4 \pi v^2 } \int_{0}^{4 m^2_{i\chi} v^2} 
  \mbox{\rm d} q^2 G_{i\chi}^2(q^2) , 
\end{equation} 
where $ m_{i\chi} $ is the neutralino-nucleus reduced mass, $q$ is the 
momentum transfer and $G_{i\chi}(q^2) $ is the effective 
neutralino-nucleus vertex. One may write 
\begin{equation} 
  G^2_{i\chi}(q^2) = A_i^2 F^2_{SI}(q^2) G_{SI}^2 + 
  4 \lambda_i^2 J(J+1) F^2_{SD}(q^2) G_{SD}^2 , 
  \label{detrate1} 
\end{equation} 
which shows the coherent enhancement factor $A_i^2$ for the 
spin-independent cross section. A reasonable approximation for the 
gaussian scalar and axial nuclear form factors is  
\begin{equation} 
  F_{SI}(q^2) = F_{SD}(q^2) = 
  \exp(-q^2R_i^2/6\hbar^2) , 
\end{equation} 
\begin{equation} 
  R_i = ( 0.3 + 0.89 A_i^{1/3} )\  {\rm fm} , 
\end{equation} 
which gives good approximation to the integrated 
detection rate \cite{ellisflores} (but is less accurate for 
the differential rate \cite{engel}). Here $\lambda_i$ is related to 
the average spin of the nucleons making up the nucleus. For the relation 
between $G_{SI}$, $G_{SD}$ and $f_{SI}$, $f_{SD}$ as well as a discussion of 
the several Feynman diagrams which contribute to these 
couplings, see e.g.~\cite{bsg,nojiri,bailin}. 
One should be aware that these expressions are at best approximate. A more 
sophisticated treatment (see discussion and references in \cite{jkg}) 
would, however, plausibly change the values by much less than the spread due to 
the unknown supersymmetric parameters.

For a target consisting of $N_i$ nuclei the differential 
scattering rate per unit time and unit recoil energy $E_R$ 
is given by 
\beq 
S_0(E_R) = 
{dR\over dE_R}=N_i{\rho_\chi\over m_\chi} 
\int \,d^3v\,f(\vec v)\,v {d\sigma_{i\chi}\over dE_R}(v,E_R). \label{eq:diffscatt} 
\eeq 
The nuclear recoil energy $E_R$ is given by 
\beq 
E_R={m_{i\chi}^2v^2(1-\cos \theta^*)\over {m_i}} 
\eeq 
where $\theta^*$ is the scattering 
angle in the center of mass frame. 
The range and slope of the recoil energy spectrum is essentially 
given by non-relativistic kinematics. For a low-mass $\chi$, the spectrum is 
steeply falling with $E_R$; interaction with a high-mass $\chi$ 
gives a flatter spectrum with higher cutoff in $E_R$. 
 
The total predicted rate integrated over recoil energy above a given 
generally (detector-dependent) 
threshold can be compared with upper limits coming from various 
direct detection experiments. In this way, limits on the 
$\chi$-nucleon cross section can been obtained as a function of the mass 
 $m_\chi$ 
 \cite{directlimits}. The cross section on neutrons is usually 
very similar to that on protons, so in general only the latter 
is displayed.  
below.  
Major steps forward have been taken in recent years. 
For example, the CDMS-II experiment \cite{cdms-ii} and XENON100 \cite{xenon100} 
have been pushing the limits down by a large factor, reaching now 
$10^{-44}$ cm$^2$ for masses around 50 GeV. 
 This together 
with a  larger detector mass (for XENON, 1 tonne is presently being installed)  and other improvements will enable 
a thorough search well beyond the present range of WIMP-nucleon cross sections. 
In Europe there are several other ambitious endeavours underway, 
such as DARWIN, a large liquid noble gas detector, and EURECA, a solid state 
detector. 
 
The  rate in (\ref{eq:diffscatt}) is strongly dependent 
on the velocity $v$ of the neutralino with respect to the target 
nucleus. Therefore 
an annual modulation of the counting rate 
is in principle possible, due to the motion of the Earth around 
the Sun \cite{freese}. One can thus write 
\beq 
S(E_R,t) = S_0 (E_R) + S_m (E_R) \cos \left[\omega (t-t_0)\right], 
\eeq 
where $\omega = 2\pi/365$ days$^{-1}$. Starting to count time 
in days from January 1$^{\rm st}$, the phase  is $t_0 = 153$ days 
since the maximal signal occurs when the direction of 
motion of the Earth 
around the Sun and the Sun around the galactic center coincide maximally, 
which happens on June 2$^{\rm nd}$ every year \cite{freese}. Similarly, the 
counting rate is expected to be the lowest December 2$^{\rm nd}$ every 
year. Here $S_0 (E_r)$ is the average 
 differential scattering rate  in 
(\ref{eq:diffscatt})and $S_m (E_R)$ is the 
modulation amplitude of the rate. 
The relative size of $S_m (E_R)$ and $S_0 (E_R)$ 
depends on the target and neutralino mass as well as on $E_R$. Typically $S_m (E_R)$ 
is of the order of a few percent of $S_0 (E_R)$, but 
may approach 10 \%  for small $m_\chi$ (below, say, 50 GeV) and small $E_R$ 
(below some 10 keV).

Since the basic couplings in the MSSM are between neutralinos and 
quarks, there are uncertainties related to the hadronic physics 
step which relates quarks and gluons with nucleons, as well the step from 
nucleons to nuclei. These uncertainties are substantial, and can 
plague all estimates of scattering rates by at least a factor of 
2, 
maybe even by an order of magnitude \cite{bottinonew}. 
The largest rates, which as first shown in \cite{bsg} could be 
already ruled out by contemporary  experiments, are generally obtained 
for mixed neutralinos, i.e. with $Z_g$ neither very near 0 nor very near 1, 
and for relatively light Higgs masses (since Higgs bosons mediate 
a scalar, spin-independent exchange interaction).  
 
The experimental situation is becoming interesting as several 
direct detection experiments after many years of continuing 
sophistication are starting to probe interesting parts of 
the parameter space of the MSSM, given reasonable, central values 
of the astrophysical and nuclear physics parameters. Perhaps 
most striking is the 8 $\sigma$ evidence for an annual modulation 
effect claimed to be seen in the NaI experiment DAMA/LIBRA \cite{dama} (see Section~\ref{ch:4} where present data are summarized). 
 
Many of the present day detectors are severely hampered by a large 
background of various types of ambient radioactivity or cosmic-ray induced 
activity (neutrons are a particularly severe problem since 
they may produce recoils which are very similar to the expected 
signal). A great improvement in sensitivity would be acquired if 
one could use directional information about the recoils  
There 
are some very interesting developments also along this line, 
but a full-scale detector is yet to be built. 
 
Direction-sensitive detectors would have an even bigger 
advantage over pure counting experiments if the dark matter 
velocity distribution 
is less trivial than the commonly assumed maxwellian.

\subsection{Indirect Searches} 
 
Besides these possibilities of direct detection of supersymmetric dark 
 matter 
 (with even present indications of the existence of a  signal \cite{dama}), 
 one also has the possibility of indirect detection through 
 neutralino annihilation 
in the galactic halo. This is becoming a promising method thanks 
 to very powerful new detectors for 
cosmic gamma rays  and neutrinos planned and under construction. 
Also, with time more has become known about the distribution of dark matter thanks to very ambitious N-body simulations \cite{aquarius,diemand}, and a large amount of substructure has been found. This would enhance indirect detection, as it 
is proportional to the line-of-sight integral of the square of the number density.

There has  been a  balloon-borne  detection experiment 
\cite{heat}, 
with increased sensitivity to eventual positrons from neutralino annihilation, 
where an excess of positrons over that expected from ordinary sources 
was found. However, due to the rather poor quality of the data, it was not very conclusive.  
 
In 2008, however, this changed completely when the data on the positron to electron ratio, rising with energy, from the satellite PAMELA was presented \cite{pamela}. Somewhat 
later, FERMI-LAT reported a rise above the expectation from secondary production (by cosmic rays) also for the sum of positrons and electrons \cite{fermi_e}. 
 
 An unexpectedly high ratio of positrons over electrons was measured by PAMELA, in particular in the region between 10 and 100 GeV, where previously only 
weak indications of an excess had been seen \cite{oldpos}. This new precision 
measurement of the cosmic ray positron flux, which definitely disagrees with  
a standard background \cite{moskstrong} has opened up a whole new field of 
speculations about the possible cause of this positron excess.  
Simultaneously, other data from PAMELA indicate that the antiproton 
flux is in agreement with standard expectations \cite{pamela2}. 
 
There are a variety of astrophysical models proposed for the needed extra primary component of positrons, mainly based on having nearby  
pulsars as a source \cite{pulsars}. Although pulsars with the required  
properties like distance, age, and energy output are known to exist, it turns out not to be trivial to fit both FERMI and  PAMELA data with these models 
(see, for example, \cite{profpuls,cholpuls}). For this and other reasons, 
 the dark matter interpretation, 
which already had been applied to the much more uncertain HEAT data  
\cite{heat} was one of the leading hypotheses. 
 
It was clear from the outset that to fit the PAMELA positron data and FERMI's sum of positrons and electrons with a dark matter  model a high mass is needed (on the order of 600 GeV to several TeV). However, since the local average dark matter density 
is well-known to be around 0.4 GeV/cm$^3$ \cite{catena}, the number density decreases 
as $1/M_X$ and therefore the  annihilation rate goes as   
$1/M_X^2$ with $M_X$ the mass of  
the annihilating particle. This means that with $\langle\sigma v\rangle=3\cdot 10^{-26}$ cm$^3$/s, which is the standard value of the annihilation rate  
in the halo for  
thermally produced WIMPs (see (\ref{eq:wimp})), the rate of positrons, 
even for a contrived model which annihilates to $e^+e^-$ with unit branching 
ratio is much too small to explain the measured result. 
 
To a good approximation, the local electron plus  positron flux for  
such a model 
is  given by, assuming an energy loss of $10^{-16} E^2$ GeVs$^{-1}$  
(with $E$ in GeV) 
from inverse Compton and synchrotron radiation, 
\begin{equation} 
E^3{d\phi\over dE}=6\cdot 10^{-4}E 
\left({1\ {\rm TeV}\over M_X}\right)^2 
\theta(M_X-E)B_{\rm tot}\   
{\rm m}^{-2} 
{\rm s}^{-1}{\rm sr}^{-1}{\rm GeV}^2, 
\end{equation} 
which means that the boost factor $B_{\rm tot}\sim 200$  
for a 600 GeV particle,  
that may otherwise explain the positron excess. 
Similar boost factors seem to be generic, also for supersymmetric models giving 
$e^+e^-$ through internal bremsstrahlung \cite{bring3}. 
 
Such a boost factor can in principle be 
given by a large inhomogeneity which has to be very local, since positrons and 
electrons of several hundred GeV do not diffuse very far before  
losing essentially 
all their energy. Although not excluded, this would seem to be  extremely 
 unlikely in most structure 
formation scenarios. Therefore, most models rely on the Sommerfeld enhancement factor. This means 
 a non-negligible amount of fine-tuning of the mass  
spectrum, in particular 
also for the degeneracy between the lightest and next-to-lightest particle  
in the new sector. For a detailed discussion of the required model-building, 
see \cite{sommerfeld2}. Similar fine-tuning is needed for the decaying 
dark matter scenario, where the decay rate has to be precisely tuned  
to give the measured flux. Since the antiproton ratio seems to be normal 
according to the PAMELA measurements \cite{pamela2}, the final states should be mainly 
leptons (with perhaps intermediate light new particles decaying into leptons). 
For an interesting such model, which may in fact contain an almost standard 
axion, see \cite{thaler}.  
 
It seems that at present it is possible to construct 
models of the Sommerfeld enhanced type \cite{tracy} which do marginally not  
contradict present data. However, constraints are getting severe and the dark matter solution to the positron excess is currently not as fashionable as a couple  of years ago. It will be interesting, however, to see the first results  
from the AMS-02 experiment \cite{ams-02} on the International Space Station, which should appear in the summer of 2012.

A very rare process in proton-proton collisions, antideuteron production, 
may be less rare in neutralino annihilation \cite{fiorenza}. However, 
the fluxes are so small that the possibility of detection seems marginal 
even in the AMS-02 experiment, and probably a dedicated space probe has to be  
employed \cite{gaps}. 
 
\subsection{Indirect Detection by $\gamma$-rays from the Halo}

With the problem of a lack of clear signature of positrons and antiprotons, one would expect that 
the situation 
 of gamma rays and neutrinos is similar, if they only arise from 
secondary decays in the 
annihilation process. For instance, the gamma ray spectrum arising from 
the fragmentation of fermion and gauge boson final states is quite 
featureless and gives the bulk of the gamma rays at low energy where the 
cosmic gamma ray background is severe. However, an advantage is the directional 
information that photons carry in contrast to charged particles which 
random walk through the magnetic fields of the Galaxy \cite{contgammas}. 
 
For annihilation into quark-antiquark pairs, or $W$ and $Z$ bosons, the continuous energy spectrum one gets after fragmentation into SM particles can rather well
and conveniently be parametrized as
\beq
{dN_{\rm cont}(E_\gamma)/dE_\gamma}=(0.42 / m_\chi)e^{-8x} / (x^{1.5}+0.00014),\label{eq:contappr}
\eeq
 where $m_\chi$ is the WIMP mass and $x=E_\gamma/m_\chi$.
For more detailed spectra, one may for instance use standard particle physics
codes like PYTHIA \cite{pythia} (as is done in \cite{ds}). One should note that 
for $\tau$ leptons in the final state (\ref{eq:contappr}) is not a good
approximation, as this gives a harder spectrum.
 
\subsubsection{Gamma-ray Lines}\label{subs:lines}  

An early idea was to look for a spectral feature, a line, in the 
radiative annihilation process to a charm-anticharm bound state 
$\chi\chi\to (\bar c c)_{\rm bound} +\gamma$ \cite{STS}. 
However, as the experimental lower bound on the lightest neutralino 
became higher it was shown that form factor suppression rapidly 
makes this process unfeasible \cite{berg-snell}. The  surprising 
discovery was made that 
the loop-induced annihilations 
$\chi\chi\to\gamma\gamma$ \cite{berg-snell,gammaline} and $\chi\chi\to Z\gamma$ 
\cite{zgamma} do not suffer from any form factor suppression.
 
The rates of these processes are difficult to estimate because of 
uncertainties in 
the supersymmetric parameters, cross sections and halo density profile. However, 
in contrast to the other proposed detection methods they have 
the virtue of giving  very 
distinct, ``smoking gun'' signals of 
monoenergetic photons with energy $E_\gamma = m_\chi$ 
(for $\chi\chi\to\gamma\gamma$) or $E_\gamma = m_\chi 
(1-m_{Z}^2/4m_{\chi}^2)$ (for $\chi\chi\to Z\gamma$) 
emanating from annihilations in the halo.

The detection probability of a gamma ray signal, either continuous or line, will of course depend 
sensitively on the density profile of the dark matter halo. 
To illustrate this point, let us consider the characteristic 
angular dependence of 
the $\gamma$-ray line intensity from neutralino annihilation $\chi\chi\to\gamma\gamma$ 
in the galactic halo. 
Annihilation of neutralinos in an isothermal halo 
with core radius 
$a$ leads to a $\gamma$-ray flux along the line-of-sight direction $\hat n$ of 
$$
     {d{\cal F} \over {d \Omega}}\left(\hat n \right)\simeq \, 
     (0.94\times10^{-13} {\rm cm}^{-2} {\rm s}^{-1} {\rm sr}^{-1})\times
$$
\beq
    \left({\sigma_{\gamma\gamma} v\over 10^{-29}\ {\rm cm}^{-3}{\rm 
    s}^{-1}}\right) 
     \left({\rho_\chi\over 0.3\ {\rm GeV}\,{\rm cm}^{-3}}\right)^2 
     \left({100\,{\rm GeV}\over m_\chi}\right)^2 \left({R\over 8.5\ {\rm kpc}}\right)J(\hat n)
\eeq 
where 
 $\sigma_{\gamma\gamma} 
v$ is the annihilation rate, 
$\rho_\chi$ is the 
local neutralino halo density 
and $R$ is the distance 
to the galactic center. 
The integral $J(\hat n)$ is given by 
\beq 
J(\hat n)={1\over R\rho_\chi^2}\int_{\rm 
line-of-sight}\rho^2(\ell)d\ell(\hat n),\label{eq:j} 
\eeq 
and is evidently very  sensitive to local density variations along the 
line-of-sight path of integration. In the case of a 
smooth halo, its value ranges from a few at 
high galactic latitudes to several thousand for a small angle 
average towards the galactic center in the NFW model \cite{BUB}. 
 
 Since the 
neutralino velocities in the halo are of the order of 10$^{-3}$ of the 
velocity of light, the 
annihilation can be considered to be at rest. The resulting gamma ray 
spectrum is a line 
at $E_\gamma=m_\chi$ of relative linewidth 10$^{-3}$ (coming from the Doppler effect caused by the motion of the WIMP) which in 
favourable cases 
will stand out against background. 

Detection of a $\gamma$-rate line signal would need a detector with very
good energy resolution, like one percent or better. This is not achieved 
by FERMI (although data on a line search have been presented \cite{fermilines}).
 However, the Russian satellite GAMMA-400 \cite{gamma-400} seems to have very
promising characteristics for this type of dark matter search, when it is launched by the end of this decade. This could be a very interesting new instrument in the
search for $\gamma$-ray lines from annihilation (or decay) of dark matter.

The calculation of the $\chi\chi\to\gamma\gamma$ cross section is 
technically quite 
involved with a large number of loop diagrams contributing. A full calculation in the MSSM 
was performed in \cite{newgamma}. 
Since the 
different contributions all have to be added coherently, there may be 
cancellations or 
enhancements, depending on the supersymmetric parameters. 
The process $\chi\chi\to Z\gamma$ 
is treated analogously  and has a  similar rate 
\cite{zgamma}. 

An important contribution, especially for neutralinos that contain 
a fair fraction of a higgsino component, is from virtual $W^+W^-$ 
intermediate states. 
This is  true both for the $\gamma\gamma$ and $Z\gamma$ final 
state for very massive neutralinos \cite{zgamma}. In fact, thanks to 
the effects of coannihilations \cite{coann}, neutralinos as heavy as 
several TeV are allowed without giving a too large $\Omega$. These 
extremely heavy dark matter candidates (which, however, 
would require quite a 
degree of finetuning in most supersymmetric models) 
are predominantly higgsinos 
and have a remarkably large branching ratio into the loop-induced 
$\gamma\gamma$ and $Z\gamma$ final states (the sum of these can be as 
large as 30\%). If there would exist such heavy, stable neutralinos, 
the gamma ray line annihilation process may be the only one which could reveal 
their existence in the foreseeable  future (since not even LHC would be sensitive 
to supersymmetry if the lightest supersymmetric particle weighs several 
TeV). In fact the high branching ratio for higgsino annihilation to $2\gamma$ 
was the reason that Hisano et al. \cite{sommerfeld} took a closer look at
the process and discovered the effect of Sommerfeld enhancement.

\subsubsection{Internal Bremsstrahlung}
The $\gamma\gamma$ process appears in a closed loop meaning that it is suppressed by powers of the electromagnetic coupling constant.  An amusing effect appears, however, for Majorana fermions at even lower order. It
was early realized that there could be       important spectral 
features \cite{lbe89}, and
recently it has been shown that internal bremsstrahlung (IB) from 
produced charged particles
in the annihilations could yield a detectable "bump" near the 
highest energy for heavy
gauginos or Higgsinos annihilating into $W$ boson pairs, such 
as expected in split supersymmetry
models \cite{heavysusy}.
 In \cite{birkedal}, it was furthermore pointed out that 
IB often can be estimated
by simple, universal formulas and often gives rise to a very prominent step in the spectrum at  
photon energies of $E_\gamma=m_\chi$ (such as in LKP models \cite{Bergstrom:2004cy}).
The IB process was thoroughly treated in \cite{fsr},
and here we summarize the main results. 

In  \cite{lbe89} it was shown that
the  radiative process $\chi^0\chi^0\to f\bar f\gamma$  
may circumvent the chiral suppression, i.e., 
the annihilation
rate being proportional to $m_f^2$. This is normally what one would get for 
annihilation into a fermion pair
from an $S$-wave initial state \cite{goldberg}, as is the case in lowest order 
for non-relativistic
dark matter Majorana particles in the Galactic halo (see also \cite{baltz_bergstrom}). Since this enhancement 
mechanism is most prominent in cases where the neutralino is close to 
degenerate with charged sleptons, it is of special importance in the 
so-called stau coannihilation region.

A fermion final state containing an additional photon, $f\bar f\gamma$, is thus not subject to  a helicity suppression. The full analytical expressions lengthy,but simplify in the limit of $m_f\rightarrow0$. Then one finds
$$
  \frac{\mathrm{d}N_{f^+f^-}} {\mathrm{d}x} = 
$$
\beq
\lambda\times\Big\{\frac{4x}{\mu(\mu-2x)}
 -\frac{2x}{(\mu-x)^2}
   -\frac{\mu(\mu-2x)}{(\mu-x)^3}\log\frac{\mu}{\mu-2x}\Big\}\,, \label{eq:ib}
\eeq
with 
$$ \lambda = (1-x) \alpha_\mathrm{em}Q^2_f\frac{\left|\tilde g_R\right|^4+\left|\tilde g_L\right|^4}{64\pi^2} \Big(m_\chi^2 \langle\sigma v\rangle_{\chi\chi\rightarrow f\bar f}\Big)^{-1}.
$$
where $\mu\equiv m_{{\tilde f}_R}^2/m_\chi^2+1 =m_{{\tilde f}_L}^2/m_\chi^2+1$ and $\tilde g_RP_L$ ($\tilde g_LP_R$) is the coupling between neutralino, fermion and right-handed (left-handed) sfermion. This confirmed the result found in \cite{lbe89} for photino annihilation. Note the large enhancement factor $m_\chi^2/m_f^2$ due to the lifted helicity suppression (from ${\langle\sigma v\rangle}_{\chi\chi\rightarrow f\bar f}\propto m_f^2m\chi^{-4}$), and another large enhancement that appears at high photon energies for sfermions degenerate with the neutralino.

Internal bremsstrahlung from the various possible final states of neutralino annihilations is included in \ds\ \cite{ds}. The total $\gamma$-ray spectrum  is given by
\beq
  \frac{dN^{\gamma,\mathrm{tot}}}{dx}=\sum_f B_f\left(\frac{dN_f^{\gamma,\mathrm{sec}}}{dx}+\frac{dN_f^{\gamma,\mathrm{IB}}}{dx} + \frac{dN_f^{\gamma,\mathrm{line}}}{dx}\right)\,,
\eeq
where $B_f$ denotes the branching ratio into the annihilation channel $f$. The last term in the above equation gives the contribution from the direct annihilation into photons, $\gamma\gamma$ or $Z\gamma$, which result in a sharp line feature \cite{newgamma,zgamma}. The first term is the contribution from secondary photons from the fragmentation of the fermion pair. 
This ``standard'' part of the total $\gamma$-ray yield from dark matter annihilations shows a feature-less spectrum with a rather soft cutoff at $E_\gamma=m_\chi$.

\begin{figure}[!htb] 
\begin{center} 
\includegraphics[width=10cm]{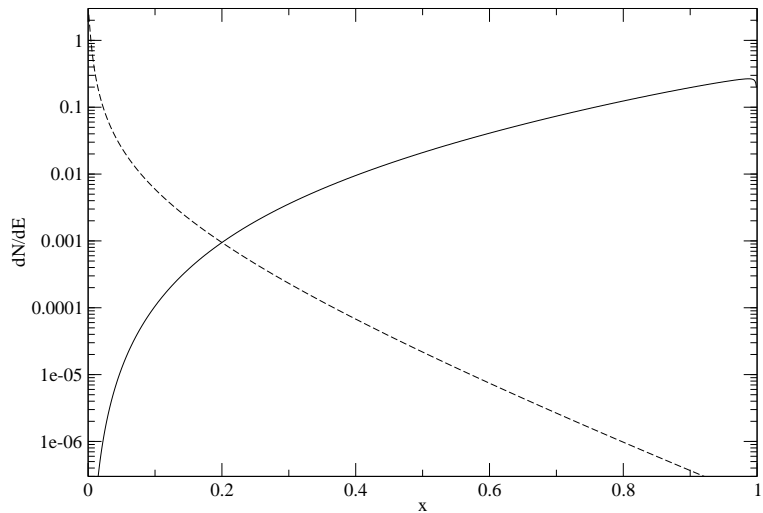} 
\end{center} 
\caption{The distribution of $\gamma$-rays from the internal
bremsstrahlung process $\chi^0\chi^0\to f\bar f\gamma$  is shown as the solid line, and compared
to the standard case, (\ref{eq:contappr}) (dashed line). As can be seen, the internal bremsstrahlung process gives a very hard spectrum, which may counteract the fact that
radiation of a photon always is suppressed by factor $\sim \alpha_{em}/\pi$.\label{fig:ib}} 
\end{figure}

In Fig.~\ref{fig:ib} an example of the energy distribution of photons given by
(\ref{eq:ib}) is shown.

\subsubsection{Density Profile and $\gamma$-ray Detection}
 
To compute $J(\hat n)$ in (\ref{eq:j}), a 
model of the dark matter halo has to be chosen. 
 The universal halo profile found in 
 simulations by Navarro, Frenk and White \cite{nfw} 
 has a rather significant enhancement $\propto 1/r$ 
near the halo centre,
\beq
\rho_{NFW}=\frac{c}{r(a+r)},
\eeq
where $c$ is a concentration parameter and $a$ a typical length scale for the halo. In fact, more detailed later simulations have given a slightly
different shape, the so-called Einasto profile,
\beq
\rho_{Einasto}=\rho_s e^{-\frac{2}{\alpha}\left[\left(\frac{r}{\alpha}\right)^\alpha-1\right]},
\eeq
with $\alpha\sim 0.17$ for the Milky Way. Except near $r=0$, this profile is 
actually quite similar to the NFW profile, and it has slightly higher density outside
the very center. The local dark matter density near the solar system 
can be quite well determined \cite{catena} and is $\rho_0\simeq 0.4$ GeV/cm$^3$.  
If these forms of the density can be applied to the Milky Way, this 
 would lead to a much enhanced annihilation 
rate towards the galactic centre, and also to a very characteristic 
angular dependence of the line signal. This would be very beneficial 
when discriminating against the  extragalactic $\gamma$ 
ray background, and Imaging Air Cherenkov Telescope Arrays (IACTAs) are be eminently 
suited to look for these signals since they have an angular 
acceptance which is well matched to the angular size of the Galactic central 
region where a cusp is likely to be. Both H.E.S.S. \cite{hess}, MAGIC \cite{magic} and Whipple \cite{veritas} have searched for a signal at the galactic center or in other dark matter concentrations, but are still a couple
of orders of magnitude above the flux which would correspond to the canonical WIMP flux, (\ref{eq:wimp}). Maybe with the planned CTA project \cite{cta0} one may get to the interesting region of parameter space for supersymmetric or other WIMPs.  

\begin{figure}[!htb] 
\begin{center} 
\includegraphics[width=0.9\linewidth]{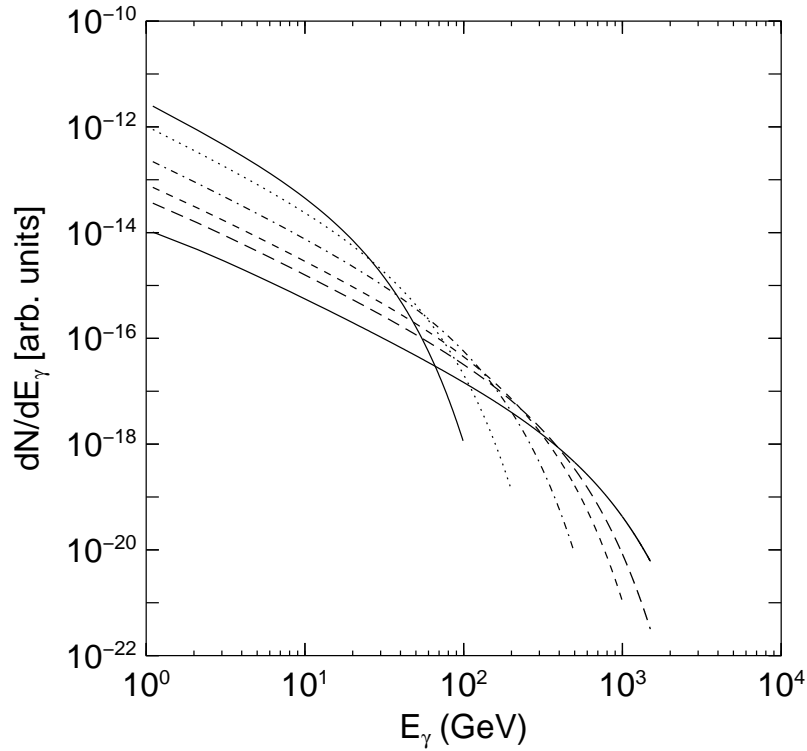} 
\end{center} 
\label{fig:5at5} 
\caption{The energy distribution of $\gamma$-rays from WIMP dark matter annihilation
into a $b\bar b$ pair, for a dark matter particle mass of 100, 200, 500, 1000, 1500, and 3000 GeV, respectively. One can see that the bulk of the signal is at
low energies. (Here the line signals from $\gamma\gamma$ and $Z\gamma$ have not been included.)}
\end{figure}

Also the energy threshold of present-day IACTAs is too high (of the order of 50 GeV or higher) to be very useful for WIMPs of 100 GeV or lighter. There have been discussions about a high-altitude detector with lower threshold, perhaps as 
low as 5 GeV \cite{aharonian}, which would be very beneficial for dark matter detection, see 
Fig.~\ref{fig:5at5}

Space-borne gamma ray detectors, like the FERMI 
satellite  
 have a much smaller area (on the order of 1 
m$^2$ instead of $10^4-10^5$ m$^2$ for IACTAs), but a 
correspondingly larger angular acceptance so that the integrated 
sensitivity is in fact similar. This is at least true if the Galactic 
center does not have a very large dark matter density enhancement which 
would favour IACTAs. The total rate expected in FERMI can be 
computed with much less uncertainty because of the angular 
integration \cite{preglast}. Directional information is obtained and can 
be used to discriminate against the diffuse extragalactic 
background. A line signal can be searched for with high precision, 
since the energy resolution of FERMI is at the few percent 
level.

\subsubsection{Indirect Detection through Neutrinos} 
 
The density of 
neutralinos in the halo is not large enough to give a measurable flux 
of secondary neutrinos, unless the dark matter halo is very clumpy \cite{clumpy}. 
In particular, the central Galactic black hole may have interacted with the dissipationless 
dark matter of the halo so that a spike of very high dark matter density 
may exist right at the Galactic centre \cite{gondolosilk}. However, the existence 
of these different forms of density enhancements are very uncertain and depend 
extremely sensitively on presently completely unknown aspects of the formation 
history of the Milky Way. 
 
More model-independent predictions (where essentially only the relatively 
well-determined local halo dark matter density is of importance) 
can be made for neutrinos  from the centre 
of the Sun or Earth, where 
neutralinos may have been gravitationally trapped and therefore their density 
enhanced.  As they annihilate, many of the possible final states 
(in particular, $\tau^+\tau^-$ lepton pairs, heavy quark-antiquark pairs 
and, if kinematically allowed, $W^\pm H^\mp$, 
$Z^0H_i^0$, $W^+W^-$ or $Z^0Z^0$ pairs) give 
after decays and perhaps hadronization energetic neutrinos which 
will propagate out from the interior of the Sun or Earth. 
(For neutrinos from the Sun, 
energy loss of the hadrons 
in the solar medium and the energy loss of neutrinos 
have to be considered \cite{ritzseckel,jethesis}). In particular, 
the muon neutrinos are useful for indirect detection of 
neutralino annihilation processes, since muons 
have a quite long range in a suitable detector medium like ice or water. 
Therefore they can be detected through their Cherenkov radiation after 
having been produced 
at or near the detector, through the action of 
a charged current weak interaction 
$\nu_\mu + A \to \mu + X$. 
 
Detection of neutralino annihilation into neutrinos is 
one of the most promising indirect detection methods, 
and  will be subject to  extensive experimental investigations in view 
of the new neutrino telescopes (IceCube, ANTARES, KM3NET) 
planned or under construction \cite{halzen}. The advantage shared with 
gamma rays is that neutrinos keep their original direction. A high-energy 
neutrino signal in the direction of the centre of the Sun or Earth 
is therefore an excellent experimental signature which may stand up against 
the background of neutrinos generated by cosmic-ray interactions in the 
Earth's atmosphere.

The differential neutrino flux from neutralino annihilation is 
\beq 
\frac{dN_\nu}{dE_\nu} = 
\frac{\Gamma_A}{4\pi D^2} \sum_{f} 
B^{f}_{\chi}\frac{dN^f_\nu}{dE_\nu} 
\eeq 
where $\Gamma_A$ is the annihilation rate, 
$D$ is the distance of the detector from the source (the 
central region of the Earth or the Sun), $f$ is the neutralino pair 
annihilation final states, 
and $B^{f}_{\chi}$ are the branching ratios into the final state $f$. 
 $dN^f_\nu/dE_{\nu}$ are the energy 
distributions of  neutrinos generated by the final state $f$. 
Detailed calculations of these spectra 
can be made using Monte Carlo 
methods \cite{bottnuflux,jethesis,BEG}.  Effects of neutrino oscillations have
also been included \cite{tommy}.

The neutrino-induced muon flux may be detected in a neutrino telescope 
by measuring the muons that come from the direction of the centre 
of the Sun or Earth. For a shallow detector, this usually has to 
be done in the case of the Sun by looking (as always the case for 
the Earth) at upward-going muons, since there is a huge background 
of downward-going muons created by cosmic-ray interactions in the 
atmosphere. 
The flux of muons at the detector is  given by 
 
\beq 
\frac{d N_\mu}{d E_\mu} 
= N_A \int^\infty_{E_\mu^{\rm th}} d E_\nu 
\int_0^\infty d\lambda \int_{E_\mu}^{E_\nu} 
d {E'_\mu }\,\, 
P(E_\mu,E'_\mu; \lambda)\,\, 
\frac{d \sigma_\nu (E_\nu,E'_\mu)}{d E'_\mu} \,\, 
\frac{d N_\nu}{d E_\nu}\, , 
\label{eq:muflux} 
\eeq 
where $\lambda$ is the muon range in the medium (ice or water 
for the large detectors in the ocean or at the South Pole, 
or rock which surrounds the smaller underground detectors), 
$d \sigma_\nu (E_\nu,E'_\mu) / d E'_\mu$ is 
the weak interaction cross section for production of a muon of 
energy $E'_\mu$ from a parent neutrino of energy $E_\nu$, and 
$P(E_\mu,E'_\mu; \lambda)$ is the 
probability for a muon of initial energy $E'_\mu$ 
to have a final energy $E_\mu$ after passing 
 a path--length $\lambda$ inside the detector medium. 
$E_\mu^{\rm th}$ is the detector threshold energy, which for 
``small'' 
neutrino telescopes like Baksan, MACRO and Super-Kamiokande is 
around 1 GeV. 
Large area neutrino telescopes in the ocean  or in Antarctic ice 
typically 
have thresholds of the order of tens of GeV, which makes them 
sensitive mainly to heavy neutralinos (above 100 GeV) 
\cite{begnu2}. Convenient approximation formulas relating the observable 
muon flux to the neutrino flux at a given energy exist \cite{halzenreview}. 
 
The integrand in (\ref{eq:muflux}) is weighted towards high 
neutrino energies, both because the cross section $\sigma_\nu$ 
rises approximately linearly with energy and because the average 
muon energy, and therefore the range $\lambda$, also grow 
approximately linearly with $E_\nu$. Therefore, final states 
which give a hard neutrino spectrum (such as heavy quarks, $\tau$ 
leptons and $W$ or $Z$ bosons) are usually more important 
than the soft spectrum arising from light quarks and gluons.

The rate of change of the number of  neutralinos $N_\chi$ in the Sun or 
Earth is governed by the equation 
\beq 
\dot N_\chi=C_C-C_AN_\chi^2\label{eq:ca} 
\eeq 
where $C_C$ is the capture rate and $C_A$ is related to the annihilation rate $\Gamma_A$, 
$\Gamma_A=C_AN_\chi^2$. 
This has the solution 
\beq 
\Gamma_A={C_C\over 2} \tanh^2\left({t\over \tau}\right), 
\eeq 
where the equilibration time scale $\tau=1/\sqrt{C_CC_A}$. In most 
cases for the Sun, and in the cases of observable fluxes for the 
Earth, $\tau$ is much smaller than a few billion years, and 
therefore equilibrium is often a good approximation ($\dot N_\chi=0$ 
in (\ref{eq:ca})). This means that it is the capture rate 
which is the important quantity that determines the neutrino flux.

The capture rate induced by scalar (spin-independent) interactions between 
the neutralinos and the nuclei in the interior of the Earth or Sun is 
the most difficult one to compute, since it depends sensitively on 
Higgs mass, form factors, and other poorly known quantities. However, 
this 
spin-independent capture rate calculation is the same as for 
direct detection. Therefore, 
there is a strong correlation between the  neutrino flux expected from 
the Earth (which is mainly composed of spin-less nuclei) and the signal 
predicted in direct detection experiments \cite{begnu2,kamsad}. It 
seems that even the large (kilometer-scale) neutrino telescopes 
planned will not be competitive with the next generation of direct 
detection experiments when it comes to detecting neutralino dark 
matter,  searching for 
annihilations from the Earth. However, the situation concerning 
the Sun is more favourable. Due to the low counting rates for 
the spin-dependent interactions in terrestrial detectors, high-energy 
neutrinos 
from the Sun constitute  a competitive and complementary neutralino dark matter 
search. Of course, even if a neutralino is found through direct 
detection, it will be extremely important to confirm its identity 
and investigate its properties through indirect detection. In 
particular, the mass can be determined with reasonable accuracy 
by looking at the angular distribution of the detected 
muons \cite{EG,BEK}. 
 
For the  the Sun, dominated by hydrogen, 
the axial (spin-dependent) cross section is  important and 
relatively easy to compute. A good approximation 
is given by \cite{jkg} 
$$ 
     {C^{\rm sd}_\odot\over (1.3\cdot 10^{23}\, {\rm s}^{-1})} =\ \ \ \ \ \ \ \ \ \ \ \ \ \ \ \ \ \ \ \ \ \ \ \ \ \ \ \ \ \ \ \ \ \ \ \ \ \ \ \ \ \ \ \ \ \ \ \ \hfill$$
 
\beq\left({\rho_\chi\over 0.3\ {\rm GeV}\,{\rm cm}^{-3}}\right) 
     \left({100\,{\rm GeV}\over m_\chi}\right) 
\left({\sigma_{p\chi}^{\rm sd}\over 10^{-40}\ {\rm 
     cm}^2}\right) 
\left(270\ {\rm km/s}\over \bar v\right), 
\eeq 
where $\sigma_{p\chi}^{\rm sd}$ is the cross section for 
neutralino-proton elastic scattering via the axial-vector interaction, 
$\bar v$ is the dark-matter 
velocity dispersion, and 
$\rho_\chi$ is the local dark matter mass. 
The capture rate in the Earth is dominated 
by scalar interactions, where there may be kinematic and other 
enhancements, in particular if the mass of the neutralino almost 
matches one of the heavy elements in the Earth. For this case, 
a more detailed analysis is called for, but convenient approximations 
are available \cite{jkg}. In fact, also for the Sun the 
spin-dependent contribution can be important, in particular 
iron may contribute non-negligibly.

A neutrino 
telescope of area around 1 km$^2$, which is roughly the size of IceCube, 
has discovery potential for a range of  supersymmetric models, which cannot easily be probed using other methods, see \cite{begnu2}.

To conclude this section on detection methods of WIMPs, we have seen that supersymmetric particles, which are the theoretically most plausible WIMPs have many interesting features which may make them detectable in the nor too distant future.
Supersymmetry, in particular MSSM, invented already in the 1970's, and obtained as a phenomenological manifestation of the most realistic string theories, has since the early 1980's, when the $CDM$ paradigm first won universal acclaim, been the prime template for a WIMP \cite{goldberg,ellis}. 

Even in the MSSM, however, there are in principle more than a hundred free parameters, meaning that for practical reasons the templates, for instance used at the LHC experiments, are drastically simplified versions (like CMSSM or the even more constrained mSUGRA), which do not, in contrast to the full MSSM, correspond very well to more recent thinking about supersymmetry breaking  \cite{seiberg}. This has to be kept in mind when discussing the impressive LHC limits discussed extensively at this conference. Even in still simplified versions, like the 19 to 24-parameter "phenomenological MSSM", pMSSM \cite{pmssm}, the bounds on particle masses given, e.g., by fulfilling the WMAP relic density, are not very constraining at the moment \cite{abdussalam}. Of course, the outlook for the MSSM would be much bleaker if a light Higgs (with mass below roughly 130 GeV) were not to be established by the end of the 7 TeV run, in 2012.  
 
With the freely available \cite{joakim_ds} DarkSUSY package \cite{ds}, one can compute in detail the relic density, not only for supersymmetric models, but since 
the package has a modular design, one can insert any favourite model one has 
for WIMP-like dark matter. Of course, DarkSUSY is mostly used for the supersymmetric case, and it has been  originally set up for a general pMSSM model, with 
large freedom in the choice of parameters.  

\subsection{Antimatter Detection of Dark Matter} 
Antimatter does not seem to be present in large quantities in the universe, as can be inferred from the absence of $\gamma$-ray radiation that would have been created in large amounts if astrophysical anti-objects would annihilate on their matter counterparts (this would also cause deviations from the pure black-body form of the cosmic microwave background, something which is very severely limited by WMAP data and will be further probed by the PLANCK satellite). In fact, both the analysis of primordial nucleosynthesis and the CMB, give a  
non-zero number around $10^{-10}$ for the baryon-antibaryon asymmetry, which means  
that matter dominated over antimatter already in the very early universe. On the other hand, dark matter annihilation in almost all models occurs from  
a matter-antimatter symmetric initial state and thus equal amounts of matter and antimatter is created. This leads to an interesting possible new primary source of positrons and antiprotons (i.e. the stable anti-particles of protons) in the cosmic rays of dark matter halos, including the one where the Milky Way resides. (There is always a small amount of antimatter produced as secondary particles in collisions with galactic gas and dust by ordinary cosmic rays, of course.) As discussed extensively at conferences in 2009 (see, e.g., \cite{strumia}) this was an extremely hot topic then. This was due to the PAMELA and FERMI collaborations just having discovered an anomalously high ratio of positrons over electrons up to 100 GeV \cite{pamela}, and sum of positrons and electrons up to 1 TeV \cite{fermi_e}, respectively. During the last two years, this anomaly, although possible to explain by dark matter annihilation, needs such large boost factors (e.g., from Sommerfeld enhancement to be discussed below), and  somewhat contrived, leptophilic models, that these models are feeling severe pressure from other detection methods, e.g, $\gamma$-rays from the central parts of the Galaxy \cite{bertone-lb}. Alternative astrophysical explanations are on the other hand possible with quite standard assumptions. One cannot say that the dark matter explanation is yet completely ruled out, but it is in strong  tension from other measurements. 
 
Returning to more standard WIMP models, there have recently been improvements in the computations of the annihilation rate at low velocity as is the case in galaxies, where $v/c\sim 10^{-3}$. An amusing effect is caused due to the suppression of the $^3S_1$ for an initial initial state of two Majorana spinors (such as neutralinos) at zero velocity, due to the requirement of Fermi statistics. Namely, one cannot have  two identical fermions  in the same spin state.  This means that annihilation only occurs from the pseudoscalar $^1S_0$ state where one of the particles has spin up, the other spin down. This causes for instance the annihilation amplitude into a light fermion-antifermion pair, like $e^+e^-$,  to be suppressed by an explicit helicity  factor of the fermion mass (as in the limit of zero mass, the vertices are helicity-preserving, and to cause a spin flip a mass term is needed). Direct annihilation into $e^+e^-$ was thus thought to be very subdominant. However, it was realized \cite{bbe} (building on an old idea \cite{b}), that a spin-flip by one of the Majorana fermions caused by emitting a photon could first of all relieve the helicity suppression of the  process to a mere $\alpha/\pi$ ordinary radiative factor. And, in addition, the spectral shape of the emitted photon is very favourable for detection, causing a shoulder which peaks close to the dark matter particle mass. In particular, for heavy (TeV-scale) WIMPs this could be quite important, and using the radiative peak would help extracting the signal over background \cite{desy}. Recently, these radiative processes have been generalized also to emission of other gauge bosons, and have been shown to be quite important generally \cite{radiative}.  
 
\subsubsection{The Sommerfeld Effect} 
The possibility of an enhanced annihilation rate due 
to DM halo substructure has been realized for a long 
time \cite{clumpy}. 
However, it seems hard to produce a boost 
factor of the order of a few hundred to a thousand in 
the solar neighborhood, as would be needed to explain the PAMELA and FERMI excesses. This is because substructure survives in  numerical 
simulations mostly in the outer portions 
of the halo, due to tidal stripping in 
the inner part.  
 
Another potentially very important effect, Sommerfeld enhancement, 
which may explain the large boost had been found 
a few years earlier. This effect, was computed for electromagnetism 
by Arnold Sommerfeld many years ago \cite{sommerfeld_o}, 
 but it was 
rediscovered \cite{sommerfeld,sommerfeld2} in the quantum field theory of very heavy dark matter particles  
in the limit when the gauge particles, $\gamma$, $Z^0$ and $W^\pm$ are essentially massless, or at least have a Compton wavelength that is sufficiently large 
compared to the would-be bound state caused by the attractive gauge forces. (Of course, a bound state is never really formed due to the fast time scale of annihilation.) 
 
\begin{figure}[!htb] 
\begin{center} 
\includegraphics[width=0.35\linewidth]{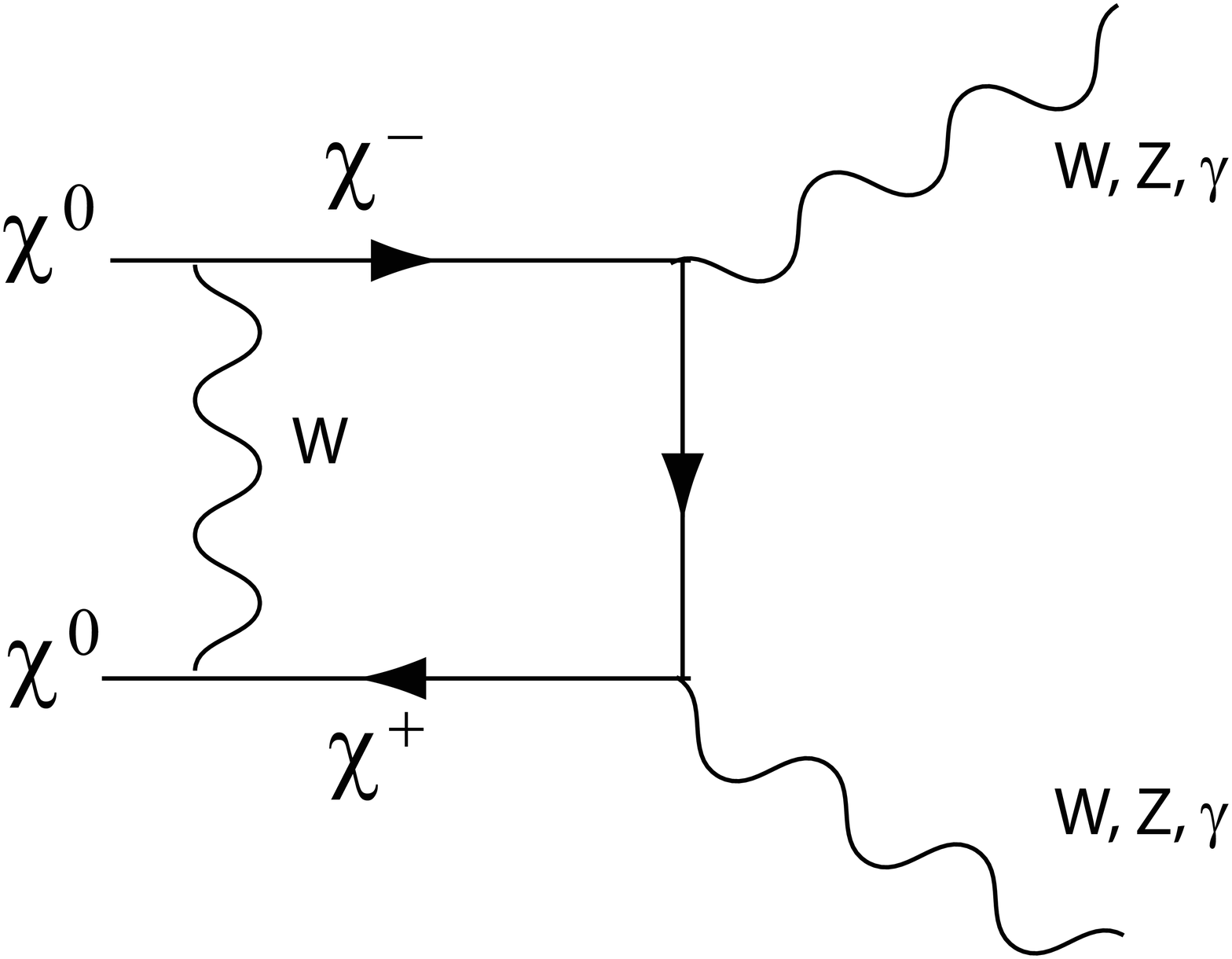} 
\includegraphics[width=0.5\linewidth]{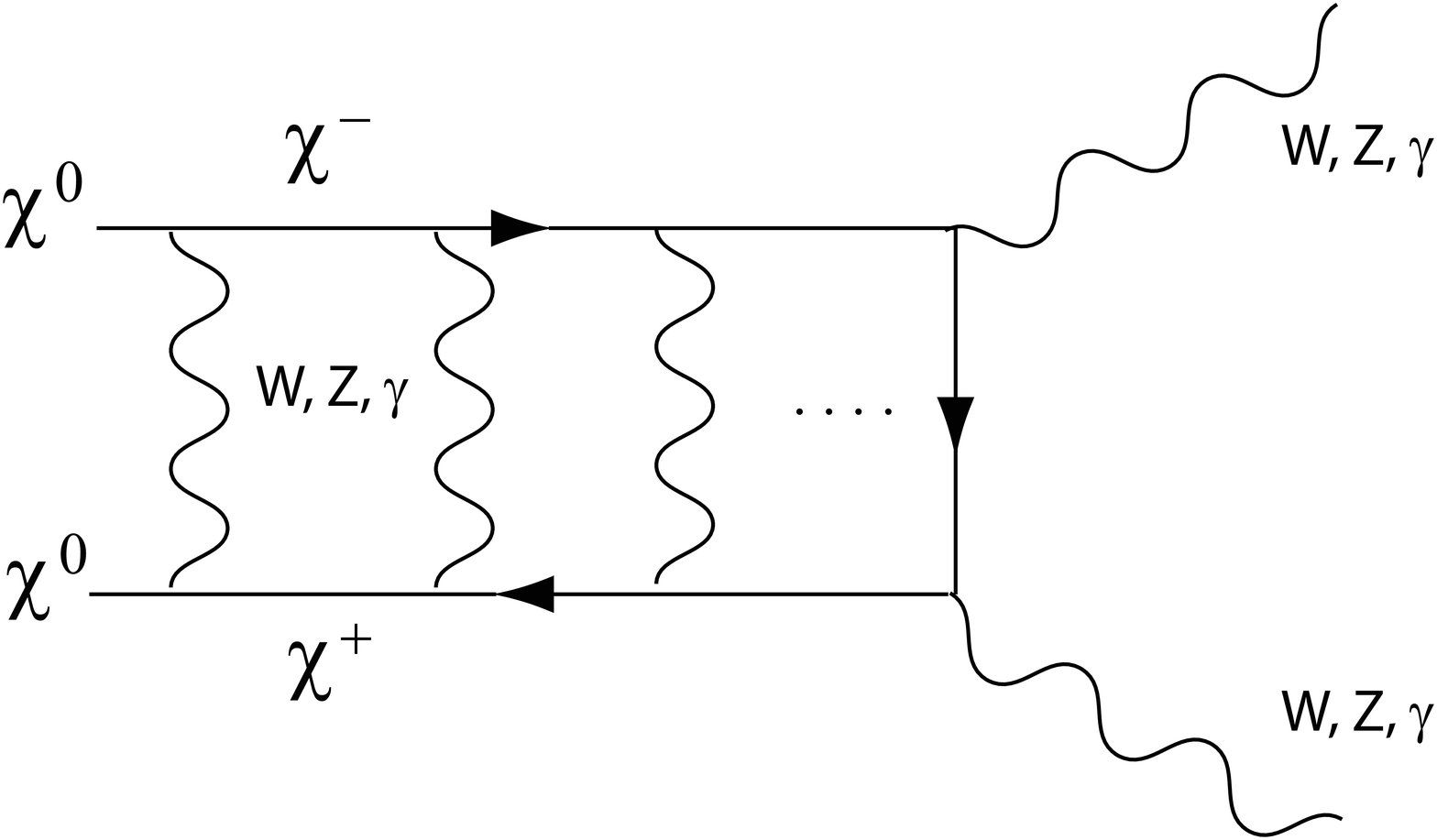} 
\end{center} 
\label{fig:ladder} 
\caption{Diagrams illustrating the field-theoretical reason for the Sommerfeld enhancement. The figure is drawn for a supersymmetric neutralino (which is the case where the effect was first found in dark matter physics \protect\cite{sommerfeld}), but similar diagrams apply for any dark matter candidate which first of all is heavy compared to the exchanged particle in the $t$-channel (i.e. in the ``ladder"), and where there is a near degeneracy between the neutral state being the dark matter and the virtual states (in this case charged particles, charginos). In (a) is shown the lowest order contribution, which gets very important for large masses, and which is further enhanced by the ladder diagrams of the type shown in (b). The net result could be an ``explosive annihilation", to quote \protect\cite{sommerfeld}.} 
\end{figure}

In the quantum mechanical calculation of electron scattering and 
$e^+e^-$ annihilation, Sommerfeld enhancement is caused by the distortion 
of the plane wave describing the relative motion of 
the annihilating particle pair through the near formation 
of a bound state caused by photon exchange. In the 
so-called ladder approximation for QED (where one sums only certain types of Feynman diagrams), one obtains this Sommerfeld 
effect, and the square of the  wave function  
at the origin in relative coordinates $r_1-r_2$, which enters into the probability for 
the short-distance process of annihilation, is increased by 
the factor \cite{sommerfeld2} 
\beq  
S=\frac{|\Psi(0)|^2}{|\Psi_{(0)}(0)|^2}= 
\frac{\left(\frac{\pi\alpha}{\beta}\right)}{1-e^{-\left(\frac{\pi\alpha}{\beta}\right)}}, 
\eeq 
with $\alpha$ the fine-structure constant, and $\beta$ the 
relative velocity. This can be expanded to $S_{QED} = \pi\alpha/\beta$ for 
small relative velocities. In the Milky Way halo, velocities are typically  
$\beta/c\sim 10^{-3}$, so this limit is certainly relevant. For smaller galaxies 
or DM substructure, velocities (as measured by velocity dispersions) are even smaller. Of course, there is no direct photon exchange between DM particles, since they are electrically neutral. However, if there are charged states nearby in mass, the neutral pair may momentarily, before annihilation, transform into a charged pair which in turn may exchange a photon between then. These are the basic processes that have to be summed to all orders in the ladder approximation, and which lead to Sommerfeld enhancement 
(see Fig.~\ref{fig:ladder}).

One could of course also have a  Yukawa-like particle (i.e., spinless) of mass $m_Y$, 
mediating a weak attractive force with coupling constant 
$\alpha_Y$ between DM particles of mass $m_\chi$. The small velocity 
limit of the enhancement then becomes  
\beq 
S_Y\propto \frac{\alpha_Ym_\chi}{m_Y}. 
\eeq 
 
 In some cases, depending on the detailed nature of the mediating particles, 
the enhancement factor $S$ can indeed be as high 
as several hundred to a few thousand, depending 
on the exact parameters. The effect is 
generally strongly velocity-dependent, depending on velocity 
as $1/\beta$ or even (near resonance) $1/\beta^2$ but in the Yukawa case the $1/\beta$ scaling is valid only for 
$\beta > m_Y/m_\chi$. At smaller velocities and outside resonances, 
the effect saturates at $m_Y/m_\chi$ \cite{kam_prof} 
 
Important bounds comes from $\gamma$-rays, but also from the non-observation of energy distortions in the cosmic microwave background.  
It may still be possible to (marginally) fit the PAMELA/FERMI excess, if one takes astrophysical uncertainties into account \cite{tracy}. 
 
It should be noted that the Sommerfeld effect  has a solid  
theoretical backing and is important, if the mass and coupling parameters are in the right range. 
For supersymmetric models, however, it occurs only for very heavy neutralinos 
(generally higgsinos) and the phenomenology has only been partly investigated 
\cite{hryczuk}. 
 
\section{Particular Dark Matter Candidates}\label{ch:6} 
\subsection{WIMP models} 
The particle physics connection is particularly striking in the WIMP scenario, namely that for typical gauge couplings and a mass at the weak interaction scale of a few hundred GeV, the relic density computed using standard big bang thermodynamics, as we saw in Section~\ref{ch:3}. This is rather well as tested by the  calculation of the abundances of hydrogen and helium in the early universe, through big bang nucleosynthesis. This calculation of these abundances turns out to be in amazingly good agreement with the measured ones. 
Using the same early universe thermodynamics and solving the Boltzmann equation for hypothetical dark matter particles of mass $m_\chi$, we found that the annihilation rate $<\sigma v>$ needed to explain $\Omega_{\chi}h^2\sim 0.11$ (as determined by WMAP), naturally appears for ordinary gauge couplings and a mass between around 20 GeV to a few TeV - a WIMP. 
  
Although this is not a completely convincing argument for WIMP dark matter -- it may perhaps be a coincidence -- it nevertheless gives WIMP candidates a flavour of naturalness. For non-WIMP candidates there is, on the other hand, usually a finetuning involved, or use of non-standard cosmology, to obtain the correct relic density. Even limiting oneself to WIMP models for dark matter, the literature is extensive, and among some recent developments, which cannot be discussed in this review in any detail, can be mentioned: 
\subsection{Dark Stars} Since cosmological structure in WIMP models occurs hierarchically, starting from scales  as small as $(10^{-12}-10^{-6})m_\odot$ \cite{bringmann}, the idea has been put forward that the earliest dense, small structures created by dark matter may play a role in star formation and if the dark matter particles  annihilate within the stars, unusual stellar evolution may result \cite{darkstars}. 
\subsection{Inelastic Dark Matter} These are dark matter candidates which may be excited to a state with slightly higher mass and therefore cause a higher than usual direct detection rate \cite{inelastic}, and also relieve the tension between the different direct detection experiments. 
\subsection{Dynamical Dark Matter} As it is not obvious that there is only one type of particle making up the dark matter (neutrinos should, for example contribute up to a few percent), an extreme solution could be to have a very large number, with different spins, masses, etc. \cite{dynamical}. 
\subsection{Leptophilic Dark Matter} As we have mentioned, there was  
an almost explosion of suggestions of this kind of models in 2009, when the dark matter interpretation of the anomalous positron ratio measured by PAMELA\cite{pamela} and FERMI \cite{fermi_e} was proposed to be explained by dark matter annihilation. Leptophilic means that these dark matter particles  annihilate mainly to leptons, for example by proceeding through axion-like particles below the pion mass \cite{leptophilic}.  Although the original motivation for these models has 
become somewhat weaker, the concept has established itself in the dark matter  
community. 
\subsection{Supersymmetric Models Beyond the MSSM} \cite{bmssm}. Of course, even though the minimal supersymmetric version of the standard model, the MSSM, has more than 100 free parameters, models having, e.g., motivation from new scenarios of supersymmetry breaking, are of course logically possible. These ``beyond the MSSM'' or BMSSM models may among other things give a higher Higgs mass than the limit of 130 GeV given by minimal SUSY models. In the summer of 2011, this was perhaps a favoured scenario, as the first indications of the Higgs mass was around 140 GeV. However, with more data, the preferred range (not yet significant enough by the end of 2011 to be called a discovery) is now 124-126 GeV which is more easily encompassed in the MSSM. 
\subsection{Asymmetric Dark Matter} This is a class of dark matter models which may also explain the baryon (or lepton) asymmetry of the universe \cite{asymmetric}. This generally only works for masses around or below 10 GeV, and this mass range has been in focus recently due to a (possible) signal in direct detection experiments \cite{dama,cogent,cresst}, and maybe also in $\gamma$-ray detection in the direction near the Galactic centre \cite{hooper-linden}. However, it remains to see whether these indications will stand the test of time. A similar model is ``emergent dark matter''. This is a recent version of asymmetric DM with larger possible parameter range, such as DM mass up to 100 GeV \cite{asymmetric}. 
 
\subsection{Kaluza-Klein Models}
A candidate for dark matter, the so-called LKP  (for lightest Kaluza-Klein particle)  has been identified. This appears in theories with extra dimensions, and has a rich phenomenology which we will not enter into here (for a review,
see \cite{prof_hoop}). The main difference with supersymmetry is that the 
dark matter candidate has spin-1, and can give the correct relic density for a mass in the range 600 GeV to 1 TeV.

\subsection{Inert Higgs Doublet}
Interesting are also versions of the Standard Model with an enlarged Higgs sector. If there would be, for instance, a second Higgs doublet which does not couple  directly to Standard Model particles (an ``inert doublet''), there turns out to be a stable spin-0 state which then would be the dark matter particle (see \cite{inert}, and references therein).  

\subsection{Non-WIMP Models}  
WIMPs are arguably the leading candidates for Dark Matter, due to lack of 
fine-tuning to get correct relic density. In most models, the annihilation 
cross section which sets the relic density also implies observable rates in 
various DM detection experiments. 
 
A word of caution is in place here, however. There are many non-WIMP models that also have good particle 
physics motivation, and may be detectable, like:  axions, gravitinos, superWIMPS, non-thermal dark matter, decaying dark matter, sterile Neutrinos, Q-balls\ldots 
The literature is extensive, but a good summary of both WIMP and non-WIMP models has recently appeared, namely, 700-page book giving details of most dark matterscenarios \cite{bertone}. 
  
\subsection{The Axion}  
Another, rather different candidate \cite{axions} for dark matter is provided by the axion, a hypothetical 
light boson which was introduced for theoretical reasons to explain the 
absence of $CP$ violation in the strong interactions (as far as we  
know, $CP$ violations only take place in the weak interactions). It turns 
out that for a mass range between $10^{-6}$ and $10^{-3}$ eV, the 
axion could give a sizable contribution to $\Omega_M$. It couples very 
weakly to ordinary matter, but it may be converted into a photon in 
a cavity containing a strong magnetic field (the basic coupling is to 
two photons, but here the magnetic field takes the role of one photon). 
Experiments in the USA and Japan are currently probing parts of the interesting 
mass region. A section about the axion should always be inserted when 
describing dark matter candidates, since the axion does, as does the lightest supersymmetric particle, have a good particle physics motivation for its existence.

\section{Dark Matter Detection: Status}\label{ch:4} 
As we have mentioned, there are basically three different, and complementary methods for detecting WIMPs. First, the dark matter particle may be directly produced at accelerators, in particular at the LHC, which today is the only high-energy accelerator running (although data from Fermilab's Tevatron collider will still be analyzed and may give surprises in the coming year or so). Of course, it is not clear that the particle will be kinematically allowed, and even if it is produced, one will not know that the lifetime is of the required cosmological order of magnitude. Anyway, detecting a candidate and determining its mass would be a great gain when combining with the other two search methods of dark matter, namely direct and indirect detection. In particular, direct detection experiments have seen an impressive gain of sensitivity during the last few years. The idea is to register rare events giving a combination of scintillation, ionization and nuclear recoil signals in chunks of matter shielded from cosmic rays in underground sites.  
 
In indirect detection, one rather registers products of dark matter annihilation from regions in the surrounding universe with a high dark matter  
density like the galactic centre, dwarf spheroidal galaxies, or the interior of the Earth or the Sun.  An interesting feature of indirect detection is that the expression for the local annihilation rate of a pair of DM particles $\chi$ (here assumed, like in supersymmetry, to be self-charge-conjugate, of relative velocity $v_{rel}$ 
\begin{equation} 
\Gamma_{ann}\propto n^2_{\chi}\sigma_{ann}(v_{rel}) v_{rel} 
\end{equation} 
is the dependence on the square of the number density. Also, the cross section may depend in non-trivial ways on the relative velocity. In particular, for low velocities the rate may be much higher than at high velocity, for models containing an attractive force between the annihilating particles. This is in particular true for models with  so-called Sommerfeld enhancement \cite{sommerfeld}, a resonant enhancement by in some cases orders of magnitude. This means that dwarf galaxies (dark matter subhalos) may be particularly interesting objects to study, as they are completely dark matter dominated with low rate of cosmic ray-induced $\gamma$-rays, and their low mass means a relatively low velocity dispersion, meaning higher possible rates if Sommerfeld enhancement is active.  
 
So far, indirect methods have not been as competitive as direct detection, but recently the FERMI collaboration has started to probe the interesting WIMP region by stacking data from several dwarf galaxies \cite{maja}. 
 
For non-WIMP dark matter, like sterile neutrinos (warm DM), the production rate in the early universe generally has to be tuned to give the observed relic density, but phenomenologically warm DM is possible, and according to some analyses even preferred in cosmological data \cite{hamann}. However, the significance is weak and may be influenced by statistical bias \cite{verde}. Ordinary, active neutrinos have too small mass to contribute significantly to the dark matter density, although in the extreme case may contribute a couple of percent to the critical density today. 
 
A very interesting effect for direct detection of dark matter WIMPs in 
terrestrial detectors comes about due to the motion of the solar system 
in the Galaxy \cite{freese}. This circular speed is around 200 km/s, and the direction 
of the ``wind'' of dark matter particles varies in between seasons. This is due  
to the detector following the Earth's motion around the Sun and sometimes (actually around June 2) having ``headwind'' of WIMPs and sometimes (December 2) ``tailwind''. As the cross section between a WIMP and the detector target depends strongly on their relative velocity, this causes a few percent annual modulation 
of the detection rate, something that is a very distinct signature. 
The DAMA/LIBRA experiment in the Gran Sasso tunnel \cite{dama} has in fact seen an annual modulation, which has a statistical significance of more than 8 standard deviations. 
However, since no other experiment has found the same effect (see Table~\ref{tab:exp}), the effect can still not be taken as an established  detection of dark matter.    
There have been attempts to 
interpreted the DAMA signal as possibly being due to a neutralino of 
the MSSM \cite{damabott,nath2}. It seems premature, however, to 
draw strong conclusions from this experiment alone. Besides 
some cloudy  experimental issues, the implied 
scattering rate seems somewhat too high for the MSSM or any other canonical WIMP, given the  strong Higgs mass bounds from LEP and LHC  
unless one  stretches the astrophysical  and nuclear 
physics quantities. Also, it is disturbing that neither XENON100 nor CDMS-II  
see an effect despite their nominally higher sensitivity. 
Clearly, even more sensitive 
experiments, preferably also using NaI, seem to be needed to settle this issue. 
An interesting idea, DM-Ice \cite{dmice}, uses the IceCube site to deploy
crystals of NaI with ice as a very calm surrounding medium. If an annual modulation could be measured also there one could check whether it has the same phase as that of DAMA, or if it rather follows the seasons (which are opposite on 
the southern hemisphere).

\vskip .5cm 
\begin{table} 
\caption{Some of the recent experimental claims for possible dark matter detection, and a comment on the present status.\label{tab:exp}} 
\begin{center} 
\begin{tabular}{|p{5cm}|p{5cm}|} 
\hline 
Experiment & Status of claim\\ \hline 
\hline DAMA/LIBRA annual modulation\hfill\protect\cite{dama} & Unexplained at the moment; not confirmed by other experiments \cite{cdms,xenon100}\\ 
\hline CoGeNT excess events and annual modulation \cite{cogent} & Tension with other data \cite{cdms,xenon100}\\ 
\hline EGRET excess of GeV photons \protect\cite{egret,wim}& Due to instrument error (?) -- not confirmed by FERMI \protect\cite{fermi_diff}\\ 
\hline INTEGRAL 511 keV $\gamma$-line from galactic centre region \protect\cite{integral}& Does not seem to have spherical symmetry -- shows an asymmetry which follows the disk (?) \protect\cite{integral_new}\\   
\hline PAMELA: Anomalous ratio of cosmic ray positrons/electrons \protect\cite{pamela}& May be due to DM \cite{pamela_th}, or pulsars \cite{pulsars} 
-- energy signature not unique for DM\\ 
\hline FERMI positrons + electrons \protect\cite{fermi_e}& May be due to DM \cite{pamela_th}, or pulsars \cite{pulsars} -- energy signature not unique for DM\\ 
\hline FERMI $\gamma$-ray excess towards galactic centre \protect\cite{hooper_goodenough}& Unexplained at the moment -- astrophysical explanations possible \cite{felix,boyarski}, no statement from the FERMI collaboration\\ 
\hline WMAP radio ``haze'' \cite{fink}& Has a correspondence in  ``FERMI bubbles'' \cite{su} -- probably caused by outflow from the galactic center\\    
\hline      
\end{tabular} 
\end{center} 
\end{table} 
\vskip .5cm 
 
There have recently been a number of claimed possible detections of dark matter, see Table~\ref{tab:exp}.  
Of the items in Table~\ref{tab:exp}, 
it seems that only the positron excess at high energy (20 GeV - 1 TeV) and the $\gamma$-ray excess towards the galactic center, inferred by an analysis of FERMI public data \cite{hooper_goodenough}, can be due to dark matter annihilation without tension from other data. However, they may both perhaps more naturally be explained by ordinary astrophysical processes. In addition, the DM explanation of the PAMELA and FERMI data as we have seen needs a leptophilic particle of TeV-scale mass and a very much boosted cross section. Although this may perhaps be obtained, stretching all uncertainties involved \cite{bertone_limits}, and employing Sommerfeld enhancement \cite{tracy}, the remaining window seems quite tight.  
 
The DAMA/LIBRA annual modulation is a statistically very strong signal (significance of the order of 8$\sigma$), however the lack of supporting data from other experiments is disturbing. The annual modulation hinted at by CoGeNT \cite{cogent} is statistically much weaker, and the purported excess unmodulated signal may in fact be incompatible with the level of modulated reported. Also, it seems that the DAMA/LIBRA and GoGeNT signals, if interpreted as being due to dark matter, may be in tension with each other, even if one uses freedom in isospin violation, inelastic scattering, and non-standard halo properties \cite{schwetz}. At the moment this is one of the unsolved, frequently debated issues in the dark matter community. 
    
The recent improvement of the upper limits on the WIMP-nucleon scattering cross section reported by CDMS II \cite{cdms} and, in particular, XENON100 \cite{xenon100} are truly impressive. Not only does it cast some doubt on other reported experimental results, the sensitivity is also good enough to start probing the parameter space of supersymmetric models \cite{ds}. The new calibration of the sensitivity  
to low-energy recoils of Xenon adds to the credibility of the new limits. The very good news is also that the installation of the next stage, a 1 ton liquid Xenon detector, has already started in the  Gran Sasso experimental halls in Italy. 
 
Of course, a much more decisive claim of detection of dark matter would result if any of the other methods, like a suitable new particle candidate being detected at the LHC, or a signature in $gamma$-rays from the Galactic dark matter halo would be discovered.

In the first runs at LHC, no signs of a Higgs particle, nor supersymmetry or any other of the prime candidates for dark matter, have been discovered. On the other hand, the mass region 115 - 130 GeV, interesting for the lightest Higgs boson in the simplest versions of supersymmetry, was yet to be investigated, and in fact a weak indication around 125 GeV seem to have been found. 

One possible scenario might be that such a Higgs particle is indeed found, but the particles carrying non-trivial $R$-parity  all have masses beyond reach with the LHC. This is not impossible, depending on the amount of fine-tuning one is willing to tolerate. In fact, if one puts no prior constraints on the supersymmetric parameter space other than one should have the WMAP-measured relic density, and fulfill all other experimental constraints (cf.~\cite{abdussalam}), a mass for the lightest supersymmetric neutralino in the TeV region is generic. For such heavy dark matter neutralinos, the rate for direct detection will also be small, and it would seem impossible to  test such a scenario. However, for this particular case indirect detection through gamma rays turns out to have an interesting  
advantage, as the new imaging air Cherenkov arrays like CTA 
will 
have their peak sensitivity in the energy range between a few hundred GeV to a few TeV \cite{CTA}. 

Depending on the particular model realized in nature, Sommerfeld enhancement of indirect detection may also be operative. However, 
these large arrays will be served by a large astrophysical community which will be very much interested in transient or periodic events, meaning that a ``boring'' search for a stationary dark matter spectral signature during hundreds or even thousands of hours seem out of the question. One may therefore consider a dedicated particle physics  experiment, the ``Dark Matter Array'', DMA \cite{dma} only used for dark matter search. This would have great, and complementary, potential to the large direct detection experiments that are presently being planned. In fact, we mentioned, and you heard at the lectures by F. Aharonian, that there are ideas \cite{aharonian} on how to decrease the lower threshold for detection, something that could increase the sensitivity for DM detection considerably (see Fig.~\ref{fig:5at5}). If a working prototype of this type could be built, this idea may materialize in the next decade as a new way to search for phenomena beyond the Standard Model -- with an expensive dedicated detector, still far below the cost of a new high-energy accelerator. 
 
Of course, LHC data has already started to exclude some regions of supersymmetric parameter space, although not very much. This may be surprising, but is in fact due to the relative independence of the squark and gluino sector of supersymmetry, and the neutralino sector, which hosts the dark matter candidate. In fact, 
as mentioned, there are so-called split supersymmetry models, which have this dichotomy explicitly postulated \cite{split}. 
 
\begin{figure}[!htb] 
\begin{center} 
\includegraphics[width=0.9\linewidth]{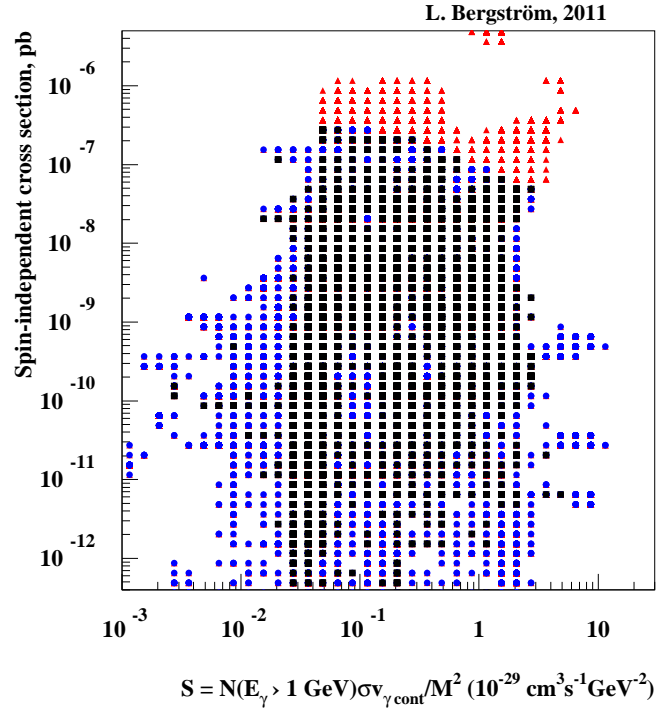} 
\end{center} 
\label{fig:compl} 
\caption{Scan of the MSSM parameter space showing the direct detection cross 
section vs.~indirect detection through gamma rays. The uppermost points are excluded by XENON100, and points which survive also the LHC 2011 data are shown in black.} 
\end{figure}

The complementarity of direct and indirect detection is shown in Fig.~\ref{fig:compl}, where also the effects on the parameter space caused by the XENON100 bounds and LHC 2011 bounds, respectively, are shown. 
 
\section{A Detailed Calculation: The Saas-Fee WIMP}\label{ch:8} 
 
An interesting question came up during the Saas-Fee Workshop:
Could there be a cosmological contribution to the $\gamma$-ray spectrum 
making up the deficit in the diffuse $\gamma$-ray emission measured by FERMI? As  we heard, this is not readily explained by  adding the well-known sources like AGNs, millisecond pulsars and star-forming galaxies described by C. Dermer.

 Here we will outline the simple steps in making the dark matter prediction 
for this flux, based on \cite{cosmdm_prl} (see \cite{cosmdm} for a much more thorough treatment). We will see how that could lead us to predict a several hundred GeV dark matter particle - the Saas-Fee particle as we named it at the Saas-Fee school it in 2010. This was only published in online slides from my talk 
(it is probably still there on the homepage of the school), and should not be taken too seriously. However, as a pedagogical example of a surprising effect of 
the accumulated dark matter structure in the universe it is quite instructive.   
 
As we have seem, in the presently most  
successful  model for structure  
formation, $\Lambda$CDM, most of the matter is in the form  
of non-relativistic cold dark matter (CDM), but with a contribution  
to the present-day energy density also from a cosmological constant  
($\Lambda$). As shown by detailed $N$-body simulations 
(see, e.g., \cite{aquarius,diemand} and references therein), 
in such a picture large structures form by the successive merging of small 
substructures, with smaller objects generally being denser. $N$-body simulations also show that the dark matter density profile 
in clusters of galaxies and in single galaxies develops an enhancement of the density near the centre, although it is at present unclear how steep this increase is, and whether it even shows a cusp 
near the center like in the popular parametrization of Navarro, Frenk and White, $\rho_{CDM}(r)\sim r^{-\alpha}$ with $\alpha$ close 
to $1$ \cite{nfw} (a very similar profile, the Einasto profile, does not show 
a singularity, but behaves rather similarly on moderate length scales). 
 
At present, it is not clear whether these $N$-body predictions are in 
agreement or not with all available data (one recently acknowledged problem 
is, for example, the apparent lack of a halo mass threshold for dwarf galaxies \cite{Ferrero:2011au}). On large scales, however, the $\Lambda$CDM scenario 
gives excellent agreement with observations.  
On smaller scales, the dynamical range of the simulations is not enough, and one of the main puzzles is how to properly include 
the non-linearities induced by baryonic matter in the form of supernova explosions and other feedback mechanisms.  
 
Let us assume that the $\Lambda$CDM picture is basically 
correct and that structure forms hierarchically, with the number 
density of halos of mass $M$ being distributed as $dN/dM\propto M^{-\beta}$ 
with $\beta\sim 1.9$ -- $2$, as predicted by Press-Schechter theory  
and also verified in $N$-body simulations. Furthermore, 
the concentration of halos grows in these simulations with decreasing 
mass. 
 
It is interesting that the averaging involved in computing the integrated signal of annihilation $\gamma$-rays of unresolved cosmological dark matter 
 gives results which are more robust to 
changes in the details of how the dark matter is distributed on small scales. (The same is actually also true for all sources which are completely encompassed by the angular resolution cone of a given $\gamma$-ray experiment, for the obvious source of the galactic centre, the prediction of fluxes differ by up to 4 orders of magnitude for different models: in particular they are very sensitive to the presence or not by a central cusp.) 
 
Let us consider annihilation of a WIMP 
such as the lightest neutralino $\chi$ of the MSSM,  as a template.  
The mass range is from around 20 GeV up to several TeV \cite{ds}.   
For the sake of pedagogy, let us start with the unrealistic case of all the dark matter is 
smoothly distributed with the density distribution being just a number,  the average density, at all redshifts. The idea is that since the dark matter was more dense in the early universe, one may get a large (red-shifted) flux by integrating over the line of sight from 0 to very high redshifts. Actually, in papers preceding \cite{cosmdm_prl} this was the only effect considered. We will soon modify the results by 
introducing the effects of structure, which indeed increases the flux by many orders of magnitude.

\subsection{The Flux in a Smooth Universe} 
The comoving number density $n_c$ of 
WIMPS, after decoupling from chemical equilibrium 
(``freeze-out'') at very large temperatures ($T\sim m_\chi/20$) is 
depleted only slightly due to self-annihilations, governed by the Boltzmann 
equation 
 \beq \dot n_c=-\langle\sigma v 
\rangle(1+z)^3n_c^2,\eeq 
 where $\langle\sigma v\rangle$  
is the thermal- and angular-averaged annihilation rate, which, to an 
excellent approximation after freeze-out, is velocity independent, 
since the neutralinos move non-relativistically, and there always in a dominant 
$S$-wave component (at least for our supersymmetric WIMP templates).

Each pair of  $\chi$ particles that disappears through annihilation 
 give rise to $N_\gamma$ photons on the average, with an energy 
distribution in the rest frame of the annihilation pair, 
\beq 
{dN_\gamma(E)\over dE}={dN_{\rm cont}\over dE}(E)+b_{\gamma\gamma} 
\delta\left(m_\chi-E\right). \label{eq:edist} 
\eeq 
Here the first term gives the average continuum gamma ray distribution per 
annihilating $\chi$ and we have also added a term for the possible $\gamma\gamma$ line 
contribution, with $b_{\gamma\gamma}$ being the branching ratio into 
$\gamma\gamma$  (one could also have a $Z\gamma$ 
channel). 
  
A $\gamma$-ray observed today, at redshift $z=0$, of energy $E_0$  would correspond 
to an energy at the emission at redshift $z$ of $E=(1+z)E_0$. We can now track, using  
the Boltzmann equation, the number of WIMPs 
that have disappeared from redshift $z$ 
until now, and fold in the energy distribution (\ref{eq:edist}). 
Thus we we get a first  estimate of the level of the diffuse  
extragalactic $\gamma$-ray flux. 
As usual, $H_0$ is the Hubble parameter, and 
we use the relation between time and redshift (see, e.g., 
\cite{BGbook}) 
$d/dt=-H_0(1+z)h(z)d/dz$ 
with 
\beq h(z)=\sqrt{\Omega_M(1+z)^3+\Omega_K(1+z)^2+\Omega_\Lambda}\sim \sqrt{\Omega_M(1+z)^3+\Omega_\Lambda}.\eeq 
Here $\Omega_M$, $\Omega_\Lambda$ and $\Omega_K= 
1-\Omega_M-\Omega_\Lambda$ are the present fractions of the critical density 
given by matter, vacuum energy and curvature. We can here use the  result from the first section that 
the universe to an excellent approximation is flat, $\Omega_K=0$. We then obtain 
\beq 
{dn_c(z)\over dz}=\frac{\langle\sigma v\rangle}{H_0}\left((1+z)^2\over h(z)\right)n_c(z)^2.\label{eq:zboltz} 
\eeq 
 
The differential energy spectrum of the number density $n_\gamma$ of photons 
generated by WIMP annihilations is then given by 
\beq 
{dn_\gamma\over dz}=N_\gamma{dn_c\over dz} 
=\int_0^{m_\chi}{dN_\gamma(E)\over dE}{dn_c\over dz}dE. 
\eeq 
Here, $dn_c/dz$ can be computed directly from (\ref{eq:zboltz}) to excellent 
accuracy, replacing 
the exact solution $n_c(z)$ by the present average number density of neutralinos 
$n_0$ on the right hand side. This we can do since the comoving number density does not change appreciably after freeze-out.  
  
Neglecting the baryon contribution (as we will see, factors of order unity will 
not make a difference here, due to much larger uncertainties in structure formation), $\Omega_\chi \sim \Omega_M$, we obtain 
\beq n_0=\rho_\chi/m_\chi=\rho_{\rm c}\Omega_M/m_\chi.\eeq Here 
$\rho_{\rm c} =1.06\cdot 10^{-5}\ h^2\ {\rm GeV/cm}^3$ and $h$ as before is  
the scaled Hubble parameter in units of 100 km s$^{-1}$ Mpc$^{-1}$, $h\sim 0.7$.  
There are a few more effects we have to include. We have to use the fact that all photons move with velocity $c$ and that the average flux is isotropic from each volume element where annihilation takes place, giving a factor $1/4\pi$ per steradian. The cross section times velocity average should, for Majorana particles, also be divided by 2, something which was missing in the original derivation \cite{cosmdm_prl}, but added in \cite{cosmdm} (see the published version).  
Some of the photons will be absorbed after travelling over cosmological distances. This can to the level of our approximate calculation be handled by introducing 
a simple energy- and redshift-dependent factor  $e^{-z/z_{\max}}$ (or the more detailed 
calculation in \cite{cosmdm} a more complicated factor depending on $z$ and $E_0$). 
 
The resulting  $\gamma$-ray flux at the detector is then given by: 
\beqa 
\phi_\gamma ={c\over 8\pi}{dn_\gamma\over dE_0}=4.2\cdot 10^{-14} 
\ {\rm cm}^{-2}{\rm s}^{-1}{\rm sr}^{-1}{\rm GeV}^{-1}\times \nonumber\\ 
{\Gamma_{26}\Omega_M^2 h^3\over m^2_{100}} 
\int_0^{z_{up}}dz {(1+z)^3e^{-z/z_{\max}}\over h(z)} 
{dN_\gamma(E_0(1+z))\over dE}.\label{eq:master} 
\eeqa 
where we defined 
$\Gamma_{26}=\langle\sigma v\rangle/(10^{-26}\ {\rm cm}^3{\rm s}^{-1})$ and 
$m_{100}$ the mass in units of 100 GeV. 
 
For the energies we are interested 
in, $1$ GeV $< E_0 < 500$ GeV, it is the starlight and 
(more poorly known) infrared background radiation which is most important,  
whereas the CMBR may be important for higher energies. 
An optical depth of order unity is reached for a redshift which in \cite{cosmdm} was  
approximated by $z^{\rm old}_{\rm max}(E_0)\sim 3.3(E_0/10\ {\rm GeV})^{-0.8}$,  
which represented  older results. However, the newer 
data discussed at length in the lectures by C. Dermer, indicate much less absorption. As a representative of the more recent evaluation of this absorption \cite{finke}, we take instead the simple approximation 
$z^{\rm new}_{\rm max}(E_0)\sim 2.3(E_0/50\ {\rm GeV})^{-1.1}$. 
  
The exponential form is a good approximation for 
small values of $z_{\rm max}$ as is dominant in most of the cases we study here. 
The upper limit of integration is given by kinematics, $z_{up}=m_\chi/E_0-1$,  
as the maximum rest frame energy of a photon in an annihilation event 
is $E=m_\chi$. The gamma line contribution to (\ref{eq:master}) is 
particularly simple, just picking out the integrand at $z+1=m_\chi/E_0$; 
it has the very distinctive and potentially observable signature of being 
asymmetrically smeared to lower energies (due to the redshift) and of 
suddenly dropping just above $m_\chi$. Unfortunately, for most models the branching ratio for this channel is too small to be measurable with present-day energy resolution, and we will drop it from now on. (This may however change when the  
high-resolution instrument GAMMA-400 \cite{gamma-400} is operational towards the end of this decade. This is specified to have an energy resolution of 1\%, which will be a perfect instrument for searching for $\gamma$ lines from annihilation, and also from models where dark matter decays radiatively \cite{wilfried}.) The continuum emission 
will produce a characteristic, although less conspicuous feature,  
a smooth ``bump'' below 
around one tenth of the neutralino mass, and may be more difficult to detect. 
One should notice that there are particular models where radiative corrections (``internal bremsstrahlung'') may give a significantly harder spectrum near 
$E_\gamma=m_\chi$, facilitating discrimination against most backgrounds \cite{bbe}. 
\subsection{Including Effects of Cosmic Structure} 
To give an example of the results (which in \cite{cosmdm_prl} contained both obsolete SUSY models and not very accurate data from the old EGRET experiment),  
we  take a generic model with mass 600 GeV, and the canonical WIMP averaged cross section times velocity of $\langle \sigma c\rangle=3\cdot 10^{-26}$ cm$^3$s$^{-1}$, 
in the ``concordance'' cosmology $\Omega_M=0.3$, $\Omega_\Lambda=0.7$, 
$h=0.7$. 
The continuous $\gamma$-ray rest frame energy distribution per annihilating 
particle comes mainly from hadronization and decay of $\pi^0$s and is 
conveniently parametrized to reasonable accuracy as  (see (\ref{eq:contappr}))
$$ dN_{\rm cont}(E)/dE=(0.42 / m_\chi)e^{-8x} / (x^{1.5}+0.00014)$$
where $x=E/m_\chi$. This is valid for most quark jet final states, except for top. Also, $\tau$ lepton decays give a somewhat harder $\gamma$-ray spectrum, and as mentioned internal bremsstrahlung may be important for certain types of models.

The most difficult, but also most important and interesting part of the calculation is to include the effects of structure formation. 
Following  \cite{cosmdm_prl}, we  consider first a halo of mass $M$ whose radial density profile can be 
 described by  
$\rho_{DM}(r)=\rho'_{DM}f\left(r/a\right)$, 
with $\rho'_{DM}$ being a characteristic density and $a$ a 
length scale. These are found in $N$-body simulations 
not to be independent parameters, as smaller halos are 
generally associated with higher densities.  
 
As a simple first model for structure formation, assume that the halo of mass $M$ accreted from a spherical 
volume of radius $R_M$, determined by requiring that the average cosmological 
density times that volume is equal to $M$, $\rho_0\cdot 4\pi R_M^3/3 =M$ 
(with $\rho_0\sim 1.3\cdot 10^{-6}\ {\rm GeV/cm}^{3}$). 
The increase of average squared overdensity per halo, which is what enters the annihilation rate, is given by: 
\beqa 
\Delta^2 &\equiv& 
\langle\left({\rho_{DM}\over\rho_0}\right)^2\rangle_{r<R_M} = \left({\rho'_{DM}\over \rho_0}\right) {I_2\over I_1}, 
\eeqa 
where $ I_n \equiv \int_0^{R_M/a}y^2dy(f(y))^n$. 
Here the dependence on the limits of integration is rather weak, 
at least for profiles less cuspy than the NFW profile. 
  
Computing $I_2/I_1$ numerically, and using values of  
$\rho'_{DM}/\rho_0$ as determined for Milky Way size halos we find values of $\Delta^2$ of  $1.5\cdot 10^{4}$  
for the Navarro-Frenk-White (NFW) 
profile\cite{nfw}, and $7\cdot 10^{3}$ for a cored, modified 
isothermal profile (modified so that the density falls as $1/r^3$ at 
large radii).  
The flux ratio, $2:1$ for these two models 
should be compared with the ratios roughly $100:1$ obtained within 
a few-degree cone encompassing the galactic center, showing the announced relative insensitivity to halo density profiles. 
  
We should now also take into account that the number density 
of halos is scaling like $\sim 1/M^{1.9}$, and that small-mass halos 
are denser. We can  resort to the highest-resolution $N$-body 
simulations available to date \cite{millenniumii} . Fitting the concentration parameter 
of  halos by 
 
\beq c\sim 100\, (M_{\rm vir}/h^{-1}M_\odot)^{-0.08},\eeq  
 
one finds to a good approximation 
 
\beq\Delta^2\sim 2\cdot 10^{5}M_{12}^{-0.2},\eeq 
 
where $M_{12}$ is the halo mass in units of $10^{12}$ solar masses. 
This means that the total flux from a halo of mass $M$ scales 
as $M^{0.8}$. Since the number density of halos goes as $M^{-2}$, 
the fraction of flux coming from halos of mass $M$ 
scales as $M^{-1.2}$. Thus the $\gamma$-ray flux will dominantly 
come from the smallest CDM halos. In simulations, substructure has 
been found on all scales (being limited only by numerical resolution). 
For very small dark matter clumps, however, no gain in overdensity 
is expected, since once the matter power spectrum enters 
the $k^{-4}$ region a constant density is expected. 
There are arguments \cite{hofmann} that structure is present in cold dark matter models all the way down to $10^{-6}$ or smaller \cite{torsten}. We conservatively set $1 M_\odot$ as the minimal scale. In a more detailed treatment, one should 
also include effects of clumps within clumps, which increase the enhancement. However, destruction of DM clumps near large central densities of halos should also be included.

Finally, regarding redshift dependencies, we  assumed in \cite{cosmdm_prl} a 
 constant 
enhancement factor $\Delta^2$ out to $z\sim 1$, and somewhat arbitrarily 
imposed quadratic growth in the enhancement factor from $z\sim 10$ to 
the fully developed epoch $z=1$.  (The computed flux  is not very 
sensitive to this assumption.)  Furthermore, in (\ref{eq:master}) 
we make the replacement $(1+z)^3 \rightarrow 1$, 
reflecting the fact that the we are now 
dealing with a clumped, rather than a smooth distribution 
with density scaling $\sim (1+z)^3$. 
 
We thus arrive at the  
following expression for the flux including structure formation 
\beqa 
\phi_\gamma ={c\over 8\pi}{dn_\gamma\over dE_0}=4.2\cdot 10^{-14} 
\ {\rm cm}^{-2}{\rm s}^{-1}{\rm sr}^{-1}{\rm GeV}^{-1}\times \nonumber\\ 
{\Gamma_{26}\Omega_M^2 h^3\over m^2_{100}} 
\int_0^{z_{up}}dz {\Delta^2(z) e^{-z/z_{\max}}\over h(z)} 
{dN_\gamma(E_0(1+z))\over dE}.\label{eq:master2} 
\eeqa 
 
\subsection{The Saas-Fee WIMP} 
We find using (\ref{eq:master2}) (see also \cite{fermi-cosmdm}) that 
the flux from small halo structure is enhanced by roughly  
a factor $(4-10)\cdot 10^{7}$ compared to the smooth case, giving in the upper range observability  
for the  annihilation parameters as used above. The uncertainties reside mainly 
in the still poorly known factor $\Delta^2(z)$ and its extrapolation to small 
halo masses (and also the effects of DM clumps within clumps, for instance).  
 
\begin{figure}[!htb] 
\begin{center} 
\includegraphics[width=\textwidth]{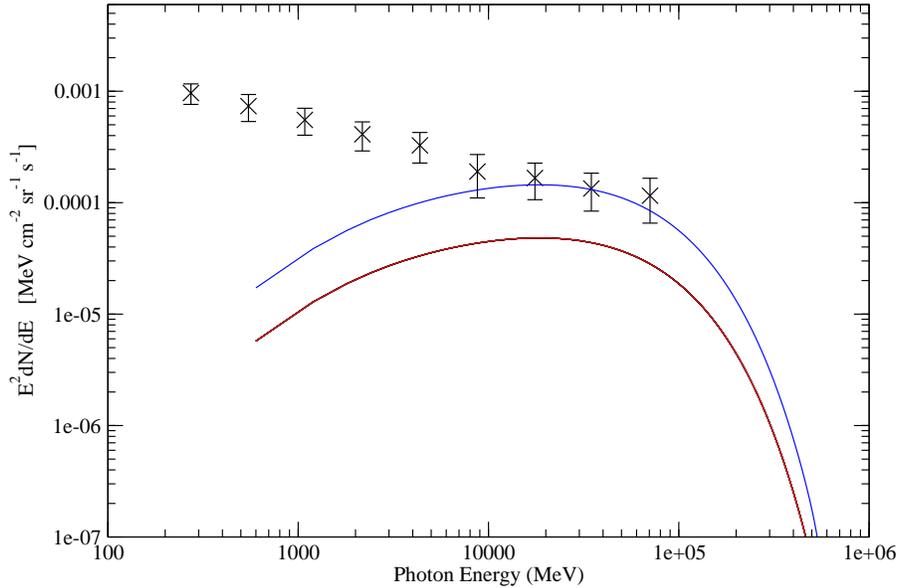} 
\end{center} 
\caption{\small The predicted diffuse extragalactic $\gamma$-ray flux computed  
using the methods described in the text, for a 600 GeV WIMP with a cross section compatible with the WMAP-inferred relic density and with different 
assumptions for the effects of structure. The diffuse extragalactic data was measured by FERMI-LAT \protect\cite{fermi_diff}. \label{fig:fermi}} 
\end{figure}

In Fig.\,~{\ref{fig:fermi}}, we show the results for this 600 GeV WIMP model. 
The results are compared with the measurements from FERMI-LAT \cite{fermi_diff}, and despite the fact that there is this uncertainty in the absolute rates, it is amusing, as I discussed at the Saas-Fee school, that the possible break in the FERMI data may be caused by a new  
contribution from 500-600 GeV mass annihilating dark matter (``The Saas-Fee WIMP'', of which there would be one per 2 litres in the lecture hall at Les Diablerets as in all other places on Earth) setting in. It will obviously be interesting to follow the development of this data set during the next few years, to see if this models survives or even becomes more convincing. 
 
It has of course to be remembered that the strength of the annihilation signal 
can be much lower than the proof-of-existence example  chosen for Fig.\,~{\ref{fig:fermi} 
in which case a detection would be correspondingly more difficult. On the other hand, there may be particle physics effects (such as Sommerfeld enhancement) which could give a higher flux. 
 
As a recent illustration of the importance of adding up all structure present, e.g., in galaxy clusters, can be mentioned the results of \cite{pinzke} and 
\cite{frenk}, where it was shown that by choosing particularly favourable, not  
too distant clusters, one is very close to the current observational limits 
from FERMI-LAT. Indeed, there may even be (weak) indications from FERMI data 
of a signal \cite{han_frenk}. 
 
A related type of analysis for the diffuse extragalactic case  
is performed in a similar way as when analyzing the angular fluctuations 
in the CMB. Also using this method of analysis, the conclusion is that with FERMI-LAT data one may be very near detection \cite{jennifer}.  
  
\section{Primordial Black Holes as Dark Matter?} 
 
Sometimes one gets the question from lay persons, when telling that we are interested in the problem of dark matter: {\em Could it be black holes?} Black holes are in some sense dark: they do not emit light from within the event horizon, so the question is not completely out of context. However, the only black holes which we are relatively certain to exist are those around 
$2-20$ solar masses, and the supermassive ones like the one residing at the  
Galactic center (of mass a few times $10^6$ $M_\odot$). We also know of even more massive ones (mass up to a few times $10^9$ $M_\odot$) making up active galactic nuclei (AGNs). However, the galactic halos in which even the most massive, ``supermassive'', black holes reside have a factor of at least 1000 more total mass.Thus their influence on the cosmic energy balance is rather marginal.   
Also the solar mass type black holes which are produced as end results of stellar evolution constitute a small fraction, by far too small to explain the abundant and omnipresent dark matter. Finally, most black holes are in practice not very dark, as their concentrated mass has an effect on surrounding matter, causing various types of radiation processes, as ordinary matter is ``eaten'' by the black hole.  
 
However, very massive black holes may be important for dark matter detection: if left undisturbed for a long time, they may adiabatically attract surrounding dark matter, 
changing the NFW-type distribution to a much more spiky cusp \cite{gondolo_silk}. As the annihilation rate grows with the square of the dark matter density, this could give a dramatically increased rate of $\gamma$-rays, and in particular 
neutrinos (which are unlikely to be absorbed by  surrounding matter). More extreme versions of this scenario are in fact ruled out already, due to the lack of unambiguous dark matter signals from the galactic centre. An interesting possibility is that intermediate mass black holes exist, where this type of signal could be close to detection, e.g., with FERMI-LAT \cite{gianfranco_ibh}. 
\subsection{Primordial Black Holes} 
There is also a small, be definite probability that small mass black holes would have been formed in the early universe. These ``primordial'' black holes  
would have to have been formed with a rather peculiar mass spectrum not to overclose the universe, and not to create too much radiation due to the interesting spontaneous emission of Hawking radiation, named after its discoverer (in theory). 
 
Let us first remind ourselves of the metric surrounding a point mass $M$,   
 
\beq 
ds^2=\left(1-\frac{r_S}{r}\right)dt^2- 
\frac{1}{1-\frac{r_S}{r}}dr^2-r^2\left(d\theta^2+\sin^2\theta d\varphi^2\right), 
\eeq 
where $r_S$ is the Schwarzschild radius, $r_S=2GM$ (here $G$ is Newton's constant, and as usual we set $c=1$). 
A radial photon trajectory is this metric is given by $ds^2=0$, which gives 
\beq dt=\frac{dr}{1-\frac{r_S}{r}},\eeq 
for $\theta=\phi=const$. The time for the photon to travel from $r_i$ to $r_f$ 
is thus  
\beq t_f-t_i=\int_{r_i}^{r_f}\frac{dr}{1-\frac{r_S}{r}}.\eeq 
This diverges as $r_S\to r_f$, so that a light ray which starts at $r<r_S$ will never reach an outside observer, we have a black hole! If we define the sphere  
at $r=r_S$ as the areas $A_{BH}$ of the black hole, we find 
\beq 
A_{BH}=4\pi r_S^2=4\pi(2GM)^2= 16\pi G^2M^2. 
\eeq  
It is interesting to contrast this with the behaviour of a solid sphere in classical physics, where $M\sim R^3$ which gives $R\sim M^{\frac{1}{3}}$, so that  
$A_{class} \sim M^{\frac{2}{3}}$. This difference is due to the strong curvature of space-time near the black hole.  
 
As we noted, known black hole candidates either have a mass of a few solar masses (probably remnants of stellar collapse, as the maximal mass of a neutron star is somewhere between 1.4 solar masses – the Chandrasekhar mass - and a few solar masses), or a few million solar masses (Milky Way centre) to billions of solar masses (AGNs). 
There is no known present formation mechanism for BHs of mass less than a solar mass, so these, if they exist,  must be primordial (PBHs), i.e. produced in the early universe, e.g. at some phase transition after inflation. There are various limits restricting formation scenarios, in general one has to ``cook up'' a power spectrum of density fluctuations which peaks at a particular mass length scale. When the horizon passes that scale, copious production of BHs may  
occur in such a scenario. An example can be found in a recent paper \cite{frampton_2010} where one tries to explain all of dark matter with PBHs, by having a power spectrum with a huge peak ($\delta\rho/\rho\sim 0.1$) at a scale corresponding to a black hole mass of $10^{-7}M_\odot.$ 
 
\subsection{Hawking Radiation} 
If PBHs exist, one may detect them through Hawking radiation, as Hawking discovered in a remarkable 1975 paper \cite{hawking} that a black hole should emit thermal radiation. This can be explained as a tunneling phenomenon \cite{parikh}. 
 
Let us make an extremely simplified heuristic version of the derivation. Let us say that we have an isolated black hole. We can then for certain say that it is  
inside the Schwarzschild radius $r_S$. This uncertainty in the position of the radiation gives an uncertainty of the time $\Delta t\sim r_S/c=2GM/c$, but the uncertainty relation between time and energy, 
$\Delta E\Delta t\sim \hbar/2$ gives  
\beq 
\Delta E\sim\frac{\hbar c^3}{4GM}\sim E_{th}=k_BT\to 
k_BT\sim\frac{\hbar c^3}{4GM}=\frac{1}{4GM}. 
\eeq  
Thus, in our units, where also $k_B=1$, the temperature $T=1/(4GM).$ This is only a factor of $2\pi$ different  from Hawking's result: 
\beq 
T_H=\frac{1}{8\pi GM}. 
\eeq 
Of course, Hawking's derivation is much more beautiful, by explaining  
the radiation by mixing of positive and negative energy states due to the strong space-time curvature near the black hole. Another way to understand the process is that for a virtual particle pair created just at the horizon, one of the particles will be dragged into the black hole, releasing gravitational binding energy 
to the other particle, which can then appear as a real propagating state outside the horizon.     
 
An interesting consequence of Hawking radiation and the equivalence principle is 
a uniformly accelerated observer, with acceleration  $a$, in “empty space” should see a thermal distribution of particles - the Unruh effect.  The Unruh temperature is $T = a/2\pi$. 
Attempts have been made to measure the Unruh effect at accelerators, but present-day accelerations are not large enough. 
It has been argued, however, that the so-called Sokolov-Ternov effect (depolarization of electrons in a storage ring due to synchrotron radiation) really has  the same origin as the Unruh effect - and it has been experimentally verified \cite{Akhmedov}. 
\subsection{Thermodynamics of Black Holes} 
If we regard the Hawking temperature as a true thermodynamical temperature $T(M) = T(E)$, there should also be an entropy (Bekenstein entropy) associated with the BH: 
 
\beq 
T(E)=\frac{1}{8\pi GE};\ \ \ dS=\frac{dE}{T(E)} 
\to S =\int_0^M8\pi GEdE=4\pi G M^2=\frac{1}{4}\frac{A_{BH}}{G}. 
\eeq 
 
If we remember that $G = 1/M_{Pl}^2 = l_{Pl}^2$, we see that each ``Planck area'' of the surface of the BH contributes one quarter unit of entropy, and one gets huge numbers. This is still mysterious – what are the degrees of freedom describing the black hole, and why does ordinary matter that forms a BH suddenly increase its entropy enormously?   
 
T describe black hole evaporation, it is useful to remember the form of a thermal  
distribution for a particle species 
\beq 
f_i(\vec p)= 
\frac{1}{e^{\frac{E_i-\mu_i}{k_BT}}\pm 1}= 
\frac{1}{e^{\frac{E_i-\mu_i}{T}}\pm 1}. 
\eeq 
 
This means that for the rate of mass loss we can analogously write \cite{MacGibbon} 
 \beq 
\frac{dM}{dt}=-\sum_j\frac{1}{2\pi}\int_{m_j}^\infty 
\Gamma_j \frac{EdE}{e^{8\pi GME}\pm 1}=\ldots=-5\cdot 10^{25} f(M)M^{-2}\ \ gs^{-1}. 
\eeq 
Here $\Gamma_i$ is the absorption rate for particle of type $j$ and the sum is  
over all particle-antiparticle pairs. This gives the evaporation time 
\beq 
\tau_{evap}\sim \int_{M_{min}}^{M_{max}}\frac{M^2}{f(M)} dM \sim \frac{6\cdot 10^{-27}}{f(M_i)}  \left(\frac{M_i}{1\ g}\right)^3\ s. 
\eeq 
 
Thus, only BHs with mass $> 10^{15}$ g are stable on cosmological time scales (so don’t worry about BHs produced at LHC – they would evaporate immediately - if they exist!) 
Upper limits of $\gamma$-ray emission from EGRET and FERMI-LAT gives the  approximate bound for light pBHs: 
\beq 
\Omega_{PBH}(M < 10^{15}\ g)< 10^{-8}. 
\eeq  
 
Actually, since the temperature  
increases with decreasing mass, all particles, even more massive than those presently 
produced at accelerators, may give a contribution in the final ``Hawking explosion''. In particular, if supersymmetry is realized in nature, the end-time evolution may have interesting differences from the scenario with only Standard Model 
particles \cite{Sahlen}. 
 
Let us discuss how primordial black holes formed in the early universe (see \cite{Carr}). The relevant length scale is the particle horizon length, so that   
\beq 
M=\gamma M_{\rm (particle\ horizon)}=2\cdot 10^5\gamma\left(\frac{t}{1\ s}\right)M_\odot, 
\eeq 
where $\gamma\sim 0.2$ depends on the detailed formation scenario. We can now  
compute the fraction of total energy density in black holes at formation: 
\beq 
\beta(M)\equiv \frac{\rho_{PBH}(t_i)}{\rho(t_i)}= 
8\cdot 10^{-29}\frac{1}{\sqrt{\gamma}}\left(\frac{g_i}{106.75}\right)^\frac{1}{4} 
\left(\frac{M}{M_\odot}\right)^\frac{3}{2}\left(\frac{n_{PBH}(t_0)}{1\ {\rm Gpc}^{-3}}\right). 
\eeq 
This means a contribution to $\Omega$ today of 
\beq 
\Omega_{PBH}=\frac{Mn_{PBH}(t_0)}{\rho_c}=\left(\frac{\beta(M)}{1.2\cdot 10^{-8}}\right) 
\sqrt{\gamma}\left(\frac{g_i}{106.75}\right)^{-\frac{1}{4}} 
\left(\frac{M}{M_\odot}\right)^{-\frac{1}{2}}. 
\eeq 
The WMAP bound (the PBHs would behave gravitationally as cold dark matter)  $\Omega < 0.25$, gives 
}\beq 
\beta(M) < 2\cdot 10^{-18} 
\frac{1}{\sqrt{\gamma}}\left(\frac{g_i}{106.75}\right)^{\frac{1}{4}} 
\left(\frac{M}{10^{15}\ g}\right)^{\frac{1}{2}}. 
\eeq 
(This is valid for BHs that have not evaporated today, i.e., for $M > 10^{15}$ g.) It is convenient to divide out the cosmology/formation factors and consider the simpler expression for the energy density limit from WMAP: 
\beq 
\beta'(M)<\sqrt{\gamma}\left(\frac{g_i}{106.75}\right)^{-\frac{1}{4}}\beta(M)= 
2\cdot 10^{-18}\left(\frac{M}{10^{15}\ g}\right)^{\frac{1}{2}}. 
\eeq 
Limits on $\beta'(M)$ can be obtained from a variety of data, from BBN and CMB in the early universe to the galactic and diffuse extragalactic $\gamma$-ray emission, gravitational lensing data and large scale structure. The limits we just computed on $\Omega_{PBH}$ is also important in the region $M\sim 10^{15}-10^{27}$ g (for a detailed summary of the situation, see \cite{Carr}).  
 
To conclude: PBHs of mass less than around $10^{15}$ g cannot be the dark matter due to important constraints from the absence of Hawking radiation in 1 - 100 MeV $\gamma$-rays, but may still be a subdominant component. It is worthwhile to look for $\gamma$-ray signatures - a discovery of Hawking radiation would be truly wonderful! 
 
At all masses, there are important bounds from a variety of methods. In principle, there are mass ranges where PBHs can still be the dark matter - all of dark matter, but one needs contrived production mechanisms such as a strongly peaked, and fine-tuned, density power spectrum.  
 
\section{Gravitational Waves}\label{ch:9} 
 
We will now, in view of the multi-messenger aspects of this lecture series, 
 discuss one more type of radiation which is  
deeply linked to the theory of general relativity on which modern  
cosmology rests: gravitational radiation.

Due to the nonlinearity of Einstein's equations, it is virtually  
impossible to find exact solutions to the metric tensor  
$g^{\mu\nu}({\bf r},t)$ corresponding to the dynamics, for example, of a  
 massive star which collapses to a black hole  near the strong  
 gravitational field of the star (using supercomputers, numerical  
 studies can, however, be made). Far from the source of the  
 gravitational field, it is on the other hand reasonable to use a  
 first-order approximation. The gravitational deformation of  
 space-time near celestial bodies like the  Earth or the Sun due to conceivable astrophysical processes happening elsewhere in the Galaxy is indeed  as we will 
see 
 extremely tiny, which justifies such a perturbative approach. The same smallness of the effect unfortunately also make detection of gravitational radiation very challenging. 
\subsection{The Gauge Choice for Electromagnetism}  
 Recall the way one derives the existence of electromagnetic  
 waves in Maxwell's theory. One 
 inserts the vector potential $A^\mu$ in the equations of motion 
 for a vanishing current $j^\mu$ (since we are dealing with propagation in  in vacuum) 
 to obtain 
 \beq 
\Box A^\mu  
-\partial^\mu\left(\partial_{\nu}A^\nu\right)=0  
 \eeq 
 Through the use of the gauge freedom $A^\mu\to A^\mu+{\partial}^\mu  
 f$, we can choose $A^\mu$ to fulfill  $A^0=0$ and also the co-called  
 Lorentz condition $\partial_{\nu}A^\nu=0$. This leads to the  wave  
 equation 
 \beq 
 \Box A^\mu =0 
 \eeq 
 with solutions of the form  
 \beq 
 A^\mu({\bf r},t)=\epsilon^\mu e^{\pm i(\omega t - {\bf k}\cdot{\bf  
r})}=\epsilon^\mu e^{\pm ik^\mu x_{\mu}} 
\eeq 
where $k^\mu k_{\mu}=0$ (light-like propagation) and  the gauge  
conditions
$A^{0}=0$ 
and $\partial_{\nu}A^\nu=\nabla\cdot {\bf A}=0$ translate into $\epsilon^0=0$ and 
${\bf k}\cdot {\bf \epsilon}=0$. This means that  the two physical  
degrees of freedom are transverse to the direction of propagation, and there is no time-like mode of propagation (this is deeply connected to the masslessness 
of the photon).  
 
\subsection{Gauge Choice for the Metric Perturbation} 
In the case of gravity waves in Einstein's theory of general relativity, we can similarly  make a first-order expansion of the  
dynamical degrees of freedom, which are the components of the metric tensor  
field $g_{\mu\nu}$, around  
the constant Minkowski metric $\eta_{\mu\nu}$: 
\beq 
g_{\mu\nu}=\eta_{\mu\nu}+h_{\mu\nu},\label{eq:smallexp} 
\eeq 
 and work to to first non-vanishing order in $h_{\mu\nu}$. 
 
Now we have a spin-2 field  
$ 
h_{\mu\nu} 
$ 
instead of the vector quantity $A^\mu$, but again we can use a gauge-like invariance (which in this case  rather is re-parametrization invariance)   
\beq 
x_\mu\rightarrow x_\mu + \xi_\mu(x) 
\eeq 
translating into  
\beq 
h_{\mu\nu}\rightarrow h_{\mu\nu}-\partial_\mu\xi_\nu-\partial_\nu\xi_\mu. 
\eeq 
Using this we may impose the so-called traceless gauge condition 
\beq 
h^{\mu}_{\ \mu}=0 
\eeq 
 
The analogy of $A_0 = 0$ is 
\beq 
h_{0\nu}=h_{\nu 0}=0, 
\eeq

and of the transversality condition     
 
\beq\nabla\cdot {\bf A}=0\eeq     is  
\beq\nabla_i h^{i\nu}=\nabla_i h^{\nu i}=0.\eeq 
 
The Einstein equation (neglecting back-reaction, i.e. the contribution to the energy-momentum tensor by the gravitational field itself) becomes simply 
\beq\Box h_{\mu\nu}=0.\eeq 
 
\subsection{Solutions to the Wave Equation} 
Exactly like for photons we can write for the wave solutions to Einstein's equation 
\beq h_{\mu\nu}=E_{\mu\nu}e^{\pm i\left(\omega t-{\bf k\cdot x}\right)}\eeq 
with    $k^2=\omega^2$, i.e. massless propagation, with the speed of light. (There have been brave attempts to replace Einstein's gravity with a massive theory, with the extra component having extremely small mass. This would lead to many interesting differences, perhaps even explaining the small value of the cosmological constant. So far, there has not appeared any generally accepted way to this this, however.)  
  
\begin{figure}[!htb] 
\begin{center} 
\includegraphics[width=0.9\textwidth]{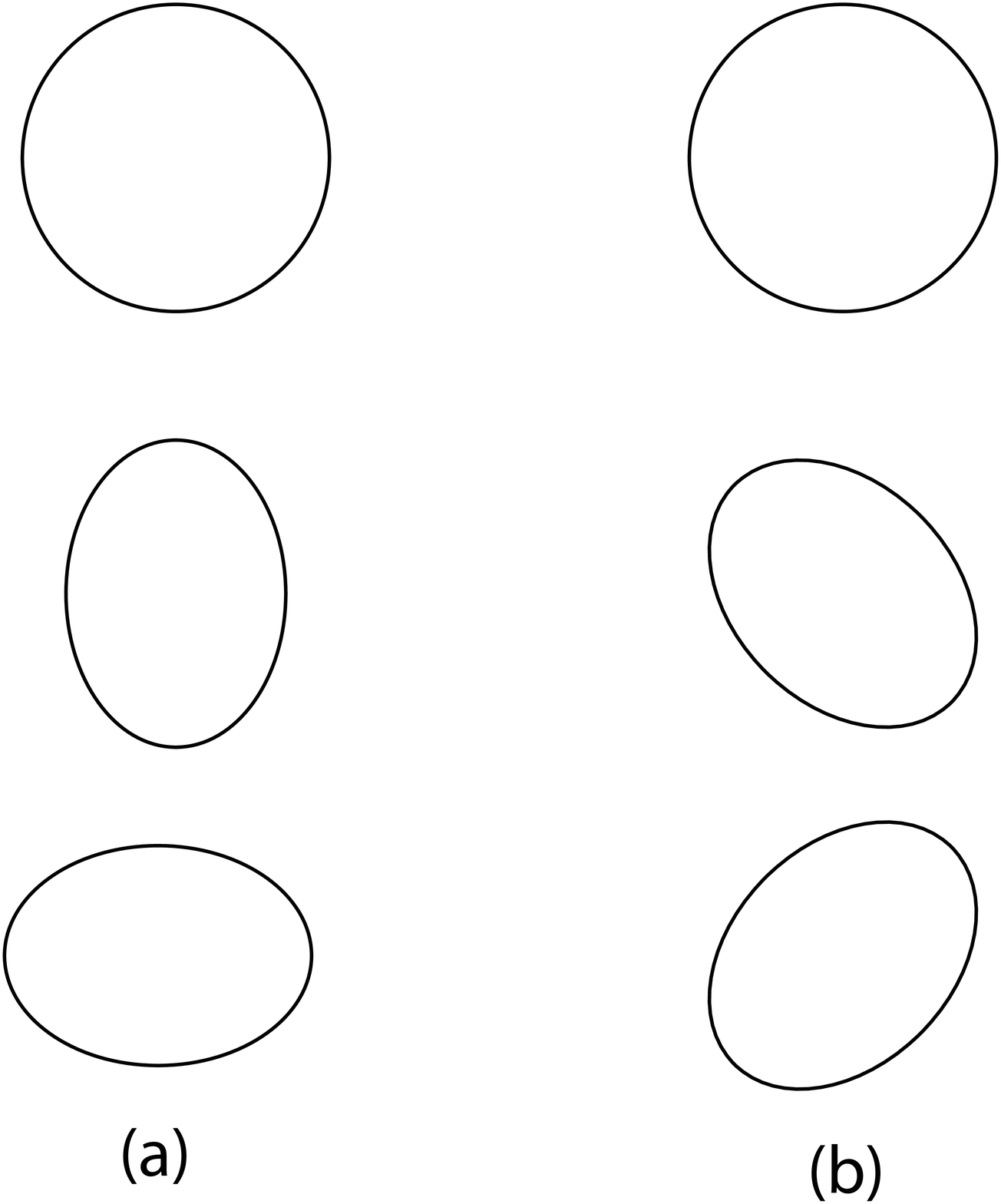}\label{fig:gr1} 
\caption{ (a) The deformation of the unit circle caused by gravity  
waves proportional to  the polarization amplitude $h_{+}$. Shown are 
the unperturbed circle and the maximally stretched configurations  
along the two axes of symmetry, the $x$ and $y$ axes. 
(b) The corresponding pattern for the orthogonal polarization state 
described by the amplitude $h_{\times}$. Note that the axes along  
which stretching and compression occur form 45-degree angles to the  
$x$ and $y$ axes.\label{fig:gravwplus}}\end{center} 
\end{figure}

We can represent $E_{\mu\nu}$ by a $4\times 4$ matrix, which, exactly like for $A_\mu$, should reflect the gauge choice. We know already that the $E_{0\nu}$  row and $E_{\mu 0}$  columns are zero. Also $E$ has to be symmetric in the two indices (since the metric is). Further, $k^iE_{i\nu} = k^jE_{\mu j} = 0$, meaning that also the elements of the $E_{3\nu}$ column and $E_{\mu 3}$ row are zero for a wave propagating in the $z$-direction. So, we really just have zeros for our perturbative solution apart from a symmetric, traceless $2\times 2$ matrix. A general such matrix is a linear combination of  
\beq\label{eq:hplusdef} 
E^+_{\mu\nu}=\left( 
\begin{array}{cccc} 
        0 & 0 & 0 & 0  \\ 
        0 & 1 & 0 & 0  \\ 
        0 & 0 & -1 & 0  \\ 
        0 & 0 & 0 & 0 
\end{array} 
\right) 
\eeq 
and  
\beq 
E^\times_{\mu\nu}=\left( 
\begin{array}{cccc} 
        0 & 0 & 0 & 0  \\ 
        0 & 0 & 1 & 0  \\ 
        0 & 1 & 0 & 0  \\ 
        0 & 0 & 0 & 0 
\end{array} 
\right) 
\eeq 
For a given value of the 3-component $z$, and at time $t$, we can then write 
\beq 
E_{\mu\nu}(t)=h_+(t)E_{\mu\nu}^+ +h_\times(t)E_{\mu\nu}^\times. 
\eeq 
Look at the case  
\beq 
h_+\ne 0,\ \ h_\times=0 .  
\eeq 
At a given time, we have in the unperturbed case 
\beq 
\Delta s^2=\eta_{\mu\nu}\Delta x^\mu \Delta x^\nu=(\Delta t)^2- 
\sum_i (\Delta x^i)^2=-\sum_i (\Delta x^i)^2 
\eeq 
For two diametrically opposed points on the unit circle,  
\beq 
\Delta x^i=(2\cos\theta,2\sin\theta,0) 
\eeq 
and their distance is

\beq 
d_0=\sqrt{-\eta_{ij} (\Delta x^i)(\Delta x^j)}=2\sqrt{\sin^2\theta +\cos\theta^2}=2. 
\eeq 
In the perturbed case (i.e., if a gravity wave passes) 
\beq 
d_+=\sqrt{-(\eta_{ij}+h_+E_{ij}^+)\Delta x^i\Delta x^j}= 
\sqrt{4-h_+(t)4(\cos^2\theta-\sin^2\theta)}\simeq  
\eeq 
\beq 
2-h_+(t)(cos^2\theta-\sin^2\theta)= 2-h_+(t)\cos2\theta. 
\eeq 
For simplicity, we may work with real h by combining as usual the waves with the two signs in the exponential, giving 
\beq h^+_{\mu\nu}=E^+_{\mu\nu}h_+(t)=E^+_{\mu\nu}\cos\left(\omega t - {\bf k}\cdot{\bf r}\right)\eeq 
and we see that the unit circle will be successively ”compressed” or ”squeezed” depending on the sign  of the last factor (see Fig.~\ref{fig:gr1}, where the corresponding  deformation caused by $h_\times$ is also shown). 
 
These are the two independent quadrupole deformations of a circle. This means that the source of the gravitational field giving gravity waves has to have a quadrupole moment. 
From dimensional reasoning, 
\beq 
h\sim \frac{G_N\ddot Q}{d}\sim \frac{4G_NE_{\rm kin}}{d}, 
\eeq 
which is obtained by the crude estimate 
\beq 
Q\sim Ml^2\Rightarrow \dot Q=M2l\dot l= 2Mlv\Rightarrow \ddot Q \sim 2Mv^2 = 4E_{kin}. 
\eeq 
For objects in the Milky Way, typically $d\sim 10$ kpc, and with $E_{\rm kin}\sim M_\odot$ we find 
\beq 
 h\sim 10^{-17}. 
\eeq 
On the other hand, for the  distance appropriate for the Virgo galaxy cluster, 
\beq 
 h\sim 10^{-20}. 
\eeq 
 
These extremely tiny deformations is the reason for the non-detection so far of gravitational radiation, although the are promising objects  
like coalescing neutron stars which should have amplitudes nearing the experimental upper limits.  
 
In a sense, gravity waves have already been indirectly detected, however,  by comparing the slowing-down of the rotation rate of the binary pulsar system PSR 1913-16 by Hulse and Taylor (Nobel Prize 1993): 
\beq 
\frac{dP}{dt}=(-2.4225\pm 0.0056)\cdot 10^{-12}, 
\eeq 
with the general relativistic calculation (with energy loss due to gravitational radiation): 
 
\beq 
\frac{dP_{GR}}{dt}=-2.40\cdot 10^{-12}. 
\eeq  
This excellent agreement has put severe limits on possible  modifications of Einstein gravity. But effects of gravity waves have so far never been detected directly on Earth,  despite an impressive increase in sensitivity of the LIGO experiment in the US, and VIRGO in Italy. Actually by combining several experiments and searching for time-coincident effects, one may both decrease various noise sources and increase the sensitivity for a signal. This is presently done by LIGO, VIRGO and GEO600 in Germany. All three detector are presently being upgraded to more advanced versions. However, it may be that a space experiment, LISA, will be needed to detect a significant signal. Its status is, however, at present unclear due to the difficult  financial situation in most countries of the world.  
 
We finally remind that there is also a possibility of  detecting 
gravitational waves that are relics of dramatic processes in the early  
universe, such as during the epoch of inflation or during the  
formation of cosmic strings, if such exist. In that case, the most  
promising method is through analyzing the imprints they have made in  
the cosmic microwave background radiation (CMBR). As 
gravitational waves carry a quadrupole moment it is possible to  
distinguish their effects through studies of CMBR  polarization. With  
the planned Planck satellite there will be a possibility of searching  
for gravitational waves of very long wavelength generated through  
these hypothetical processes.  Results are expected in early 2013. 
 
\section{Conclusions} 
This finishes our trip through the universe, looking at fundamental processes 
of similar interest to particle physicists, cosmologists, astrophysicists and 
astroparticle physicists alike. As hopefully has becomes clear, by combining 
information from all messengers that we have available: photons of all wavelengths, neutrinos, antimatter and perhaps gravitational waves, we may study from the Earth some of the most energetic and interesting processes in the universe. If we are lucky, we may even solve the problem of the nature of the dark matter, which has been with us since the times of Fritz Zwicky. Let us remind ourselves of his prophetic words from 1933 \cite{zwicky}, after observing the rapid movement of the galaxy members of  
the Coma cluster, which pointed to an overdensity of matter in the cluster: 
 
{\em    
If this over-density is confirmed we would arrive at the astonishing conclusion that dark matter is present with a much greater density than luminous matter\ldots 
} 
\section{{Acknowledgements}} 
The author is  grate\-ful to several colleagues, including 
G. Bertone, T. Bringmann, J. Conrad, J.~Edsj\"o, P. Gondolo, A. Goobar, P.O.~Hulth,  
E.~M\"ortsell, F. Aharonian and C. Dermer as well as the participants of the 2010 Saas-Fee Winter School for many useful suggestions and discussions. A particular thanks to the organizers of the School, chaired by Roland Walter, for making the School such an enjoyable and memorable event.

\end{document}